\documentclass[superscriptaddress,onecolumn,pre,nofootinbib]{revtex4}

\usepackage{amsmath}
\usepackage{graphicx,color}
\usepackage{amssymb}
\usepackage{amsfonts}
\usepackage{bm}

\newcommand{\be}{\begin{equation}}
\newcommand{\ee}{\end{equation}}
\newcommand{\bea}{\begin{eqnarray}}
\newcommand{\eea}{\end{eqnarray}}

\begin{document}
\sloppy


\title{Generalized Euler, Smoluchowski and Schr\"odinger
equations admitting\\
self-similar solutions with a Tsallis invariant profile}

\author{Pierre-Henri Chavanis}
\affiliation{Laboratoire de
Physique Th\'eorique, Universit\'e de Toulouse, CNRS, UPS, France }

\begin{abstract}

The damped isothermal Euler equations, the Smoluchowski equation and the damped
logarithmic Schr\"odinger equation with a harmonic potential admit stationary
and self-similar solutions with a Gaussian profile. They satisfy an
$H$-theorem for a free energy functional involving the von Weizs\"acker 
functional and the Boltzmann functional. We derive generalized
forms of these equations in order to obtain  stationary
and self-similar solutions with a Tsallis profile. In particular,
we introduce a nonlinear Schr\"odinger equation involving a
generalized kinetic term characterized by an index $q$ and a power-law
nonlinearity characterized by an index $\gamma$. We derive  an $H$-theorem
satisfied by a generalized free energy functional involving a generalized von
Weizs\"acker  functional (associated with $q$) and a Tsallis functional
(associated with
$\gamma$). This leads to a notion of generalized
quantum mechanics
and 
generalized thermodynamics. When $q=2\gamma-1$, our nonlinear Schr\"odinger
equation admits an exact self-similar solution with a Tsallis invariant profile.
Standard  quantum mechanics (Schr\"odinger) and standard 
thermodynamics (Boltzmann) are recovered for
$q=\gamma=1$.

\end{abstract}

\maketitle

\section{Introduction}

In the standard theory of  Brownian motion initiated by Einstein
\cite{einstein}, the
probabilistic evolution of a free particle (or an ensemble of noninteracting
particles) in the overdamped limit is governed
by the ordinary
diffusion equation. When the particle is submitted to an external potential, one
gets the Smoluchowski \cite{smoluchowski} equation. These are both
particular
cases of the Fokker-Planck \cite{fokker0,fokker,planck} equation. These
equations are associated with
standard thermodynamics \cite{risken} in the canonical ensemble based on
the Boltzmann entropy
and on the isothermal
equation of state $P=\rho k_BT/m$. They satisfy an $H$-theorem associated with
the Boltzmann
free energy and relax towards the Boltzmann distribution at
statistical equilibrium (when an
external potential is present to counteract the effect of diffusion). The
equilibrium state minimizes the Boltzmann free energy at fixed mass. On the
other hand, the diffusion equation and the Smoluchowski equation
with a harmonic potential admit a self-similar solution with a Gaussian
invariant profile \cite{einstein,smoluchowski,uo,revuechandra}.

In the laste decades, several
authors
\cite{hc,reyni,mittal,st,sm,tsallis,barenblatt,spohn,kq,pp,tb,csr,abe,stariolo,
martinez,borland,br,k0,fd,k1,frank1,cn,gen,gen2,nobre,kl,naudts,chavcourt,cll,
lang,banach,k2,cst,frank,bose,degrad,cras,csbio,logotrope,langcrit,sw1,sw2,nfp,
sw3,tsallisbook,entropy} have
tried to generalize standard thermodynamics and Brownian theory
(see, e.g., \cite{frank,tsallisbook,nfp,entropy} for
reviews).
For example,
one can encounter situations in which the diffusion coefficient of the
particles depends on their density as a power-law. In the
absence of an external potential one gets the anomalous diffusion equation which
was first introduced in the context of porous media \cite{barenblatt}. When an
external potential
is present one gets the polytropic Smoluchowski equation \cite{pp}. These are
both
particular cases of nonlinear Fokker-Planck
equations \cite{frank,tsallisbook,nfp,entropy}.
These equations are associated with a notion of generalized thermodynamics in
the canonical ensemble based
on the Tsallis entropy \cite{tsallis} and on the polytropic equation of state
$P=K\rho^{\gamma}$. 
They satisfy
an $H$-theorem associated with the Tsallis
free energy and relax towards the Tsallis distribution at statistical
equilibrium
(when an
external potential is present to counteract the effect of
diffusion). The
equilibrium state minimizes the Tsallis free energy at fixed
mass.\footnote{Initially, nonlinear Fokker-Planck equations were introduced in
relation to the Fermi-Dirac \cite{kq,csr}, Bose-Einstein  \cite{kq} and 
Tsallis \cite{pp,tb} entropies. This gave the impression that these entropies
 were special and that the generalized thermodynamical
formalism was valid  only for them. However, it was shown later  by
\cite{martinez,k0,fd,k1,frank1,gen,gen2,degrad,sw1,sw2,nfp,sw3,entropy}
that a generalized thermodynamical formalism could be developed for an arbitrary
form of entropy and for an arbitrary barotropic equation of state. In that case,
the
diffusion coefficient that appears in the corresponding Fokker-Planck equation
is a nonlinear function of the density determined by the form of entropy or by
the
equation of state (see Refs. \cite{nfp,entropy}).} On the other hand, it has
been shown that the anomalous diffusion equation and the polytropic
Smoluchowski equation with a harmonic potential admit a self-similar solution
with a Tsallis invariant profile \cite{barenblatt,pp,tb,logotrope}.  

In this paper, we first show that these properties remain valid for the Euler
and damped Euler equations with an isothermal or a polytropic equation of
state. The
Smoluchowski equation is recovered in the strong friction limit
($\xi\rightarrow +\infty$). These equations admit self-similar
solutions with a Gaussian or a Tsallis invariant profile. Interestingly, the
differential equation determining the evolution of the system's size bears
some analogies with the Friedmann equation in cosmology determining the
evolution of the radius (scale factor) of the Universe.
We then generalize these results to a quantum
mechanics context
related to generalized damped Gross-Pitaevskii (GP) equations
\cite{ggp}. We
first consider the  damped logarithmic GP equation. Using the 
Madelung transformation \cite{madelung}, we show that this equation is
equivalent to the
quantum damped isothermal Euler equations.\footnote{In the absence of friction
($\xi=0$) we get the quantum
isothermal Euler
equations and in the  strong friction limit ($\xi\rightarrow +\infty$) we get
the quantum Smoluchowski equation \cite{qs}. This provides
an interesting formal
connection between quantum mechanics and Brownian theory \cite{ggp}.} These
equations admit a 
self-similar solution with a Gaussian invariant profile. This is because the
standard quantum potential present in the  quantum Euler
equations has a structure compatible with the linear equation of state giving
rise to the Gaussian solution (they ``marry well''). We then consider the 
damped power-law GP
equation. Using the 
Madelung transformation \cite{madelung}, we show that this equation is
equivalent to the
quantum damped polytropic Euler equations. In the Thomas-Fermi (TF) limit,
where
the quantum
potential can be neglected, they reduce to the damped polytropic Euler
equations and therefore admit a self-similar solution with a Tsallis invariant
profile. However, when the standard quantum potential is taken into account,
this property is lost because the
structure of the standard quantum potential present in the  quantum Euler
equations is not compatible with the polytropic equation of state giving
rise to the Tsallis solution (they do not  ``marry well''). We search and find a
generalized form of quantum potential
that allows us to restore the Tsallis self-similar solution. This leads to a
generalized Schr\"odinger equation associated with a notion of generalized
quantum (or wave) mechanics.\footnote{Generalized
Schr\"odinger equations have also been proposed recently by
Nobre {\it et al.} \cite{nrt1,nrt2}. However, their approach is substantially
different from ours (see the Conclusion).} More
precisely, we introduce a generalized
damped power-law  GP equation involving a generalized
kinetic term characterized by an index $q$ and a power-law
nonlinearity characterized by an index $\gamma$. This equation is
associated with a notion of generalized quantum mechanics (through the index
$q$) and generalized Tsallis thermodynamics (through the index $\gamma$).
Indeed, an
$H$-theorem is satisfied by a generalized free energy functional involving a
generalized von
Weizs\"acker  functional (associated with $q$) and a Tsallis functional
(associated with $\gamma$). The equilibrium state of
this equation minimizes the  generalized free energy functional under the
normalization condition. Finally, this equation
admits a self-similar
solution with a Tsallis invariant profile provided that the condition
$q=2\gamma-1$ is fulfilled. Standard  quantum mechanics (Schr\"odinger) and
standard 
thermodynamics (Boltzmann) are recovered for
$q=\gamma=1$.

In the main part of the paper, we explain our approach, present the basic
equations under study, and exhibit their self-similar solutions.  General
results of our formalism and technical details are given in the Appendices. A
detailed study of the self-similar solutions obtained in our paper is
postponed to a forthcoming contribution \cite{prep}.

\section{Self-similar solution of the Euler equations}
\label{sec_euler}

\subsection{Euler equations}
\label{sec_eulerseul}

Let us consider the Euler equations
\begin{eqnarray}
\label{e1}
\frac{\partial\rho}{\partial t}+\nabla\cdot (\rho {\bf u})=0,
\end{eqnarray}
\begin{eqnarray}
\label{e2}
\frac{\partial {\bf u}}{\partial t}+({\bf u}\cdot \nabla){\bf
u}=-\frac{1}{\rho}\nabla P-\nabla\Phi_{\rm ext}
\end{eqnarray}
with a barotropic equation of state $P=P(\rho)$ and an external potential
$\Phi_{\rm ext}({\bf r})$ in a space of dimension $d$. Eq.
(\ref{e1}) is the
equation of continuity which expresses the local conservation of mass.
Eq. (\ref{e2}) is the momentum equation.
In the following, we specifically
consider a harmonic (quadratic) potential 
\begin{eqnarray}
\label{e3}
\Phi_{\rm ext}=\frac{1}{2}\omega_0^2r^2
\end{eqnarray}
and a polytropic equation of state of the form
\begin{eqnarray}
\label{e4}
P=K\rho^{\gamma},
\end{eqnarray}
where $\gamma$ is the polytropic index and $K$ is the polytropic constant
(sometimes called the polytropic temperature\footnote{A physical
justification of this terminology is given in Appendix \ref{sec_t}.}). This
includes, as a particular
case, the isothermal
equation of state
\begin{eqnarray}
\label{e5}
P=\rho\frac{k_B T}{m}
\end{eqnarray}
corresponding to $\gamma=1$ and $K=k_B T/m$. In most applications, we will
assume that $\omega_0^2$, $K$ and $T$ are
positive constants ($\ge 0$).

{\it Remark:} We may also consider the logotropic equation of
state \cite{logotrope}:
\begin{eqnarray}
\label{logo}
P=A\ln\rho.
\end{eqnarray}
Because of the identities  $A\nabla
\ln\rho=(A/\rho)\nabla\rho$ and $K\nabla
\rho^{\gamma}=K\gamma\rho^{\gamma-1}\nabla\rho$, the logotropic equation of
state can be viewed as a limit of the  polytropic equation of
state $P=K\rho^{\gamma}$ when $\gamma\rightarrow 0$, $K\rightarrow +\infty$ with
$K\gamma=A$ 
\cite{logotrope,ggp}.

\subsection{Self-similar solution}
\label{sec_seus}

The
Euler equations (\ref{e1})-(\ref{logo}) admit a
self-similar solution
of the form (see Appendix \ref{sec_sss})
\begin{eqnarray}
\label{e6}
\rho({\bf r},t)=\frac{M}{R(t)^d} f\left\lbrack \frac{{\bf
r}-\chi(t){\bf r}_0}{R(t)}\right
\rbrack,\qquad
{\bf u}({\bf r},t)=H(t){\bf r}+B(t){\bf r}_0
\end{eqnarray}
with
\begin{eqnarray}
\label{e6b}
H=\frac{\dot R}{R},\qquad B=\dot \chi-H\chi,
\end{eqnarray}
where $R(t)$ is the typical size (radius) of the system and $f({\bf
x})$ with ${\bf x}=[{\bf r}-\chi(t){\bf r}_0]/R(t)$ is the invariant density
profile. The density profile is spherically symmetric and   contains all the
mass ($\int \rho({\bf r},t)\, d{\bf r}=M$) so that  $\int f({\bf x})\, d{\bf
x}=1$. The quantity $\langle {\bf
r}\rangle(t)=\chi(t){\bf r}_0$ represents the position of the center of the
distribution (see Appendix \ref{sec_wf}). We take $\chi(0)=1$ so that ${\bf
r}_0$
represents the initial position of the center of the
distribution. The velocity field ${\bf u}({\bf r},t)$ is an affine function of
${\bf r}$ with
time-dependent factors $H(t)$ and $B(t)$. When $B=0$, the velocity field
is proportional to ${\bf r}$ with a proportionality factor $H(t)=\dot R/R$.
We note the formal analogy with the Hubble constant in cosmology, where $R$
plays the role of the scale factor. The  invariant density profile is
given by the Tsallis
distribution of index $\gamma$, namely
\begin{eqnarray}
\label{e8}
f(x)=\frac{1}{Z}\left\lbrack 1-(\gamma-1)x^2\right\rbrack_+^{1/(\gamma-1)},
\end{eqnarray}
where $[x]_+=x$ if $x\ge 0$ and $[x]_+=0$ if $x\le 0$. When $\gamma>1$,
the density profile has a compact support, vanishing at $x_{\rm
max}=1/\sqrt{\gamma-1}$. When $\gamma<1$, the density extends to
infinity and decreases at large distances as $f(x)\sim x^{-2/(1-\gamma)}$.
When $\gamma\rightarrow 1$, $f(x)$ tends towards the Gaussian distribution
(see below). The normalization
constant $Z$ is given by 
\begin{eqnarray}
\label{e8a}
Z=\frac{\pi^{d/2}\Gamma\left (\frac{\gamma}{\gamma-1}\right
)}{(\gamma-1)^{d/2}\Gamma\left (\frac{d}{2}+\frac{\gamma}{\gamma-1}\right
)}\qquad (\gamma\ge
1),
\end{eqnarray}
\begin{eqnarray}
\label{e8b}
Z=\frac{\pi^{d/2}\Gamma\left (\frac{1}{1-\gamma}-\frac{d}{2}\right
)}{(1-\gamma)^{d/2}\Gamma\left (\frac{1}{1-\gamma}\right
)}\qquad \left (\frac{d-2}{d}<\gamma\le 1\right ).
\end{eqnarray}
The distribution is normalizable provided that $\gamma>
(d-2)/d$. For the logotropic equation of state (\ref{logo}) in $d=1$, the
invariant profile $f(x)$ is the Lorentzian
corresponding to Eq.  (\ref{e8}) with $\gamma=0$ \cite{logotrope}. The
differential
equation
determining
the evolution of $\chi(t)$ is given by Eq. (\ref{chi1}) and the differential
equation
determining
the evolution of the radius $R(t)$ is given by
\begin{eqnarray}
\label{e9}
\ddot R+\omega_0^2
R=2Z^{1-\gamma}K\gamma\frac{M^{\gamma-1}}{R^{d\gamma-d+1}}.
\end{eqnarray}
For the logotropic equation of state (\ref{logo}) in $d=1$,
the term on the right hand side of Eq. (\ref{e9}) is a constant.
In the particular case of an
isothermal (linear) equation of state, the invariant profile is the Gaussian 
\begin{eqnarray}
\label{e10}
f(x)=\frac{1}{\pi^{d/2}}e^{-x^2}
\end{eqnarray}
and the differential equation (\ref{e9}) determining the evolution of the
radius reduces to 
\begin{eqnarray}
\label{e11}
\ddot R+\omega_0^2 R=\frac{2k_B T}{mR}.
\end{eqnarray}
Eqs. (\ref{e10}) and (\ref{e11}) are limiting cases of Eqs.
(\ref{e8})-(\ref{e9}) for $\gamma\rightarrow 1$.  In most applications, we will
assume that $R=0$ at $t=0$ so that $\rho({\bf r},0)=M\delta({\bf
r}-{\bf r}_0)$.

{\it Historical note:} As far as we
know, these self-similar solutions have not been given previously in the context
of the Euler
equations.

\subsection{Steady state}
\label{sec_stse}

A steady state of the Euler equations
(\ref{e1}) and (\ref{e2}) satisfies the condition of hydrostatic equilibrium
\begin{eqnarray}
\label{e2b}
\nabla P+\rho\nabla\Phi_{\rm ext}={\bf 0}.
\end{eqnarray}
The time-independent distribution (\ref{e6}), corresponding to
$\chi(t)=0$, $R(t)=R_e={\rm cst}$ and ${\bf u}={\bf 0}$, when it exists, is a
steady state of the Euler equations (\ref{e1}) and (\ref{e2}) with a harmonic
potential. For the
polytropic equation of state (\ref{e4}), assuming  $K>0$, 
$\omega_0^2>0$ and $\gamma\ge {\rm max}\lbrace
0,(d-2)/d\rbrace$, we obtain the Tsallis distribution 
\begin{eqnarray}
\label{e12}
\rho({\bf r})=\frac{M}{Z R_e^d}\left\lbrack
1-(\gamma-1)\frac{r^2}{R_e^2}\right\rbrack_{+}^{\frac{1}{\gamma-1}}, 
\end{eqnarray}
where
\begin{eqnarray}
\label{e12bis}
R_e=\left (\frac{2K\gamma M^{\gamma-1}}{Z^{\gamma-1}\omega_0^2}\right
)^{\frac{1}{2+d(\gamma-1)}}
\end{eqnarray}
is the stationary solution of Eq. (\ref{e9}). 
For the isothermal equation of state (\ref{e5}), assuming $T>0$ and 
$\omega_0^2>0$, we obtain the Boltzmann
distribution
\begin{eqnarray}
\label{e12b}
\rho({\bf r})=\frac{M}{\pi^{d/2} R_e^d}e^{-r^2/R_e^2}, 
\end{eqnarray}
where 
\begin{eqnarray}
\label{e12bbis}
R_e=\left
(\frac{2k_BT}{m\omega_0^2}\right
)^{1/2}
\end{eqnarray}
is the stationary solution of Eq. (\ref{e11}). By directly solving
the differential equation (\ref{e2b}) corresponding to the condition of
hydrostatic
equilibrium  one can show that all the steady
states of the Euler equations (\ref{e1}) and (\ref{e2}) are
of that form (see Appendix \ref{sec_eq}). The dynamical stability of a steady
state of the Euler
equations (\ref{e1}) and (\ref{e2}) can be determined by studying the dynamical
stability of the steady state of the differential equations (\ref{e9}) and
(\ref{e11}) \cite{prep}.  It can be shown that the steady states defined by Eqs.
(\ref{e12})-(\ref{e12bbis}) are stable.

\section{Self-similar solution of the damped Euler equations and of
the generalized Smoluchowski equation}
\label{sec_de}

\subsection{Damped Euler equations and generalized Smoluchowski
equation}

In the recent years, generalized thermodynamics and nonlinear Fokker-Planck
(NFP)
equations have played an important role in physics (see the reviews 
\cite{frank,tsallisbook,nfp,entropy} and references therein). NFP equations can
be
introduced in the following
heuristic manner which generalizes the arguments given by
Einstein \cite{einstein} in his seminal paper on Brownian theory. We
start from the Euler equations (\ref{e1}) and (\ref{e2})
and introduce a
linear friction force $-\xi {\bf u}$ in the momentum equation (\ref{e2}). This
leads to the damped Euler equations
\begin{eqnarray}
\label{de1}
\frac{\partial\rho}{\partial t}+\nabla\cdot (\rho {\bf u})=0,
\end{eqnarray}
\begin{eqnarray}
\label{de2}
\frac{\partial {\bf u}}{\partial t}+({\bf u}\cdot \nabla){\bf
u}=-\frac{1}{\rho}\nabla P-\nabla\Phi_{\rm ext}-\xi{\bf u}.
\end{eqnarray}
These hydrodynamic equations describe the motion of a fluid of
particles (like Brownian particles) experiencing a friction with another
fluid. In his theory of Brownian motion, Einstein considers an isothermal
equation of state, the so-called osmotic
pressure $P=\rho k_B T/m$, in agreement with standard (Boltzmann)
thermodynamics. Let us be
more
general and consider an arbitrary barotropic
equation of state of the
form $P=P(\rho)$. In the
strong friction
limit $\xi\rightarrow +\infty$, we can neglect the inertia of the particles,
i.e., the left hand side of Eq. (\ref{de2}). In that case, the current of
particles is given by 
\begin{eqnarray}
\label{de2b}
\rho{\bf u}=-\frac{1}{\xi}(\nabla
P+\rho\nabla\Phi_{\rm ext}).
\end{eqnarray}
Combining this relation with the continuity equation (\ref{de1}) we obtain the
generalized
Smoluchowski equation 
\begin{equation}
\label{de3}
\xi\frac{\partial\rho}{\partial t}=\nabla\cdot\left (\nabla
P+\rho\nabla \Phi_{\rm ext}\right ).
\end{equation}
For the isothermal equation of state (\ref{e5}), we recover the standard
Smoluchowski \cite{smoluchowski} equation
\begin{equation}
\label{de4}
\xi\frac{\partial\rho}{\partial t}=\nabla\cdot\left (\frac{k_B T}{m}\nabla
\rho+\rho\nabla \Phi_{\rm ext}\right )
\end{equation}
which is a particular Fokker-Planck equation
\cite{fokker0,fokker,planck}. For the polytropic equation of state (\ref{e4}),
we obtain  the polytropic
Smoluchowski
equation
\begin{equation}
\label{de5}
\xi\frac{\partial\rho}{\partial t}=\nabla\cdot\left (K\nabla
\rho^{\gamma}+\rho\nabla \Phi_{\rm ext}\right ).
\end{equation}

{\it Historical note:} The generalized Smoluchowski equation (\ref{de3}) was
introduced in
\cite{csr,gen,gen2}. It can be derived, in a strong friction
limit $\xi\rightarrow +\infty$, either from the damped  Euler
equations
(\ref{de1}) and (\ref{de2}) as we have done above (see also Sec. 4.9 of
\cite{nfp}) or from the generalized
Kramers equation (see a rigorous derivation
based on the Chapman-Enskog expansion in \cite{cll} and a derivation based on
hydrodynamic moment equations in Sec. 4.8 of \cite{nfp}).  We stress
that the damped Euler
equations and the generalized Kramers equation are not equivalent to each
other so
they
describe different physical systems.\footnote{From the  generalized Kramers
equation one can derive a hierarchy of hydrodynamic equations for the moments of
the velocity distribution, but this hierarchy is not closed. It can be closed by
making a local thermodynamic equilibrium (LTE) approximation leading to the
damped Euler equations (\ref{de1}) and (\ref{de2}) \cite{gen}. However, it
can be shown
\cite{physica} that this LTE approximation is not rigorously justified except in
the strong friction limit
$\xi\rightarrow +\infty$ in which the  damped Euler
equations (\ref{de1}) and (\ref{de2}) reduce to the generalized Smoluchowski
equation (\ref{de3}).} In this paper, we consider systems described by the
damped Euler equations.

\subsection{Generalized thermodynamics}
\label{sec_gtl}

Equations (\ref{de1})-(\ref{de5}) are associated with a
generalized
thermodynamic formalism. In particular, they satisfy an $H$-theorem for the
generalized free energy (see  Refs. \cite{nfp,ggp} and
Appendix
\ref{sec_gw})
\begin{eqnarray}
\label{de5a}
F=\int\rho \frac{{\bf u}^2}{2}\, d{\bf r}+\int
V(\rho)\, d{\bf r}+\int\rho\Phi_{\rm ext}\, d{\bf r},
\end{eqnarray}
where the potential $V(\rho)$ is determined by the equation of state $P(\rho)$
through the relation\footnote{We note that
$u(\rho)\equiv V(\rho)/\rho$
represents the density of
internal energy ($U=\int V(\rho)\, d{\bf r}=\int \rho u\, d{\bf r}$). It
satisfies the first principle of thermodynamics
$du=-Pd(1/\rho)$, or equivalently, $u'(\rho)=P(\rho)/\rho^2$. The density of
enthalpy is defined by $h(\rho)=u(\rho)+P(\rho)/\rho$ (Euler relation) and it
satisfies the Gibbs-Duhem relation  $dh=dP/\rho$, or equivalently,
$h'(\rho)=P'(\rho)/\rho$. We note that $V'(\rho)=h(\rho)$. Defining
$C(\rho)=V(\rho)/T_{\rm eff}$, where $T_{\rm eff}$ is an effective temperature,
we can write the free energy (\ref{de5a}) as $F=E_*-T_{\rm eff} S$ where
$S=-\int C(\rho)\, d{\bf r}$ is a generalized entropy and $E_*=\Theta_c+W_{\rm
ext}$ is the sum of the macroscopic kinetic energy and the external energy (see
Appendix \ref{sec_t}). It
can be shown \cite{forthcoming} that this expression of the free energy arises
naturally from {\it ordinary} thermodynamics in the canonical
ensemble.} 
\begin{equation}
\label{de5binv}
V(\rho)=\rho\int^{\rho}\frac{P(\rho')}{{\rho'}^2}\, d\rho'.
\end{equation}
Inversely, the equation of state $P(\rho)$ is related to the potential $V(\rho)$
by
\begin{equation}
\label{de5b}
P(\rho)=\rho
V'(\rho)-V(\rho)=\rho^2\left\lbrack
\frac{V(\rho)}{\rho}\right\rbrack'.
\end{equation}

An extremum of free energy at fixed mass determines
a steady state of the (damped) Euler equations (\ref{de1}) and (\ref{de2})
satisfying the
condition of hydrostatic equilibrium (\ref{e2b}). A minimum of free energy is
stable while a maximum or a saddle point is unstable. When $\xi>0$ we can use
the $H$-theorem ($\dot F\le 0$)  to prove that the system relaxes towards a
stable steady state (when it exists) for $t\rightarrow +\infty$. The typical
relaxation time is $t_{\rm relax}\sim \xi^{-1}$. When
$\xi=0$ the free energy is
conserved ($\dot F= 0$). 

For the isothermal equation of state (\ref{e5}) the
free energy can be written as $F=E_*-TS_B$ where $S_B$ is the Boltzmann entropy
(\ref{t1}). For the polytropic equation of state
(\ref{e4}) the generalized free energy can be written
as $F=E_*-KS_{\gamma}$ where $S_{\gamma}$ is the Tsallis entropy (\ref{t3}) of
index
$\gamma$. For the logotropic equation of state
(\ref{logo}) the generalized free energy can be written
as $F=E_*-AS_{L}$ where $S_{L}$ is the logarithmic entropy (\ref{logent}). The
extremization of $F$ at fixed mass $M$ leads to
the Boltzmann  and Tsallis distributions (\ref{e12b}) and
(\ref{e12}) respectively (see Refs. \cite{nfp,ggp} and Appendix
\ref{sec_gw}).

\subsection{Isothermal case}

The damped Euler equations (\ref{de1}) and (\ref{de2}) with the isothermal
equation of state (\ref{e5}) and the  harmonic potential (\ref{e3}) admit
a self-similar solution
of the form (\ref{e6}) with the Gaussian invariant profile (\ref{e10}) (see
Appendix \ref{sec_sss}). The
differential equation determining the evolution of $\chi(t)$ is given by Eq.
(\ref{chi1})  and the
differential equation determining the evolution of the
radius $R(t)$ is given by
\begin{eqnarray}
\label{de6}
\ddot R+\xi \dot
R+\omega_0^2 R=\frac{2k_B T}{mR}.
\end{eqnarray}
In the strong friction limit $\xi\rightarrow +\infty$, corresponding to the
Smoluchowski
equation (\ref{de4}), the differential equation determining the evolution of
$\chi(t)$ is given by Eq.
(\ref{chi5}) and the differential equation determining the evolution of the
radius $R(t)$ is given by
\begin{eqnarray}
\label{de7}
\xi \dot
R+\omega_0^2 R=\frac{2k_B T}{mR}.
\end{eqnarray}
When $T>0$ and 
$\omega_0^2>0$, the steady states of the damped Euler equations and of the
Smoluchowski equation are the same as those discussed in Sec.
\ref{sec_stse}.
They are given by Eqs. (\ref{e12b}) and (\ref{e12bbis}) and they are stable.
When $\xi>0$, the
system relaxes towards these equilibrium states.

\subsubsection{Without external potential}

In the absence of external potential, Eq. (\ref{de4})
reduces to the standard
diffusion
equation
\begin{equation}
\label{de8}
\frac{\partial\rho}{\partial t}=D\Delta\rho,
\end{equation}
where the diffusion coefficient is given by the Einstein \cite{einstein}
relation
\begin{equation}
\label{de9}
D=\frac{k_B T}{\xi m}.
\end{equation}
The standard
diffusion
equation has a Gaussian self-similar solution.
The differential equation (\ref{chi5})  determining the evolution of
$\chi(t)$ reduces to Eq. (\ref{chi7}). Its solution is given by Eq.
(\ref{chi8}). 
The differential equation (\ref{de7}) determining
the evolution of the
radius $R(t)$ reduces to
\begin{eqnarray}
\label{de10}
\dot R=\frac{2D}{R}.
\end{eqnarray}
Its solution is given by
\begin{eqnarray}
\label{de11}
R(t)=(4Dt)^{1/2}.
\end{eqnarray}
Combining Eqs. (\ref{e6}), (\ref{e10}), (\ref{de11}) and
(\ref{chi8}) we get 
\begin{eqnarray}
\label{de11b}
\rho({\bf r},t)=\frac{M}{(4\pi Dt)^{d/2}}e^{-\frac{({\bf r}-{\bf r}_0)^2}{4Dt}}.
\end{eqnarray}
Using $\langle r^2\rangle =(d/2)R^2+r_0^2$ obtained
from Eqs.
(\ref{chi8}), (\ref{rr2}) and (\ref{rr3}), and using Eq. (\ref{de11}), we find
that 
\begin{eqnarray}
\label{de10j}
\langle
r^2\rangle(t)=2dDt+r_0^2.
\end{eqnarray}
This result
can be obtained immediately from the virial theorem (\ref{vir10}). We also
have $\langle {\bf r}(t)\rangle={\bf r}_0$ from Eqs. (\ref{chi8}) and
(\ref{rr2n}).

{\it Historical note:} The diffusion equation (\ref{de8}) was derived by
Einstein
\cite{einstein} starting from a Markovian equation and expanding this equation
in powers of the increment $\Delta {\bf r}$ in the position of the Brownian
particle.\footnote{This is the original method that
led, later, to the Fokker-Planck equation \cite{fokker0,fokker,planck} when
one takes into account an
additional deterministic drift term, like the
effect of an external potential, in the evolution equation of the particle 
\cite{revuechandra}.} He gave the
solution of this equation [see Eq.
(\ref{de11b})] and, from it, obtained  the formula (\ref{de10j}) giving the
temporal evolution of the arithmetic mean of the squares of displacements of the
Brownian particles. He also derived the Einstein relation (\ref{de9})
between the diffusion
coefficient, the friction coefficient, and the temperature. Actually, the same
relation was previously obtained by Sutherland \cite{sutherland} using
essentially the same arguments.\footnote{For that
reason, Eq. (\ref{de9}) was originally called the Sutherland-Einstein relation
(see Ref. \cite{smoluchowski16}, P. 569).} A similar
relation was also derived by Smoluchowski \cite{smoluchowski06}. On the
other hand, Eq. (\ref{de10j}) was derived by Langevin
\cite{langevin} from the Langevin equation.

\subsubsection{With a harmonic potential}

The  Smoluchowski equation (\ref{de4}) with the harmonic external potential
(\ref{e3}) can be written as
\begin{equation}
\label{de12}
\xi\frac{\partial\rho}{\partial t}=\nabla\cdot\left (\frac{k_B T}{m}\nabla
\rho+\rho\omega^2_0 {\bf r}\right ).
\end{equation}
It has a Gaussian self-similar solution.
The solution of the differential equation (\ref{chi5})  determining the
evolution of
$\chi(t)$ is given by Eq.
(\ref{chi6}). The solution of the 
differential equation (\ref{de7}) determining the
evolution of the radius $R(t)$ is given by 
\begin{eqnarray}
\label{de13}
R(t)=\left (\frac{2k_B T}{m\omega_0^2}\right )^{1/2}\left
(1-e^{-2\omega_0^2t/\xi}\right )^{1/2}.
\end{eqnarray}
For $t\rightarrow
+\infty$, we recover the equilibrium radius (\ref{e12bbis}).
Combining Eqs. (\ref{e6}), (\ref{e10}), (\ref{de13}) and (\ref{chi6}) we
get
\begin{eqnarray}
\label{de13b}
\rho({\bf r},t)=M\left\lbrack \frac{m\omega_0^2}{2\pi k_B T
\left (1-e^{-2\omega_0^2t/\xi}\right )}\right\rbrack^{d/2}e^{-\frac{m\omega_0^2
\left ({\bf r}-e^{-\omega_0^2t/\xi}{\bf r}_0\right )^2}{2k_B T
\left (1-e^{-2\omega_0^2t/\xi}\right )}}.
\end{eqnarray}
For $t\rightarrow +\infty$, this equation
tends towards the
stationary solution
\begin{eqnarray}
\label{de13c}
\rho({\bf r})=M\left (\frac{m\omega_0^2}{2\pi k_B T}\right
)^{d/2}e^{-\frac{m\omega_0^2
r^2}{2k_B T}}.
\end{eqnarray}
equivalent to Eq. (\ref{e12b}). It corresponds to the Boltzmann
distribution of statistical equilibrium for
an ensemble of harmonic oscillators (comparing the steady state
of the Smoluchowski equation with the Boltzmann
distribution is the direct manner to establish the Einstein relation (\ref{de9})
\cite{revuechandra}).
Using Eq. (\ref{de13}) and  $\langle r^2\rangle
=(d/2)R^2+{\rm
exp}(-2\omega_0^2t/\xi) r_0^2$ obtained from Eqs.
(\ref{chi6}), (\ref{rr2}) and (\ref{rr3}), we find that
\begin{eqnarray}
\label{de13j}
\langle r^2\rangle(t)=\frac{dk_B T}{m\omega_0^2}+\left
( r_0^2-\frac{dk_B T}{m\omega_0^2}\right
)e^{-2\omega_0^2t/\xi}.
\end{eqnarray}
This result
can be obtained immediately from the virial theorem
(\ref{vir7}). We also have $\langle {\bf r}(t)\rangle=e^{-\omega_0^2 t/\xi}\,
{\bf r}_0$ from Eqs. (\ref{chi6}) and (\ref{rr2n}). For
$t\rightarrow +\infty$, we get $\langle r^2\rangle_{\infty}={dk_B
T}/{m\omega_0^2}$ and $\langle {\bf r}\rangle_{\infty}={\bf 0}$.

{\it Historical note:} The Smoluchowski equation (\ref{de4}) was introduced by
Smoluchowski in 1915 \cite{smoluchowski}. It can be viewed as a generalization
of the diffusion equation considered by Einstein \cite{einstein} when an
external potential is acting on a Brownian particle.\footnote{Actually,
drift-diffusion equations similar to the Smoluchowski equation, and coupled to
the Poisson equation by a mean field potential, were previously introduced by
Nernst \cite{nernst1,nernst2} and Planck \cite{planck2} in the context of
electrolytes (see also Debye and H\"uckel \cite{dh}).} This is also a
particular form of the Fokker-Planck equation that was introduced later
by Fokker \cite{fokker0,fokker} and Planck \cite{planck}.  The
solutions (\ref{de13b}) and (\ref{de13j}) were first obtained by 
Smoluchowski \cite{smoluchowski13,smoluchowski}. The
solution (\ref{de13b}) was rederived later by Uhlenbeck
and Ornstein  \cite{uo} by different methods: (i) from the Langevin 
equation, (ii)  from the Smoluchowski equation using Lord
Rayleigh's \cite{lr} method, and (iii)  from the
Smoluchowski
equation
using a trick due to Kramers (see Appendix II of \cite{uo}).
They also derived
Eq. (\ref{de13j}) from the Langevin equation.\footnote{Surprisingly, Langevin
\cite{langevin} is not 
quoted by Uhlenbeck and Ornstein \cite{uo} suggesting that his work was not
well-known at that time (see in this respect the comment in the Introduction of
\cite{ovw}).}

\subsubsection{Analogies with  Lord Rayleigh's kinetic theory}
\label{sec_lr}

It is usually considered that the theory of Brownian motion
started with Einstein's seminal paper
\cite{einstein}. Actually, more than a decade
before, Lord
Rayleigh \cite{lr} studied the dynamics of massive
particles bombared by
numerous small
projectiles. Although he did not explicitly refer to Brownian motion, his paper
may be considered as the first  theory of Brownian
motion with the
important difference that Lord Rayleigh \cite{lr} considered the
velocity distribution $f({\bf v},t)$ of homogeneously distributed
particles while Einstein \cite{einstein} considered the spatial
density $\rho({\bf r},t)$  of Brownian particles in the strong
friction limit $\xi\rightarrow +\infty$ (or, equivalently, for large times
$t\gg \xi^{-1}$). In his paper,  Lord
Rayleigh \cite{lr} derived  a partial
differential
equation for the temporal  evolution of the velocity distribution function
$f({\bf v},t)$ of the particles which, in modern notations, may be written as
\begin{eqnarray}
\label{lr1}
\frac{\partial f}{\partial t}=\xi\frac{\partial}{\partial {\bf v}}\cdot \left
(\frac{k_B T}{m}\frac{\partial f}{\partial {\bf v}}+f{\bf v}\right ).
\end{eqnarray}
 This equation includes a diffusion term in velocity
space and a friction term. This is a
particular form of the general Fokker-Planck equation
\cite{fokker0,fokker,planck}. An equation related to Eq.
(\ref{lr1}) but including an advection term in phase space taking into account
the presence of an external potential was introduced later by Klein
\cite{klein}, Kramers \cite{kramersbrown} and Chandrasekhar
\cite{revuechandra}, and is usually called the Kramers equation. The
Rayleigh equation (\ref{lr1}) is formally equivalent
to the Smoluchowski equation (\ref{de12}) with the harmonic external potential
(\ref{e3}). In this analogy, the velocity ${\bf v}$ plays the role of the
position ${\bf
r}$ and the
linear friction the role of the linear harmonic force. In the absence of
friction (or for sufficiently short times), Eq. (\ref{lr1}) reduces to the
diffusion equation in
velocity space
\begin{eqnarray}
\label{lr2}
\frac{\partial f}{\partial t}=D_*\frac{\partial^2f}{\partial {\bf v}^2},
\end{eqnarray}
with a diffusion coefficient given by
\begin{eqnarray}
\label{lr3}
D_*=\frac{\xi k_B T}{m}.
\end{eqnarray}
This can be viewed as the appropriate form of Einstein
relation in the present context \cite{revuechandra}.
Lord Rayleigh \cite{lr} obtained the solution of Eq. (\ref{lr2}), given by
\begin{eqnarray}
f({\bf v},t)=\frac{M}{V(4\pi D_* t)^{d/2}}e^{-\frac{({\bf v}-{\bf
v}_0)^2}{4D_*t}},
\end{eqnarray}
which is formally equivalent to Einstein's solution (\ref{de11b}). He refered
to it as Fourier's solution. He gave the result $\langle
v^2\rangle =2dD_*t+v_0^2$
which is formally equivalent to the Einstein result (\ref{de10j}). He
also obtained the solution of Eq. (\ref{lr1}), given by
\begin{eqnarray}
\label{lr4}
f({\bf v},t)=\frac{M}{V}\left\lbrack \frac{m}{2\pi k_B T
\left (1-e^{-2\xi t}\right )}\right\rbrack^{d/2}e^{-\frac{m
\left ({\bf v}-e^{-\xi t}{\bf v}_0\right )^2}{2k_B T
\left (1-e^{-2\xi t}\right )}},
\end{eqnarray}
which is formally equivalent to the Smoluchowski (or Uhlenbeck-Ornstein)
solution (\ref{de13b}). He
pointed out that
Eq. (\ref{lr4}) relaxes towards the stationary solution 
\begin{eqnarray}
\label{lr5}
f({\bf v})=\frac{M}{V}\left ( \frac{m}{2\pi k_B T}\right
)^{d/2}e^{-\frac{mv^2}{2k_B T}},
\end{eqnarray}
corresponding to Maxwell's distribution in velocity space
(comparing the steady state
of the Rayleigh equation with the Maxwell-Boltzmann
distribution is the direct manner to establish the Einstein relation (\ref{lr3})
\cite{revuechandra}). This is the
analogue of Eq. (\ref{de13c}).  He also pointed out
the fact that the
friction  is necessary to avoid the divergence of the kinetic energy
of the particles since  $\langle
v^2\rangle \sim 2dD_*t\rightarrow +\infty$ when the friction term is ignored. A
similar
argument was invoked later by Chandrasekhar \cite{chandra} in his
famous paper on the
{\it Dynamical
Friction}.\footnote{Surprisingly, the paper of
Lord
Rayleigh \cite{lr} is not mentioned at
that occasion.} When the friction is taken into account, one
finds that
\begin{eqnarray}
\label{de13janal}
\langle v^2\rangle(t)=\frac{dk_B T}{m}+\left
(v_0^2-\frac{dk_B T}{m}\right
)e^{-2\xi t},
\end{eqnarray}
which is the analogue of
Eq. (\ref{de13j}). We also have $\langle {\bf v}(t)\rangle=e^{-\xi t}\,
{\bf v}_0$. For
$t\rightarrow +\infty$, we get $\langle v^2\rangle_{\infty}={dk_B
T}/{m}$ and $\langle {\bf v}\rangle_{\infty}={\bf 0}$.

\subsection{Polytropic case}

The damped Euler equations (\ref{de1}) and (\ref{de2}) with the polytropic
equation of state (\ref{e4}) and the  harmonic potential (\ref{e3}) admit
a self-similar solution
of the form (\ref{e6}) with the Tsallis invariant profile (\ref{e8}) (see
Appendix \ref{sec_sss}). The
differential equation determining the evolution of $\chi(t)$ is given by Eq.
(\ref{chi1}) and the differential equation determining the evolution of the
radius $R(t)$ is given by
\begin{eqnarray}
\label{de14}
\ddot R+\xi \dot
R+\omega_0^2 R=2Z^{1-\gamma}K\gamma\frac{M^{\gamma-1}}{R^{d\gamma-d+1}}.
\end{eqnarray}
In the strong friction limit $\xi\rightarrow +\infty$, corresponding to the
polytropic Smoluchowski
equation (\ref{de5}), the differential equation determining the evolution of
$\chi(t)$ is given by Eq.
(\ref{chi5}) and the differential equation determining the evolution of the
radius $R(t)$ is given by
\begin{eqnarray}
\label{de15}
\xi \dot
R+\omega_0^2 R=2Z^{1-\gamma}K\gamma\frac{M^{\gamma-1}}{R^{d\gamma-d+1}}.
\end{eqnarray}
When $K>0$, 
$\omega_0^2>0$ and $\gamma>{\rm max}\lbrace
0,(d-2)/d\rbrace$, the steady states of the damped Euler equations and of the
polytropic Smoluchowski
equation are the same as those discussed in Sec. \ref{sec_stse}. They are given
by Eqs. (\ref{e12}) and (\ref{e12bis}) and they are stable. When $\xi>0$, the
system relaxes
towards these equilibrium states.

\subsubsection{Without external potential}

In the absence of external potential, Eq.
(\ref{de5}) reduces to the anomalous
diffusion
equation
\begin{equation}
\label{de16}
\frac{\partial\rho}{\partial t}=\frac{K}{\xi}\Delta\rho^{\gamma}
\end{equation}
originally introduced in the physics of porous media \cite{barenblatt}.
It has a Tsallis self-similar solution.
The differential equation (\ref{chi5})  determining the evolution of
$\chi(t)$ reduces to Eq. (\ref{chi7}). Its solution is given by Eq.
(\ref{chi8}). The differential equation (\ref{de15}) determining
the evolution of the
radius $R(t)$ reduces to
\begin{eqnarray}
\label{de17}
\xi \dot
R=2Z^{1-\gamma}K\gamma\frac{M^{\gamma-1}}{R^{d\gamma-d+1}}.
\end{eqnarray}
When $\gamma>{\rm max}\lbrace
0,(d-2)/d\rbrace$, the solution of Eq. (\ref{de17}) is given by 
\begin{eqnarray}
\label{de18}
R=\left\lbrack 2(d\gamma-d+2)Z^{1-\gamma}K\gamma
M^{\gamma-1}\frac{t}{\xi}\right\rbrack^{1/(d\gamma-d+2)}.
\end{eqnarray}
When $\gamma>d(d+2)$ the variance $\langle r^2\rangle$
is given by Eq. (\ref{rr2})
with Eqs. (\ref{chi8}), (\ref{rr4}),  and (\ref{de18}). We note that the
exponent in Eq. (\ref{de18}) is always positive since $\gamma>(d-2)/d$. The
dynamics is
superdiffusive when
$\gamma<1$ and subdiffusive when
$\gamma>1$. 

When $(d-2)/d<\gamma<0$ (this supposes $d<2$), the solution of Eq. (\ref{de17})
is given by 
\begin{eqnarray}
\label{de18b}
R(t)=\left\lbrack 2(d\gamma-d+2)Z^{1-\gamma}K|\gamma|
M^{\gamma-1}\frac{t_{\rm coll}-t}{\xi}\right\rbrack^{1/(d\gamma-d+2)}.
\end{eqnarray}
We note that the radius $R(t)$ vanishes in a finite time $t_{\rm coll}$
obtained by setting $t=0$ in Eq. (\ref{de18b}). Correspondingly, the density
becomes infinite at $t_{\rm coll}$. This corresponds to a finite time collapse. 
Note that the variance $\langle r^2\rangle$ is not defined when $\gamma<0$ (see
Appendix \ref{sec_rr}).

{\it Historical note:} The self-similar solution  of the anomalous diffusion
equation (\ref{de16}), given  by Eqs. (\ref{e6}),
(\ref{e8}), (\ref{de18}), and
(\ref{chi8})
was discovered by Barenblatt \cite{barenblatt} in the context of porous media.
It was later realized that the invariant profile is a Tsallis
distribution  (see Refs. \cite{frank,tsallisbook} and the
discussion in Sec. VI.A. of \cite{langcrit}).

\subsubsection{With a harmonic potential}

The polytropic Smoluchowski equation (\ref{de5}) with the harmonic external
potential
(\ref{e3}) can be written as
\begin{equation}
\label{de19}
\xi\frac{\partial\rho}{\partial t}=\nabla\cdot\left (K\nabla
\rho^{\gamma}+\rho\omega_0{\bf r}\right ).
\end{equation}
It has a Tsallis self-similar solution. The solution of the
differential equation (\ref{chi5})  determining the
evolution of
$\chi(t)$ is given by Eq.
(\ref{chi6}). When $\gamma>{\rm max}\lbrace
0,(d-2)/d\rbrace$,  the solution of the 
differential equation (\ref{de15}) determining the
evolution of the radius $R(t)$ is given by 
\begin{eqnarray}
\label{de20}
R=\left (\frac{2Z^{1-\gamma}K\gamma M^{\gamma-1}}{\omega_0^2}\right
)^{1/(d\gamma-d+2)}\left\lbrack
1-e^{-(d\gamma-d+2)\omega_0^2t/\xi}\right\rbrack^{1/(d\gamma-d+2)}.
\end{eqnarray}
When $\gamma>d(d+2)$ the variance $\langle r^2\rangle$
is given by Eq. (\ref{rr2})
with Eqs.  (\ref{chi6}), (\ref{rr4}), and (\ref{de20}). For $t\rightarrow
+\infty$, we recover the equilibrium state from Eqs. (\ref{e12}) and
(\ref{e12bis}).
When $(d-2)/d<\gamma<0$ (this supposes $d<2$), the solution of Eq.
(\ref{de15}) is given by 
\begin{eqnarray}
\label{de20c}
R(t)=\left (\frac{2Z^{1-\gamma}K|\gamma| M^{\gamma-1}}{\omega_0^2}\right
)^{1/(d\gamma-d+2)}\left\lbrack
e^{(d\gamma-d+2)\omega_0^2(t_{\rm
coll}-t)/\xi}-1\right\rbrack^{1/(d\gamma-d+2)}.
\end{eqnarray}
We note that the radius $R(t)$ vanishes in a finite time $t_{\rm coll}$
obtained by setting $t=0$ in Eq. (\ref{de20c}). Correspondingly, the density
becomes infinite at $t_{\rm coll}$. This corresponds to a finite time collapse. 
Note that the variance $\langle r^2\rangle$ is not defined when $\gamma<0$ (see
Appendix \ref{sec_rr}).

{\it Historical note:} The Tsallis
self-similar solution of the polytropic Smoluchowski
equation with a harmonic potential was discovered by Plastino
and Plastino
\cite{pp}. However, they did not explicitly solve the differential equation for
$R$. Tsallis and Bukman \cite{tb} solved the differential equation for
$R$ in $d=1$.  The general solution is provided by Eqs.
(\ref{de20}) and (\ref{de20c}) above. The Lorentzian self-similar solution of
the logotropic Smoluchowski
equation (corresponding to a polytrope $\gamma=0$) has been obtained by
Chavanis and Sire \cite{logotrope}.

\section{Self-similar solution of the generalized damped Gross-Pitaevskii
equation and of the generalized quantum damped Euler equations}
\label{sec_qe}

\subsection{Generalized damped Gross-Pitaevskii equation and quantum damped
Euler
equations}

We consider the generalized damped GP equation introduced in
\cite{ggp}:
\begin{equation}
\label{qe1}
i\hbar \frac{\partial\psi}{\partial t}=-\frac{\hbar^2}{2m}\Delta\psi
+m\Phi_{\rm ext}\psi+m\frac{dV}{d|\psi|^2}\psi
-i\frac{\hbar}{2}\xi\left\lbrack \ln\left (\frac{\psi}{\psi^*}\right
)-\left\langle \ln\left (\frac{\psi}{\psi^*}\right
)\right\rangle\right\rbrack\psi,
\end{equation}
where $V(|\psi|^2)$ is the self-interaction potential of the bosons and $\xi$ is
the
friction coefficient. This equation describes the evolution of the wavefunction
$\psi({\bf r},t)$ of a  a dissipative
BEC. For $\xi=0$ and $V(|\psi|^2)=(2\pi
a_s\hbar^2/m^3)|\psi|^4$, where $a_s$ is the s-scattering length of the bosons
\cite{revuebec}, we recover the  GP
equation \cite{gross1,gross2,gross3,pitaevskii2} with a cubic nonlinearity.
Performing 
the Madelung \cite{madelung} transformation, we find (see \cite{ggp} and
Appendix
\ref{sec_gw}) that Eq. (\ref{qe1}) is
equivalent to
the quantum damped Euler equations\footnote{Since a BEC can be considered as a
quantum fluid, these
hydrodynamic equations have a truly physical nature. In particular, being
equivalent to the (generalized) damped GP equation (\ref{qe1}), there is no
viscosity in the (generalized) quantum damped Euler 
equations (\ref{qe2}) and (\ref{qe3}). As a result, they  describe a
superfluid.}
\begin{equation}
\label{qe2}
\frac{\partial\rho}{\partial t}+\nabla\cdot (\rho {\bf u})=0,
\end{equation}
\begin{equation}
\label{qe3}
\frac{\partial {\bf u}}{\partial t}+({\bf u}\cdot \nabla){\bf
u}=-\frac{1}{\rho}\nabla P-\nabla \Phi_{\rm ext}-\frac{1}{m}\nabla
Q-\xi{\bf u},
\end{equation}
where
\begin{equation}
\label{qe4}
Q=-\frac{\hbar^2}{2m}\frac{\Delta
\sqrt{\rho}}{\sqrt{\rho}}=-\frac{\hbar^2}{4m}\left\lbrack
\frac{\Delta\rho}{\rho}-\frac{1}{2}\frac{(\nabla\rho)^2}{\rho^2}\right\rbrack
\end{equation}
is the quantum potential taking into account the Heisenberg uncertainty
principle. The
barotropic equation of state $P(\rho)$ is determined by the nonlinearity
$V(|\psi|^2)$ present in the GP equation (\ref{qe1}) according to Eq.
(\ref{de5b}). Several examples of  self-interaction
potentials $V(|\psi|^2)$, and the corresponding  equations of state $P(\rho)$,
are given in
\cite{ggp}. The specific cases of self-interaction
potentials leading to isothermal and polytropic equations of state
are discussed below.

In the absence of friction ($\xi=0$), Eqs. (\ref{qe2})-(\ref{qe4}) return  the
quantum Euler equations. In the  strong friction limit ($\xi\rightarrow
+\infty$) they return the generalized quantum Smoluchowski equation \cite{qs}.
Therefore, the generalized damped
GP equation (\ref{qe1}) provides an interesting formal
connection between quantum mechanics and Brownian theory (see
\cite{ggp} and Appendix \ref{sec_gw}) since it is
equivalent to a nonlinear Schr\"odinger equation  when $\xi=0$ \cite{sulem} and
to a nonlinear Fokker-Planck equation when $\xi\rightarrow +\infty$
\cite{frank,tsallisbook,nfp,entropy}.

Equations (\ref{qe1})-(\ref{qe4}) are associated with a
generalized
thermodynamic formalism. In particular, they satisfy an $H$-theorem  (see
\cite{ggp} and Appendix
\ref{sec_gw}) for the
generalized free energy $F$ obtained by adding to the free energy defined by Eq.
(\ref{de5a}) the von
Weizs\"acker  \cite{wei} functional
\begin{eqnarray}
\label{a14}
\Theta_Q=\frac{\hbar^2}{2m^2} \int
(\nabla \sqrt{\rho})^2\, d{\bf
r}=-\frac{\hbar^2}{2m^2}\int\sqrt{\rho}\Delta\sqrt{\rho}\, d{\bf
r}=\int\rho\frac{Q}{m}\, d{\bf r}=\frac{\hbar^2}{8m^2} \int
\frac{(\nabla\rho)^2}{\rho}\, d{\bf
r}.
\end{eqnarray}
As already mentioned, for the isothermal equation of state (\ref{e5}) the
generalized free energy is
associated with the Boltzmann entropy and for  the polytropic equation of state
(\ref{e4}) the generalized free energy is
associated with the Tsallis entropy (see
\cite{ggp} and Appendix
\ref{sec_gw}).

\subsection{Isothermal case}

The damped  logarithmic GP equation\footnote{See
Ref. \cite{chavnot} for a derivation of this equation from the theory of
scale relativity.} 
\begin{equation}
\label{qe6}
i\hbar \frac{\partial\psi}{\partial t}=-\frac{\hbar^2}{2m}\Delta\psi
+m\Phi_{\rm ext}\psi+2k_B T \ln|\psi|\psi
-i\frac{\hbar}{2}\xi\left\lbrack \ln\left (\frac{\psi}{\psi^*}\right
)-\left\langle \ln\left (\frac{\psi}{\psi^*}\right
)\right\rangle\right\rbrack\psi,
\end{equation}
corresponding to a self-interaction potential
\begin{equation}
\label{lp}
V(|\psi|^2)=\frac{k_B T}{m}|\psi|^2\left (\ln |\psi|^2 -1\right ),
\end{equation}
is equivalent to the quantum damped  Euler equations (\ref{qe2}) and
({\ref{qe3})
with the
isothermal equation of state (\ref{e5}). For the harmonic external potential
(\ref{e3}), these equations admit a Gaussian
self-similar solution formed by Eqs. (\ref{e6}) and (\ref{e10}) (see Appendix
\ref{sec_sss}).
The
differential equation determining the evolution of $\chi(t)$ is given by Eq.
(\ref{chi1}) and the differential equation determining the evolution of the
radius $R(t)$ is given by
\begin{eqnarray}
\label{qe7}
\ddot R+\xi\dot R+\omega_0^2
R=\frac{2k_B T}{mR}+\frac{\hbar^2}{m^2R^3}.
\end{eqnarray}
For $\xi=T=0$, it has the general analytical solution
\begin{eqnarray}
\label{qe7a}
R^2(t)=\frac{1}{\omega_0^2}\left\lbrack a\cos(2\omega_0 t)+b\sin(2\omega_0
t)+E\right\rbrack \quad {\rm with}\quad a^2+b^2=E^2-\left
(\frac{\hbar\omega_0}{m}\right )^2,
\end{eqnarray}
corresponding to a quantum harmonic oscillator (the constants $a$ and $b$ are
determined by
the initial condition) \cite{prep}. If we consider a free quantum particle
($\omega_0=0$),
we obtain
\begin{eqnarray}
\label{qe7b}
R^2(t)=R_0^2+\frac{\hbar^2 t^2}{m^2R_0^2},\qquad
H(t)=\frac{\frac{\hbar^2 t}{m^2R_0^2}}{R_0^2+\frac{\hbar^2 t^2}{m^2R_0^2}},
\end{eqnarray}
where we have assumed $\dot R(0)=0$. On the other hand, for
$T=0$ and
$\xi\rightarrow +\infty$, Eq. (\ref{qe7}) has the general analytic solution
\begin{eqnarray}
\label{qe7ab}
R^4(t)=\frac{\hbar^2}{m^2\omega_0^2}\left\lbrack 1-\left
(1-\frac{m^2\omega_0^2}{\hbar^2}R_0^4\right )e^{-4\omega_0^2t/\xi}\right\rbrack.
\end{eqnarray}
For $\omega_0=0$,
we obtain
\begin{eqnarray}
\label{qe7bb}
R^4(t)=R_0^4+\frac{4\hbar^2 t}{\xi m^2},\qquad
H(t)=\frac{\frac{\hbar^2}{\xi m^2}}{R_0^4+\frac{4\hbar^2 t}{\xi m^2}}.
\end{eqnarray}

{\it Remark:} It is remarkable that the quantum damped Euler
equations (\ref{qe2}) and (\ref{qe3}) with the
isothermal equation of state (\ref{e5}) still admit a Gaussian self-similar
solution
despite the presence of the quantum force. This
is because, when acted on the Gaussian distribution  formed by Eqs. (\ref{e6})
and (\ref{e10}), the quantum force
$-(1/m)\nabla Q$ and the isothermal pressure force $-(k_B T/m)\nabla\ln\rho$
are both proportional to ${\bf x}$ with a prefactor depending on time
but independent of $x$ (see Appendices  \ref{sec_sss} and \ref{sec_gq}).
Therefore, the quantum force and the isothermal equation of state ``marry
well''. We may
consider this striking feature as a coincidence.
Inversely, we can regard this
coincidence as being fundamental. This may reveal some connection between
standard quantum mechanics (through the quantum potential $Q$) and standard
thermodynamics
(throught the isothermal equation of state $P=\rho k_B T/m$). In other words,
the
standard quantum potential
may be linked to the isothermal equation of state. Assuming that this idea is
correct, we can now look for the  generalized quantum potential that is linked
to the
polytropic equation of state (see below).  This will lead to a notion of
generalized quantum
mechanics (through the generalized quantum potential $Q_g$) associated with
Tsallis
generalized thermodynamics (through the polytropic equation of state
$P=K\rho^{\gamma}$). Determining whether this idea is physically relevant is
beyond the scope of this paper.  In any case, it allows us to obtain a
generalized Schr\"odinger (or GP) equation that is of interest at a formal
level.

\subsection{Polytropic case}

The damped  power-law GP equation
\begin{equation}
\label{qe8}
i\hbar \frac{\partial\psi}{\partial t}=-\frac{\hbar^2}{2m}\Delta\psi+m\Phi_{\rm
ext}\psi
+\frac{K\gamma m}{\gamma-1}|\psi|^{2(\gamma-1)}\psi
-i\frac{\hbar}{2}\xi\left\lbrack \ln\left (\frac{\psi}{\psi^*}\right
)-\left\langle \ln\left (\frac{\psi}{\psi^*}\right
)\right\rangle\right\rbrack\psi,
\end{equation}
corresponding to a self-interaction potential
\begin{equation}
\label{pp}
V(|\psi|^2)=\frac{K}{\gamma-1}|\psi|^{2\gamma},
\end{equation}
is equivalent to the  quantum damped Euler equations (\ref{qe2}) and
({\ref{qe3})
with the polytropic equation of state (\ref{e4}). In the absence of quantum
force ($\hbar=0$), and for the harmonic potential (\ref{e3}), we have seen that
the damped Euler equations admit a
Tsallis self-similar solution. This is no more true for the quantum 
damped Euler equations  ($\hbar\neq 0$). We suggest therefore to look for a
generalized form of quantum potential that enables us to maintain a Tsallis
self-similar solution. We find that a suitable form of  generalized  quantum
potential is given by
(see Appendix \ref{sec_gqA})
\begin{eqnarray}
\label{qe9}
Q_{g}=-\frac{\hbar^2}{2m}\frac{\Delta\lbrack(\rho/\rho_0)^{\gamma-1/2}
\rbrack } {
(\rho/\rho_0)^{3/2-\gamma} }=-\frac{\hbar^2}{4m}(2\gamma-1)\left\lbrack
\frac{\Delta(\rho/\rho_0)}{(\rho/\rho_0)^{3-2\gamma}}-\frac{1}{2}
(3-2\gamma)\frac{\lbrack \nabla
(\rho/\rho_0)\rbrack^2}{(\rho/\rho_0)^{4-2\gamma}}\right\rbrack,
\end{eqnarray}
where $\rho_0$ is a constant with the dimension of a density that has been
introduced for dimensionality reasons. 
For $\gamma=1$, we recover the standard
quantum potential (\ref{qe4}). The generalized
GP equation leading, through the Madelung transformation, to the
generalized quantum damped Euler equations 
\begin{equation}
\label{qe2gen}
\frac{\partial\rho}{\partial t}+\nabla\cdot (\rho {\bf u})=0,
\end{equation}
\begin{equation}
\label{qe3gen}
\frac{\partial {\bf u}}{\partial t}+({\bf u}\cdot \nabla){\bf
u}=-\frac{1}{\rho}\nabla P-\nabla \Phi_{\rm ext}-\frac{1}{m}\nabla
Q_g-\xi{\bf u},
\end{equation}
with the generalized quantum potential (\ref{qe9}) and the polytropic equation
of state (\ref{e4}) is (see Appendices
\ref{sec_gw} and \ref{sec_gq})
\begin{eqnarray}
\label{qe11}
i\hbar \frac{\partial\psi}{\partial t}&=&-\frac{\hbar^2}{2m}\Delta\psi
+\frac{\hbar^2}{2m}\frac{\Delta|\psi|}{|\psi|}
\psi-\frac{\hbar^2}{2m}\frac{\Delta\lbrack
(|\psi|/\psi_0)^{2\gamma-1}\rbrack}{(|\psi|/\psi_0)^{3-2\gamma}}
\psi\nonumber\\
&+&\frac{ K\gamma m}{\gamma-1}
|\psi|^{2(\gamma-1)}\psi+m\Phi_{\rm ext}\psi
-i\frac{\hbar}{2}\xi\left\lbrack \ln\left (\frac{\psi}{\psi^*}\right
)-\left\langle \ln\left (\frac{\psi}{\psi^*}\right
)\right\rangle\right\rbrack\psi.
\end{eqnarray}
For the harmonic external potential
(\ref{e3}), these equations admit a Tsallis
self-similar solution formed by Eqs. (\ref{e6}) and (\ref{e8}) (see Appendix
\ref{sec_sss}).
The
differential equation determining the evolution of $\chi(t)$ is given by Eq.
(\ref{chi1}) and the differential equation determining the evolution of the
radius $R(t)$ is given by
\begin{eqnarray}
\label{qe12}
\ddot R+\xi\dot R+\omega_0^2
R=2Z^{1-\gamma}K\gamma\frac{M^{\gamma-1}}{R^{d\gamma-d+1}}+\frac{\hbar^2}{m^2}
\rho_0^{2(1-\gamma)}\frac{M^{
2(\gamma-1)}}{R^{ 2d(\gamma-1)+3}}\frac { 2\gamma-1} { Z^ { 2(\gamma-1) } }
\lbrack 1+d(\gamma-1)\rbrack.
\end{eqnarray}

\subsection{Steady states}

A steady state of the generalized quantum damped Euler
equations (\ref{qe2gen}) and (\ref{qe3gen}) satisfies the condition of
generalized quantum hydrostatic equilibrium
\begin{equation}
\label{eqg}
-\frac{1}{\rho}\nabla P-\nabla \Phi_{\rm ext}-\frac{1}{m}\nabla
Q_g={\bf 0}.
\end{equation}
The time-independent distribution (\ref{e6}),
corresponding to
$\chi(t)=0$, $R(t)=R_e={\rm cst}$ and ${\bf u}={\bf 0}$, when it exists, is a
steady state of the generalized quantum damped Euler equations
(\ref{qe2gen}) and (\ref{qe3gen}) with a harmonic potential. For an isothermal
equation of state this equilibrium distribution is Gaussian and for a polytropic
equation of state it is given by a Tsallis profile. The equilibrium radius of
the system is determined by the equation
\begin{eqnarray}
\omega_0^2
R_e=2Z^{1-\gamma}K\gamma\frac{M^{\gamma-1}}{R_e^{d\gamma-d+1}}+\frac{\hbar^2}{
m^2 }
\rho_0^{2(1-\gamma)}\frac{M^{
2(\gamma-1)}}{R_e^{ 2d(\gamma-1)+3}}\frac { 2\gamma-1} { Z^ { 2(\gamma-1) } }
\lbrack 1+d(\gamma-1)\rbrack.
\end{eqnarray}
When the equilibrium
state is unique and stable, and when $\xi>0$, the system relaxes towards this
steady state. However, the solution described previously is just a particular
solution of Eqs. (\ref{qe2gen}) and
(\ref{qe3gen}). This may not be the only solution. The
differential equation (\ref{eqg}) corresponding to the condition of generalized
quantum
hydrostatic
equilibrium  may have other solutions that are not given by the
Boltzmann or by the Tsallis distribution.

\section{Analogy with cosmology}
\label{sec_cosmo}

Let us consider the Euler equations (\ref{e1}) and (\ref{e2})
with the  harmonic potential (\ref{e3}) and the polytropic equation of state
(\ref{e4}) in $d=3$ dimensions. The
differential equation (\ref{e9}) governing the
evolution of the radius $R(t)$ of the system can be written as
\begin{eqnarray}
\label{n1}
\ddot
R=\frac{\kappa}{R^{3\gamma-2}}-\omega_0^2 R,
\end{eqnarray}
where we have introduced the notation $\kappa=2Z^{1-\gamma}K\gamma
M^{\gamma-1}$. This equation is formally identical to the fundamental equation
of
dynamics (Newton's equation) for a fictive particle of
unit mass in a one dimensional potential $V(R)$. Indeed, it can be rewritten as
\begin{eqnarray}
\label{n3}
\ddot R=-\frac{dV}{dR},
\end{eqnarray}
with
\begin{eqnarray}
\label{n4}
V(R)=\frac{\kappa}{3(\gamma-1)R^{3(\gamma-1)}}+\frac{1}{2}\omega_0^2R^2 \qquad
(\gamma\neq 1),
\end{eqnarray}
\begin{eqnarray}
\label{n5}
V(R)=-\kappa\ln R+\frac{1}{2}\omega_0^2R^2 \qquad (\gamma= 1).
\end{eqnarray}
The first integral of motion is
\begin{eqnarray}
\label{n6}
E=\frac{1}{2}\left (\frac{dR}{dt}\right )^2+V(R),
\end{eqnarray}
where $E$ is a constant representing the energy of the fictive particle.  The
evolution of the radius $R(t)$ of the system is therefore given by the integral
\begin{eqnarray}
\label{n7}
t=\pm\int_{R_0}^{R(t)}\frac{dR}{\sqrt{2\lbrack E-V(R)\rbrack}},
\end{eqnarray}
where $R_0$ is the value of the radius at $t=0$ and the sign in front of the
integral is $+$ when $dR/dt>0$ and $-$ when $dR/dt<0$. Substituting the
expression of the potential from Eqs. (\ref{n4}) and (\ref{n5}) into
Eq. (\ref{n7}), we obtain 
\begin{eqnarray}
\label{n8}
t=\pm\int_{R_0}^{R(t)}\frac{dR}{\sqrt{2E-\frac{2\kappa}{3(\gamma-1)R^{
3(\gamma-1) } }
-\omega_0^2R^2 } }
\qquad (\gamma\neq 1),
\end{eqnarray}
\begin{eqnarray}
\label{n9}
t=\pm\int_{R_0}^{R(t)}\frac{dR}{\sqrt{2E+2\kappa\ln
R-\omega_0^2R^2}}\qquad (\gamma= 1).
\end{eqnarray}

The first integral of motion (\ref{n6}) can be rewritten as
\begin{eqnarray}
\label{n10}
H^2=\left (\frac{\dot R}{R}\right )^2=\frac{2E}{R^2}-\frac{2V(R)}{R^2}.
\end{eqnarray}
Using the expressions (\ref{n4}) and (\ref{n5}) of the potential, we
get
\begin{eqnarray}
\label{n11}
H^2=\left (\frac{\dot R}{R}\right
)^2=\frac{2E}{R^2}-\frac{2\kappa}{3(\gamma-1)R^{3\gamma-1}}-\omega_0^2\qquad
(\gamma\neq 1),
\end{eqnarray}
\begin{eqnarray}
\label{n12}
H^2=\left (\frac{\dot R}{R}\right
)^2=\frac{2E}{R^2}+\frac{2\kappa\ln R}{R^2}-\omega_0^2\qquad (\gamma=1).
\end{eqnarray}
Equation (\ref{n10}) is similar to the Friedmann equation in
cosmology\footnote{The basic equations of cosmology needed in
the present section as well as a short
account of the early development of modern cosmology can be found in
Ref. \cite{cosmopoly1}.}
\begin{eqnarray}
\label{n13}
H^2=\left (\frac{\dot R}{R}\right )^2=-\frac{kc^2}{R^2}+\frac{8\pi
G}{3c^2}\epsilon+\frac{\Lambda}{3},
\end{eqnarray}
which determines the evolution of a universe with curvature $k$, energy
density $\epsilon$ and cosmological
constant $\Lambda$. In this analogy, $R$ plays the role of the scale factor,
$H=\dot R/R$ plays the role
of the Hubble parameter, $-2E$ plays the role of the curvature
constant $kc^2$, and $-2V(R)/R^2$ plays the role of the energy density term
$8\pi G
 \epsilon/3c^2$. We
can therefore
draw certain analogies between the evolution of the radius of our system
and the
evolution of the scale factor in the Friedmann-Lema\^itre-Robertson-Walker
(FLRW) universe.

First of all, comparing Eqs. (\ref{n11}) and (\ref{n12}) with Eq. (\ref{n13}),
we note that $-3\omega_0^2$ plays the role of the cosmological constant
$\Lambda$. A repulsive harmonic potential $\omega_0^2=-\Omega_0^2<0$ corresponds
to a positive cosmological constant $\Lambda=3\Omega_0^2>0$ (de
Sitter universe) while an attractive harmonic potential $\omega_0^2>0$
corresponds to a
negative cosmological constant $\Lambda=-3\omega_0^2<0$ (anti de
Sitter universe).

On the other hand, in a universe filled with a fluid described by a linear
equation of state
$P_{\rm cosmo}=\alpha\epsilon$, the energy density is related to the scale
factor by
$\epsilon\propto R^{-3(1+\alpha)}$ \cite{cosmopoly1}. For $\gamma\neq 1$ the
potential term  in
Eq. (\ref{n11}) is analogous  to
an energy density $\epsilon\propto R^{-(3\gamma-1)}$ in cosmology. This 
corresponds to an equation of state parameter $\alpha=\gamma-4/3$. Note that in
our system the ``energy density'' may be positive or negative (depending on the
signs of $\kappa$ and $\gamma-1$) while in cosmology the energy density is
generally positive. We also note that the energy term  $\propto R^{-2}$
in Eqs. (\ref{n11}) and (\ref{n12}), the curvature term in Eq. (\ref{n13}), and
the thermal term $\propto \ln R/R^2$ in Eq. (\ref{n12}) have the
same effect (up to a
logarithmic correction in the latter case) as a fluid described by an equation
of state $P_{\rm cosmo}=-\epsilon/3$ (gas of cosmic strings) for which
$\epsilon\propto 1/R^2$.

If we take $E=0$ and $\gamma\neq 1$, our system behaves, in the cosmological
analogy, as a fluid with an equation of state $P_{\rm
cosmo}=(\gamma-4/3)\epsilon$ in a flat
universe ($k=0$) with a cosmological constant $\Lambda$ (dark energy).
Interestingly, the index $\gamma=4/3$, implying $P_{\rm cosmo}=0$ and
$\epsilon\propto R^{-3}$, leads to the standard $\Lambda$CDM model (we need,
however, to
assume $K<0$ in order to have a positive energy
density).\footnote{We also note from Eq. (\ref{qe7}) that the standard quantum
potential term is analogous to an
energy density $\epsilon\propto -1/R^4$ in cosmology. This corresponds to
$\alpha=1/3$ like for the standard radiation. However, the energy density is
negative.} Let us now consider an arbitrary index $\gamma\neq
1$ and assume that $\omega_0=0$ (we also restrict ourselves to the case where
the system is expanding). Using the
cosmological analogy, we have the following results:

(i) For $\gamma>1$, corresponding to $\alpha>-1/3$, the expansion is
decelerating. For $\gamma<1$, corresponding to $\alpha<-1/3$, the expansion
is accelerating.

(ii) For $\gamma>1/3$, corresponding to $\alpha>-1$, the
evolution of our system is similar to the evolution of a normal universe in
which  the energy density decreases as the scale factor
increases. For $\gamma<1/3$, corresponding to $\alpha<-1$, the
evolution of our system is similar to the evolution of a phantom universe in
which  the energy density increases as the scale factor
increases. For $\gamma=1/3$, corresponding to $\alpha=-1$, i.e. $P_{\rm
cosmo}=-\epsilon$, the energy density is constant and it plays the same role as
the cosmological constant (dark energy).

These analogies with cosmology will be discussed in greater detail in a specific
paper \cite{prep}.

\section{Conclusion}

In this paper, we have introduced a nonlinear wave equation (\ref{gw6})
that
generalizes the standard Schr\"odinger equation. The Schr\"odinger equation is
recovered for $q=1$. We have also introduced a nonlinear wave equation
(\ref{gw1}) that generalizes the GP equation. The GP equation is recovered for
$q=1$, $\xi=0$, and for a cubic nonlinearity. By making the
Madelung transformation, we have shown that the generalized damped GP equation 
(\ref{gw1}) is equivalent to the generalized quantum damped Euler
equations (\ref{mad12}) and (\ref{mad13}). They include a generalized quantum
potential (\ref{gw4}), a pressure force determined by the
self-interaction potential $V(|\psi|^2)$, and a friction force. For conservative
systems ($\xi=0$), we obtain the generalized
quantum Euler equations (\ref{mad12add}) and (\ref{mad13add}). In the strong
friction limit $\xi\rightarrow +\infty$, we obtain the  generalized quantum
Smoluchowski equation (\ref{mad16}).
In the TF limit
($\hbar=0$), where the quantum potential can be neglected, we recover the Euler
equations (\ref{e1}) and (\ref{e2}) when $\xi=0$, and the generalized
Smoluchowski
equation (\ref{de3}) when $\xi\rightarrow +\infty$.

We have shown that these equations are associated with a notion of generalized
quantum mechanics, generalized Brownian theory and generalized
thermodynamics. They satisfy
an $H$-theorem for
a generalized form of free energy functional. The generalized quantum
potential (\ref{gw4}) is associated with the generalized Von Weiz\"acker
functional (\ref{gf3})-(\ref{gf6}). On the other hand, a logarithmic
potential in the generalized damped GP equation
(\ref{gw1}) leads to an isothermal equation of state (\ref{e5}) associated with
the
Boltzmann entropy (\ref{t1}) while a power-law potential  in the generalized
damped GP equation
(\ref{gw1}) leads to a polytropic equation of state (\ref{e4}) associated
with the Tsallis entropy (\ref{t3}). 

When considering the harmonic external potential (\ref{e3}) and the isothermal
equation of state (\ref{e5}), these equations admit stationary and
self-similar solutions with an invariant profile that has the form of a
Gaussian distribution.
When considering the harmonic external potential (\ref{e3}) and the polytropic
equation of state (\ref{e4}), these equations admit stationary and
self-similar solutions
with an invariant profile that has the form of a Tsallis distribution. This is
true provided that the exponent $q$ of the generalized quantum potential
(\ref{gw4}) and
the exponent $\gamma$ of the polytropic
equation of state (\ref{e4}) are related by 
\begin{eqnarray}
\label{conc1}
q=2\gamma-1.
\end{eqnarray}
This equation provides a relation between
generalized
quantum mechanics ($q$) and generalized thermodynamics ($\gamma$). For
$q=\gamma=1$, we recover standard quantum mechanics and standard thermodynamics.

Several researchers, including de Broglie \cite{dbbook}, have tried to obtain
nonlinear
wave equations generalizing the (linear) Schr\"odinger equation. They considered
that the Schr\"odinger 
equation is not the most fundamental equation of quantum mechanics and that it
may be an approximation of a more fundamental nonlinear wave equation. The
Schr\"odinger equation is the simplest (because it is linear) wave equation
that returns the relation $E=p^2/2m$ between the energy and
the
impulse of a free classical  particle if we consider plane waves of the form
$e^{i({\bf k}\cdot {\bf r}-\omega
t)}$ with the correspondances $E=\hbar \omega$ (Planck-Einstein) and ${\bf
p}=\hbar {\bf k}$ (de Broglie). But this is not the only equation enjoying
this property. We can construct nonlinear wave equations that preserve these
relations. One example has been proposed recently by Nobre {\it et al.}
\cite{nrt1,nrt2}. Their nonlinear wave equation
admits $q$-plane waves of the form $e_q^{i({\bf k}\cdot {\bf r}-\omega
t)}$ that preserve the above relations. Another example is provided by our
nonlinear wave equation (\ref{gw1n}) that preserves the above relations for
standard  plane waves of the
form $e^{i({\bf k}\cdot {\bf r}-\omega t)}$ (see the end of Appendix
\ref{sec_gwq}). In addition, it admits wave
packet solutions which, instead of having a Gaussian profile, have a
$q$-Gaussian
profile (see Appendix \ref{sec_wf} and the end of Appendix \ref{sec_oss}).
 This is interesting because one
never exactly
measures pure Gaussians, but rather $q$-Gaussians \cite{vp}.
Therefore, one cannot reject the possibility that quantum mechanics may be
described by a nonlinear wave equation of the form of Eq. (\ref{gw6}).

Several researchers
\cite{hc,reyni,mittal,st,sm,tsallis,barenblatt,spohn,kq,pp,tb,csr,abe,stariolo,
martinez,borland,br,k0,fd,k1,frank1,cn,gen,gen2,nobre,kl,naudts,chavcourt,cll,
lang,banach,k2,cst,frank,bose,degrad,cras,csbio,logotrope,langcrit,sw1,sw2,nfp,
sw3,tsallisbook,entropy}  have also tried to generalize Brownian theory and 
obtain nonlinear Fokker-Planck equations generalizing the (linear) 
Fokker-Planck
equation associated with standard (Boltzmann) thermodynamics. However, until
now, no connection was made between generalized Schr\"odinger equations and
generalized Fokker-Planck equations.  Our approach, initiated in \cite{ggp}, is
a
first step in that direction. An interest of our formalism is to connect
(generalized)  quantum mechanics, (generalized)  Brownian
motion, and  (generalized) thermodynamics. Indeed, the (generalized) damped GP
equation (\ref{gw1}) contains
the (generalized) Schr\"odinger equation of quantum mechanics and the
(generalized) Smoluchowski equation of Brownian theory as special cases. The
(generalized)  Schr\"odinger
equation is obtained in the no friction limit ($\xi=0$) while the (generalized)
 Smoluchowski
equation is obtained in the strong friction limit $\xi\rightarrow +\infty$.

The starting point of our generalization of quantum mechanics was based on the
following
remark. The logarithmic Schr\"odinger equation (\ref{qe6}) admits self-similar
solutions with a Gaussian profile like in
standard (Boltzmann) thermodynamics. This is
because the Laplacian $\Delta$ in quantum mechanics ``marries well'' with the
logarithmic term $2k_BT\ln|\psi|$ associated with the isothermal equation of
state (\ref{e5}) of standard thermodynamics. We then looked for a
generalization of quantum mechanics
such that the generalized
Laplacian
$\Delta_q$ (see Eq. (\ref{gw2b})) in generalized quantum mechanics ``marries
well'' with
the power-law term
$K|\psi|^{2(\gamma-1)}$ associated with the polytropic equation of
state (\ref{e4}) of generalized thermodynamics. This leads to the generalized
power-law Schr\"odinger equation (\ref{qe11}) that admits self-similar solutions
with a 
Tsallis profile like in generalized (Tsallis)
thermodynamics.

It is in general very difficult to find exact
analytical solutions of nonlinear partial differential equations, especially
for $d\neq 1$ (for $d=1$ several nonlinear partial differential
equations admit soliton solutions). Therefore, our
work may be of interest to mathematicians. We have introduced  a 
nonlinear wave equations (\ref{gw1}) with a generalized Laplacian ($q$) and a
power-law
nonlinearity ($\gamma$) that admits exact  analytical self-similar
solutions in the form of wave packets with a Tsallis profile (in certain limits
they reduce to soliton-like solutions that
we called $\gamma$-gaussons). These
solutions
are valid not only for a free particle but also for a particle in a harmonic
potential. These solutions are not valid for more complicated potentials but it
is probably possible to make  perturbative expansions about the harmonic
potential.

For conservative systems ($\xi=0$) we have shown that the
radius of the system $R(t)$ in the self-similar solution satisfies a
differential
equation that shares some analogies with the Friedmann equations in cosmology 
describing the evolution of the radius (scale factor) of the
Universe. This analogy will be developed in a forthcoming
contribution \cite{prep} where analytical solutions of the equations derived
in the present paper are given.

At the present stage of the development of the theory all our results
are formal. Still, there are ``exact'' in the sense that  there
is no approximation of any sort in our approach once the basic equations are
assumed. Therefore, the equations
presented in this paper have nice mathematical properties. It will be
important in future works
to determine  if our formalism can have physical
applications. We have already proposed to apply these
generalized wave equations to the physics of dark matter halos made of
Bose-Einstein condensates (BECs) \cite{hui}. In that case, they must be
coupled to the gravity through the Poisson equation, leading to the
(generalized) Gross-Pitaevskii-Poisson (GPP) equations \cite{prd1,ggp}. In the
strong friction limit, and in the TF approximation, we
recover the (generalized) Smoluchowski-Poisson  equations studied in
Ref. \cite{qs} (and references therein). Another line of research would
consist
in extending these ideas to the context of general relativity in relation to the
Klein-Gordon-Einstein (KGE) equations. Some
work in that direction has already been performed  in Ref. \cite{chavmatos}. On
the
other hand, in Ref. \cite{supernova}, we have given
an illustration of the self-similar solution of the classical Euler equations
presented in Sec. \ref{sec_euler} in relation to the explosion of a star
(supernova). These are different exemples where the equations of
the present paper, and their self-similar solutions, may find physical and
astrophysical applications. It is likely that many other applications will be
discovered in the future.

\appendix

\section{Basic properties of the generalized damped GP equation}
\label{sec_gw}

\subsection{Generalized damped GP equation}
\label{sec_gwq}

We consider a generalized damped GP equation of the form
\begin{equation}
\label{gw1}
i\hbar \frac{\partial\psi}{\partial t}=-\frac{\hbar^2}{2m}\Delta\psi
+\frac{\hbar^2}{2m}\frac{\Delta|\psi|}{|\psi|}
\psi-\frac{\hbar^2}{2m}\frac{\Delta\lbrack
(|\psi|/\psi_0)^{q}\rbrack}{(|\psi|/\psi_0)^{2-q}}
\psi+m\left\lbrack \frac{dV}{d|\psi|^2}+\Phi+\Phi_{\rm ext}\right\rbrack\psi
-i\frac{\hbar}{2}\xi\left\lbrack \ln\left (\frac{\psi}{\psi^*}\right
)-\left\langle \ln\left (\frac{\psi}{\psi^*}\right
)\right\rangle\right\rbrack\psi
\end{equation}
describing a dissipative BEC with a mass density $\rho({\bf r},t)=|\psi|^2({\bf
r},t)$, where $\psi({\bf r},t)$ is the wave function of the
condensate.\footnote{We may also regard this nonlinear Schr\"odinger equation as
describing the evolution of the wavefunction of a single quantum particle.
However, in that case, the interpretation of the different terms demands a more
delicate discussion that is left for a future contribution.} In this
equation $V(|\psi|^2)$ is the self-interaction
potential of the bosons,\footnote{The (generalized) GP equation can be derived
from the (generalized) Klein-Gordon equation in the nonrelativistic limit
$c\rightarrow +\infty$ \cite{chavmatos,ggp}. In that case, $V(|\psi|^2)$ is the
potential that appears in the KG equation.} $\Phi({\bf r},t)$ is the mean field
potential created
by the bosons, $\Phi_{\rm ext}({\bf
r})$ is a fixed
external potential, and $\xi$ is the friction
coefficient.\footnote{We may also consider a time dependent
external potential $\Phi_{\rm ext}({\bf r},t)$. It could
account for a stochastic forcing as in \cite{prd3}. Such a term is
often written under the form $\Phi_{\rm
ext}({\bf r},t)={\bf
A}(t)\cdot {\bf r}$ where ${\bf A}(t)$ is a random force \cite{kostin}.} The
brackets
denote a
space average over the entire domain: $\langle
X\rangle=\frac{1}{M}\int \rho X\, d{\bf r}$.
The 
 mean field potential can be written as
\begin{equation}
\label{gw1b}
\Phi({\bf r},t)=\int u(|{\bf r}-{\bf r}'|)|\psi|^2({\bf r}',t)\, d{\bf r}',
\end{equation}
where $u(|{\bf r}-{\bf r}'|)$ is a binary potential of interaction between the
bosons. The mean
field approximation is valid for long-range potentials of interaction, such as
the gravitational potential $u=-G/|{\bf r}-{\bf r}'|$, in a proper thermodynamic
limit where the number of
particles $N\rightarrow
+\infty$. Self-gravitating BECs have been
proposed as a model of dark matter in cosmology (see Refs. \cite{prd1,ggp} and
references therein). Equation (\ref{gw1})
involves a nonlinear Laplacian operator 
\begin{equation}
\label{gw2b}
\Delta_q\psi\equiv \Delta\psi
-\frac{\Delta|\psi|}{|\psi|}
\psi+\frac{\Delta\lbrack
(|\psi|/\psi_0)^{q}\rbrack}{(|\psi|/\psi_0)^{2-q}}
\psi,
\end{equation}
where $\psi_0$ is a
constant with the dimension of a wave function
introduced for dimensional reasons. For $q=1$, $\Delta_q$ reduces to the
ordinary Laplacian operator $\Delta$ and Eq. (\ref{gw1}) reduces to the
generalized damped
GP equation (\ref{qe1})  studied in \cite{ggp} in the context of
standard quantum mechanics. We
refer to this
paper for a detailed discussion of this equation and for its physical
interpretation. The aim of this Appendix is to summarize in a synthetic manner
the main
results of this study and to generalize them in the case where the linear
Laplacian operator $\Delta$ is replaced by the nonlinear Laplacian operator
$\Delta_q$.
This leads to a notion of generalized quantum mechanics (see Appendix
\ref{sec_gq}).
Introducing the
function
\begin{eqnarray}
\label{gw2}
h(|\psi|^2)=\frac{dV}{d|\psi|^2}, \qquad {\rm i.e.}\qquad h(\rho)=V'(\rho),
\end{eqnarray}
the  generalized damped GP equation (\ref{gw1}) can be rewritten as 
\begin{equation}
\label{gw3}
i\hbar \frac{\partial\psi}{\partial t}=-\frac{\hbar^2}{2m}\Delta\psi
+\frac{\hbar^2}{2m}\frac{\Delta|\psi|}{|\psi|}
\psi-\frac{\hbar^2}{2m}\frac{\Delta\lbrack
(|\psi|/\psi_0)^{q}\rbrack}{(|\psi|/\psi_0)^{2-q}}
\psi+m\lbrack h(|\psi|^2)+\Phi+\Phi_{\rm ext}\rbrack\psi
-i\frac{\hbar}{2}\xi\left\lbrack \ln\left (\frac{\psi}{\psi^*}\right
)-\left\langle \ln\left (\frac{\psi}{\psi^*}\right
)\right\rangle\right\rbrack\psi.
\end{equation}
It can be shown (see below) that the generalized damped GP equation conserves
the mass $M=\int\rho\, d{\bf r}$.

{\it Remark:} If we consider a free particle ($V=\Phi=\Phi_{\rm ext}=\xi=0$),
we obtain the generalized Schr\"odinger equation
\begin{equation}
\label{gw1n}
i\hbar \frac{\partial\psi}{\partial t}=-\frac{\hbar^2}{2m}\Delta\psi
+\frac{\hbar^2}{2m}\frac{\Delta|\psi|}{|\psi|}
\psi-\frac{\hbar^2}{2m}\frac{\Delta\lbrack
(|\psi|/\psi_0)^{q}\rbrack}{(|\psi|/\psi_0)^{2-q}}
\psi.
\end{equation}
Looking for a solution in the form of a plane wave $\psi({\bf r},t)\sim
e^{i({\bf
k}\cdot
{\bf r}-\omega t)}$ we obtain the dispersion relation $\omega=\hbar k^2/2m$.
Using the Planck-Einstein relation $E=\hbar\omega$ and the de Broglie
relation ${\bf p}=\hbar{\bf k}$, we recover the relation  $E=p^2/2m$ between the
energy and the impulse of a
classical free particle. Therefore, our generalized Schr\"odinger equation
is consistent with the wave-particle duality which is at the basis of quantum
mechanics (see, e.g., the introduction of \cite{chavmatos}).

\subsection{Madelung transformation}

We use the Madelung \cite{madelung} transformation to rewrite the
generalized damped GP
equation (\ref{gw3}) under the form of hydrodynamic equations.\footnote{As
discussed in \cite{ggp,chavmatos}, this hydrodynamic representation has a clear
physical
meaning in the case where the generalized damped GP
equation (\ref{gw3}) describes a BEC. Its interpretation is less clear (and
was strongly criticized by the founders of quantum mechanics like
Pauli)  when the
nonlinear Schr\"odinger equation (\ref{gw3}) describes a single quantum
particle. In that respect, we recall that the Madelung
transformation (also introduced independently by de Broglie
\cite{broglie1927a,broglie1927b,broglie1927c} in his pilot wave theory of
relativistic
particles) was rediscovered by Bohm \cite{bohm1,bohm2}  who gave it an
interpretation in
terms of particle trajectories (see Appendix \ref{sec_lag}). For that reason the
quantum potential is sometime called the
Bohm potential although it was found previously by Madelung.}
We write the
wave
function as
\begin{equation}
\label{mad1}
\psi({\bf r},t)=\sqrt{{\rho({\bf r},t)}} e^{iS({\bf r},t)/\hbar},
\end{equation}
where the phase $S({\bf r},t)=-i(\hbar/2)\ln(\psi/\psi^*)$ represents the
action. Following Madelung, we
introduce the
velocity field
\begin{equation}
\label{mad5}
{\bf u}=\frac{\nabla
S}{m}=-i\frac{\hbar}{2m}\frac{\psi^*\nabla\psi-\psi\nabla\psi^*}{|\psi|^2}.
\end{equation}
Since the velocity is potential, the flow is irrotational: $\nabla\times {\bf
u}={\bf 0}$. Substituting Eq. (\ref{mad1}) into Eq. (\ref{gw3}) and separating
real and imaginary parts, we obtain
\begin{equation}
\label{mad6}
\frac{\partial\rho}{\partial t}+\nabla\cdot (\rho {\bf u})=0,
\end{equation}
\begin{equation}
\label{mad7}
\frac{\partial S}{\partial t}+\frac{1}{2m}(\nabla S)^2+m
\left\lbrack h(\rho)+\Phi+\Phi_{\rm ext}\right\rbrack+Q_g+\xi (S-\langle
S\rangle)=0,
\end{equation}
where
\begin{eqnarray}
\label{gw4}
Q_g=-\frac{\hbar^2}{2m}\frac{\Delta\lbrack(\rho/\rho_0)^{q/2}
\rbrack } {
(\rho/\rho_0)^{1-q/2} }=-\frac{\hbar^2}{4m}q\left\lbrack
\frac{\Delta(\rho/\rho_0)}{(\rho/\rho_0)^{2-q}}-\frac{1}{2}
(2-q)\frac{\lbrack \nabla
(\rho/\rho_0)\rbrack^2}{(\rho/\rho_0)^{3-q}}\right\rbrack
\end{eqnarray}
is the generalized quantum potential (see Appendix \ref{sec_gqA}) and 
$\rho_0=\psi_0^2$ is a constant with the dimensions of a mass density. When
$q=1$ we recover the standard quantum potential from Eq. (\ref{qe4}). Eq.
(\ref{mad6})
is similar to the
equation of continuity in
hydrodynamics. It accounts for the local conservation of mass. 
Eq. (\ref{mad7}) has a form similar to the
classical
Hamilton-Jacobi equation with an additional generalized quantum potential and a
source of
dissipation. It  can also be interpreted as a generalized Bernoulli equation for
a potential flow. Taking 
the gradient of Eq. (\ref{mad7}), and using the well-known  identity of
vector analysis $({\bf u}\cdot \nabla){\bf u}=\nabla ({{\bf u}^2}/{2})-{\bf
u}\times (\nabla\times {\bf u})$ which reduces to $({\bf u}\cdot \nabla){\bf
u}=\nabla ({{\bf u}^2}/{2})$ for an irrotational flow, we obtain an equation
similar to the Euler equation with a linear friction and a generalized quantum
force
\begin{equation}
\label{mad9}
\frac{\partial {\bf u}}{\partial t}+({\bf u}\cdot \nabla){\bf u}=-\nabla
h-\nabla\Phi-\nabla \Phi_{\rm ext}-\frac{1}{m}\nabla Q_g-\xi {\bf u}.
\end{equation}
We can also write Eq.
(\ref{mad9}) under the form
\begin{equation}
\label{mad10}
\frac{\partial {\bf u}}{\partial t}+({\bf u}\cdot \nabla){\bf
u}=-\frac{1}{\rho}\nabla P-\nabla \Phi-\nabla \Phi_{\rm
ext}-\frac{1}{m}\nabla
Q_g-\xi {\bf u},
\end{equation}
where $P({\bf r},t)$ is the pressure.  Since $h({\bf
r},t)=h\lbrack \rho({\bf
r},t)\rbrack$, the pressure $P({\bf r},t)=P\lbrack \rho({\bf r},t)\rbrack$ is a
function of the density, i.e., the flow is barotropic. The equation of state
$P(\rho)$ is determined by the potential $h(\rho)$ through the relation
\begin{equation}
\label{mad11}
h'(\rho)=\frac{P'(\rho)}{\rho}.
\end{equation}
This equation, which can be viewed as the Gibbs-Duhem relation $dh=dP/\rho$,
shows that the effective potential $h$ appearing in the
generalized damped GP
equation (\ref{gw3}) can be
interpreted as an enthalpy in the hydrodynamic equations. We
shall see later (in Appendix \ref{sec_eq}) that $h$ is one component of the
chemical
potential. Equation
(\ref{mad11}) can be integrated into 
\begin{equation}
\label{mad11b}
P(\rho)=\rho h(\rho)-V(\rho)=\rho
V'(\rho)-V(\rho)=\rho^2\left\lbrack
\frac{V(\rho)}{\rho}\right\rbrack'.
\end{equation}
This determines the equation of state $P(\rho)$ as a function of the
self-interaction potential $V(\rho)$.
The speed of
sound is $c_s^2=P'(\rho)=\rho
V''(\rho)$. Inversely, the
self-interaction potential is related to the equation of state by
\begin{equation}
\label{mad5b}
V(\rho)=\rho\int^{\rho}\frac{P(\rho')}{{\rho'}^2}\, d\rho'.
\end{equation} 
We note the identities
\begin{equation}
\label{mad5bb}
V'(\rho)=\int^{\rho} \frac{P'(\rho')}{\rho'}\, d\rho'\qquad {\rm and}\qquad 
V''(\rho)=\frac{P'(\rho)}{\rho},
\end{equation} 
which result from simple integrations by parts.

 In conclusion, the generalized damped GP equation (\ref{gw3}) is
equivalent
to
the
hydrodynamic equations
\begin{equation}
\label{mad12}
\frac{\partial\rho}{\partial t}+\nabla\cdot (\rho {\bf u})=0,
\end{equation}
\begin{equation}
\label{mad13}
\frac{\partial {\bf u}}{\partial t}+({\bf u}\cdot \nabla){\bf
u}=-\frac{1}{\rho}\nabla P-\nabla \Phi-\nabla \Phi_{\rm
ext}-\frac{1}{m}\nabla
Q_g-\xi{\bf u}.
\end{equation}
For the harmonic potential defined by Eq. (\ref{e3}), we have
$\nabla\Phi_{\rm ext}=\omega_0^2{\bf r}$. Using the continuity
equation (\ref{mad12}), the Euler equation (\ref{mad13}) can be rewritten as
\begin{eqnarray}
\label{mad14b}
\frac{\partial}{\partial t}(\rho {\bf u})+\nabla(\rho {\bf u}\otimes {\bf u})
=-\nabla P-\rho\nabla\Phi-\rho\nabla \Phi_{\rm ext}-\frac{\rho}{m}\nabla
Q_g-\xi\rho {\bf u}.
\end{eqnarray}
We shall refer to these equations as
the generalized quantum damped Euler equations. We note that the hydrodynamic
equations
(\ref{mad12}) and (\ref{mad13}) do not involve viscous terms so 
they
describe a
superfluid.  When the generalized quantum
potential can be neglected, we recover the classical damped
Euler
equations. For dissipationless systems ($\xi=0$), Eqs. (\ref{mad12}) and
(\ref{mad13}) reduce to
the generalized quantum  Euler
equations
\begin{equation}
\label{mad12add}
\frac{\partial\rho}{\partial t}+\nabla\cdot (\rho {\bf u})=0,
\end{equation}
\begin{equation}
\label{mad13add}
\frac{\partial {\bf u}}{\partial t}+({\bf u}\cdot \nabla){\bf
u}=-\frac{1}{\rho}\nabla P-\nabla \Phi-\nabla \Phi_{\rm
ext}-\frac{1}{m}\nabla
Q_g.
\end{equation}
When the generalized quantum
potential can be neglected, we recover the classical
Euler
equations. On
the
other hand, in the overdamped limit $\xi\rightarrow +\infty$, we can formally
neglect the inertia of the particles in Eq. (\ref{mad13}) so that
\begin{equation}
\label{mad15}
\xi{\bf u}\simeq -\frac{1}{\rho}\nabla P-\nabla \Phi-\nabla \Phi_{\rm
ext}-\frac{1}{m}\nabla Q_g.
\end{equation}
Substituting this relation into the continuity equation (\ref{mad12}), we obtain
the generalized quantum Smoluchowski
equation
\begin{equation}
\label{mad16}
\xi\frac{\partial\rho}{\partial t}=\nabla\cdot\left (\nabla
P+\rho\nabla \Phi+\rho\nabla \Phi_{\rm ext}+\frac{\rho}{m}\nabla Q_g\right ).
\end{equation}
When the generalized quantum
potential can be neglected, we recover the generalized classical Smoluchowski
equation \cite{nfp}. Finally, if we neglect the advection term
$\nabla(\rho {\bf u}\otimes {\bf u})$
in Eq. (\ref{mad14b}), but retain the term $\partial (\rho {\bf
u})/\partial t$, and
combine the resulting equation with the continuity equation (\ref{mad12}), we
obtain the generalized quantum 
telegraph equation
\begin{equation}
\label{tmad12}
\frac{\partial^2\rho}{\partial t^2}+\xi\frac{\partial\rho}{\partial
t}=\nabla\cdot\left (\nabla P+\rho\nabla\Phi+\rho\nabla \Phi_{\rm
ext}+\frac{\rho}{m}\nabla Q_g\right ).
\end{equation}
It can be seen as a generalization of the  quantum  
Smoluchowski equation (\ref{mad16}) taking  memory effects into
account. As a result, the generalized damped GP equation (\ref{gw3}) contains
many important equations of physics as particular cases. In particular, it
provides an interesting formal connection between (generalized) quantum
mechanics and (generalized) Brownian theory \cite{ggp} since it is
equivalent to a nonlinear Schr\"odinger equation  when $\xi=0$ \cite{sulem} and
to a nonlinear Fokker-Planck equation when $\xi\rightarrow
+\infty$ \cite{frank,tsallisbook,nfp,entropy}.

{\it Remark:} When $V=\Phi=\xi=0$, the generalized damped GP equation
(\ref{gw1})
reduces to a generalized Schr\"odinger equation of the form
\begin{eqnarray}
\label{gw6}
i\hbar\frac{\partial\psi}{\partial
t}=-\frac{\hbar^2}{2m}\Delta\psi+\frac{\hbar^2}{2m}\frac{\Delta|\psi|}{|\psi|}
\psi-\frac{\hbar^2}{2m}\frac{\Delta\lbrack
(|\psi|/\psi_0)^{q}\rbrack}{(|\psi|/\psi_0)^{2-q}}
\psi+m\Phi_{\rm ext}\psi.
\end{eqnarray}
If we ignore the third term in the right hand side of Eq.
(\ref{gw6}), we get the classical wave equation (see Appendix L.3. of
\cite{chavnot})
\begin{eqnarray}
\label{gw7}
i\hbar\frac{\partial\psi}{\partial
t}=-\frac{\hbar^2}{2m}\Delta\psi+\frac{\hbar^2}{2m}\frac{\Delta|\psi|}{|\psi|}
\psi+m\Phi_{\rm ext}\psi.
\end{eqnarray}
This equation is equivalent, through the Madelung transformation, to the
continuity equation (\ref{mad12}) and to the
classical (i.e. without quantum potential)
Euler equation
\begin{equation}
\label{gw8}
\frac{\partial {\bf u}}{\partial t}+({\bf u}\cdot \nabla){\bf
u}=-\nabla \Phi_{\rm
ext}.
\end{equation}
As a result, the generalized
Schr\"odinger equation  (\ref{gw6}) is equivalent, through the Madelung
transformation, to the
continuity equation (\ref{mad12}) and to the  Euler equation with the
generalized quantum
potential
\begin{equation}
\label{gw9}
\frac{\partial {\bf u}}{\partial t}+({\bf u}\cdot \nabla){\bf
u}=-\nabla \Phi_{\rm
ext}-\frac{1}{m}\nabla Q_{g}.
\end{equation}
This is basically how we have originally obtained Eq. (\ref{gw6}): we have first
removed the contribution of the standard quantum potential, then added the
contribution of the generalized quantum potential. As another remark, we note
that when $q\rightarrow 1$ the generalized Schr\"odinger equation (\ref{gw6})
reduces to
\begin{eqnarray}
\label{gw10}
i\hbar\frac{\partial\psi}{\partial
t}=-\frac{\hbar^2}{2m}\Delta\psi
-(q-1)\frac{\hbar^2}{2m}\left\lbrack \frac{\Delta (|\psi|\ln
(|\psi|/\psi_0))}{|\psi|}+\ln\left (\frac{|\psi|}{\psi_0}\right )\frac{\Delta
|\psi|}{|\psi|}\right\rbrack\psi+m\Phi_{\rm ext}\psi.
\end{eqnarray}
This provides the first correction to the Schr\"odinger equation. 

\subsection{Generalized quantum force}
\label{sec_gqf}

The generalized quantum force can
be written as
\begin{eqnarray}
\label{gqf1}
\frac{1}{m}\partial_iQ_g=\frac{1}{\rho}\partial_jP_{ij}^g,
\end{eqnarray}
where $P_{ij}^g$ is a generalized quantum pressure tensor defined by (see
Appendix \ref{sec_gp}) 
\begin{eqnarray}
\label{gqf2}
P_{ij}^g=\frac{\hbar^2}{4m^2}q\frac{1}{k}\rho_0^{1-q}\left\lbrack
\frac{1}{\rho^{2-q}}\partial_i\rho\partial_{j}\rho-A\frac{1}{\rho^{1-q}}\delta_{
ij}\Delta\rho-B\frac{1}{\rho^{1-q}}\partial_{ij}\rho+C\frac{1}{\rho^{2-q}}
(\nabla\rho)^2\delta_{ij}\right\rbrack,
\end{eqnarray}
\begin{eqnarray}
\label{gqf2b}
A=\frac{1-k(2-q)}{q-1},\qquad B=\frac{k-1}{q-1},\qquad C=\frac{k}{2}(3-q),
\end{eqnarray}
where $k$ is an arbitrary parameter. When $q=1$ we recover the standard quantum
pressure tensor
$P_{ij}$ given by Eq. (\ref{gp12}). We note that the contracted
quantum pressure tensor
can be written as
\begin{eqnarray}
\label{gqf2c}
P_{ii}^g=\frac{\hbar^2}{4m^2}q\frac{1}{k}\rho_0^{1-q}\left\lbrack
(1+dC)\frac{(\nabla\rho)^2}{\rho^{2-q}}-(B+dA)\frac{\Delta\rho}{
\rho^ { 1-q } }\right\rbrack,
\end{eqnarray}
where a summation is implied over repeated indices.

\subsection{Generalized time-independent GP equation}
\label{sec_tigp}

If we consider a wave function of the form
\begin{equation}
\label{tigp1}
\psi({\bf r},t)=\phi({\bf r})e^{-i E t/\hbar},
\end{equation}
where $\phi({\bf r})=\sqrt{\rho({\bf r})}$ is real, and substitute Eq.
(\ref{tigp1}) into the generalized damped GP equation (\ref{gw3}), we
obtain the generalized time-independent GP equation
\begin{eqnarray}
\label{tigp2}
-\frac{\hbar^2}{2m}\frac{\Delta\lbrack
(\phi/\psi_0)^{q}\rbrack}{(\phi/\psi_0)^{2-q}}
\phi+m\lbrack \Phi+h(\phi^2)+\Phi_{\rm ext}\rbrack\phi=E\phi.
\end{eqnarray}
Equation (\ref{tigp2}) defines a nonlinear
eigenvalue problem for the 
wave function $\phi({\bf r})$ where the eigenvalue $E$ is the energy
(we call it the eigenenergy). 
Dividing Eq. (\ref{tigp2}) by $\phi({\bf r})$ and using $\rho=\phi^2$ and Eq.
(\ref{gw4}),  we get
\begin{equation}
\label{tigp4}
m\Phi+mh(\rho)+m\Phi_{\rm ext}+Q_g=E.
\end{equation}
This relation can be directly derived from the generalized quantum damped 
Hamilton-Jacobi equation (\ref{mad7}) by setting  $S=-Et$.

\subsection{Generalized quantum hydrostatic equilibrium}
\label{sec_he}

The generalized time-independent GP equation
(\ref{tigp4}) can also be obtained from the generalized
quantum damped Euler equation (\ref{mad10}) which is equivalent to
the generalized damped GP equation (\ref{gw3}). The equilibrium state of the
generalized
quantum damped Euler equation, obtained by taking $\partial_t=0$
and
${\bf u}={\bf 0}$, satisfies
\begin{equation}
\label{he1}
\nabla P+\rho\nabla\Phi+\rho\nabla \Phi_{\rm ext}+\frac{\rho}{m}\nabla Q_g={\bf
0}.
\end{equation}
This equation generalizes the usual condition of hydrostatic equilibrium by
incorporating the contribution of the generalized quantum potential.  Equation
(\ref{he1}) describes the balance between the pressure force due to the
self-interaction of the bosons, the mean field force,  the
external force, and the generalized quantum force.
This equation is equivalent to Eq. (\ref{tigp4}). Indeed,
integrating Eq. (\ref{he1}) using Eq. (\ref{mad11}), we obtain Eq. (\ref{tigp4})
where the eigenenergy  $E$ appears as a constant of integration.

\subsection{Generalized free energy}
\label{sec_gf}

The generalized free energy associated with the generalized damped GP equation
(\ref{gw1}), or equivalently with the generalized quantum damped Euler equations
(\ref{mad12}) and (\ref{mad13}),  
can be written as 
\begin{eqnarray}
\label{gf1}
F=\Theta_c+\Theta_Q^g+U+W+W_{\rm ext},
\end{eqnarray}
where the different functionals are defined below using Madelung's hydrodynamic
variables (see \cite{ggp} for a more detailed discussion). 

The first term is the  classical
kinetic energy
\begin{eqnarray}
\label{gf2}
\Theta_c=\int\rho  \frac{{\bf u}^2}{2}\, d{\bf r}.
\end{eqnarray}

The second term is the generalized quantum kinetic energy. It is given by the
equivalent expressions (see Appendix \ref{sec_gvw})
\begin{eqnarray}
\label{gf3}
\Theta_Q^g=\frac{\hbar^2}{2m^2}\frac{\rho_0}{q} \int
\lbrace\nabla \lbrack(\rho/\rho_0)^{q/2}\rbrack\rbrace^2\, d{\bf
r},
\end{eqnarray}
\begin{eqnarray}
\label{gf5}
\Theta_Q^g=\frac{\hbar^2}{8m^2}\rho_0 q \int
\frac{\lbrack\nabla (\rho/\rho_0)\rbrack^2}{(\rho/\rho_0)^{2-q}}\, d{\bf
r},
\end{eqnarray}
\begin{eqnarray}
\label{gf4}
\Theta_Q^g=-\frac{\hbar^2}{2m^2}\frac{\rho_0}{q} \int \left
(\frac{\rho}{\rho_0}\right )^{q/2}\Delta \left
\lbrack  \left
(\frac{\rho}{\rho_0}\right )^{q/2}  \right\rbrack\, d{\bf
r},
\end{eqnarray}
\begin{eqnarray}
\label{gf6}
\Theta_Q^g=\frac{1}{q}\int\rho\frac{Q_g}{m}\, d{\bf r}.
\end{eqnarray}
According to Eq. (\ref{gf6}) the generalized quantum kinetic energy can also be
interpreted as a potential energy associated
with the generalized quantum
potential $Q_g$ (this analogy is not obvious, however, since $Q_g$ is a function
of the
density, not an external potential). The
functional (\ref{gf5}) is a generalization of the von
Weizs\"acker functional
\cite{wei}. It  is related to a generalization of the  Fisher 
\cite{fisher} entropy by Eq. (\ref{gfi4}). The functional (\ref{gf4}) is a
generalization
of the  Madelung functional
\cite{madelung,madelungearly}. For $q=1$ we recover the standard expressions
given in \cite{ggp}: 
\begin{eqnarray}
\label{gf7}
\Theta_Q=\frac{\hbar^2}{2m^2} \int
(\nabla \sqrt{\rho})^2\, d{\bf
r}=\frac{\hbar^2}{8m^2} \int
\frac{(\nabla\rho)^2}{\rho}\, d{\bf
r}=-\frac{\hbar^2}{2m^2}\int\sqrt{\rho}\Delta\sqrt{\rho}\, d{\bf
r}=\int\rho\frac{Q}{m}\, d{\bf r}\qquad (q=1).
\end{eqnarray}

The third term is the internal
energy
\begin{eqnarray}
\label{gf8}
U=\int\rho\int^{\rho}\frac{P(\rho')}{{\rho'}^2}\, d\rho'\, d{\bf r}=\int
\left\lbrack \rho h(\rho)-P(\rho)\right \rbrack\, d{\bf r}=\int V(\rho)\,
d{\bf r}.
\end{eqnarray}

The
fourth term is
the mean field energy
\begin{eqnarray}
\label{gf9}
W=\frac{1}{2}\int\rho\Phi\, d{\bf r}.
\end{eqnarray}

The fifth term is the external potential energy
\begin{eqnarray}
\label{gf10}
W_{\rm ext}=\int\rho\Phi_{\rm ext}\, d{\bf r}.
\end{eqnarray}
For the harmonic potential (\ref{e3}), one has
\begin{eqnarray}
\label{gf10b}
W_{\rm ext}=\frac{1}{2}\omega_0^2 I,\qquad {\rm where} \qquad I=\int \rho r^2\,
d{\bf r}
\end{eqnarray}
is the moment of inertia. We note that $I=M\langle r^2\rangle$ where $\langle
r^2\rangle$ measures the dispersion of the particles (or the size of the BEC).

Regrouping these results, the free energy associated with the generalized damped
GP equation
(\ref{gw1}), or equivalently with the generalized quantum damped Euler equations
(\ref{mad12}) and (\ref{mad13}),  can be explicitly
written as
\begin{eqnarray}
\label{gf12}
F=\int\rho \frac{{\bf u}^2}{2}\, d{\bf r}+\frac{1}{qm}\int \rho Q_g\, d{\bf
r}+\int
V(\rho)\, d{\bf r}
+\frac{1}{2}\int\rho\Phi\, d{\bf r}+\int\rho\Phi_{\rm ext}\, d{\bf r}.
\end{eqnarray}
On the other hand, the free energy associated with the generalized quantum
Smoluchowski equation
(\ref{mad16}) is given by 
\begin{eqnarray}
\label{gf13}
F=\Theta_Q^g+U+W+W_{\rm ext}=\frac{1}{qm}\int
\rho Q_g\, d{\bf r}+\int
V(\rho)\, d{\bf r}
+\frac{1}{2}\int\rho\Phi\, d{\bf r}+\int\rho\Phi_{\rm ext}\, d{\bf r}
\end{eqnarray}
since the classical kinetic energy $\Theta_c$, which is of order $O(\xi^{-2})$,
can be 
neglected in the overdamped limit $\xi\rightarrow +\infty$.

{\it Remark:} the functionals $\Theta_c$, $U$, $W$ and $W_{\rm ext}$ are the
same as in \cite{ggp}. Their first and second order
variations
are given in Appendix C of \cite{ggp}. The functional $\Theta_Q^g$ is a
generalization of the standard von Weizs\"acker functional $\Theta_Q$.
However, its first variations satisfy the
fundamental relation
\begin{eqnarray}
\label{gf14}
\delta \Theta_Q^g= \int \frac{Q_g}{m}\delta\rho\, d{\bf r},
\end{eqnarray}
as in the standard case \cite{ggp}. Actually, we
determined the expression of $\Theta_Q^g$
precisely in order to satisfy this relation (see Appendix \ref{sec_gvw}).

\subsection{$H$-theorem}
\label{sec_gh}

The time
derivative of the generalized free energy (\ref{gf12}) satisfies the identity
\begin{eqnarray}
\label{gh1}
\dot F=-\xi\int \rho {\bf u}^2\, d{\bf r}=-2\xi\Theta_c.
\end{eqnarray}
When $\dot F=0$, we get ${\bf u}={\bf 0}$ leading to the condition of
generalized quantum hydrostatic equilibrium (\ref{he1}).
For dissipationless systems ($\xi=0$), Eq. (\ref{gh1}) 
shows that the generalized  GP equation, or the generalized quantum Euler
equations,  
conserve
the free energy ($\dot F=0$). For dissipative systems ($\xi>0$), Eq.
(\ref{gh1}) 
shows that the generalized damped GP equation, or the generalized 
quantum damped Euler
equations, decrease the free energy ($\dot F\le 0$). This provides a form of
 $H$-theorem. In the strong friction limit $\xi\rightarrow +\infty$,
the time derivative of the generalized free energy (\ref{gf13}) associated with
the generalized quantum Smoluchowski
equation satisfies the $H$-theorem
\begin{eqnarray}
\label{gh2}
\dot F=-\frac{1}{\xi}\int \frac{1}{\rho}\left (\nabla
P+\rho\nabla\Phi+\rho\nabla\Phi_{\rm ext}+\frac{\rho}{m}\nabla Q_g\right )^2\,
d{\bf r}\le 0.
\end{eqnarray}
When $\dot F=0$, the term in parenthesis vanishes leading to the condition
of
generalized quantum hydrostatic equilibrium (\ref{he1}). A more detailed
discussion of these results is given
in Sec. 3.3 of \cite{ggp}.

 \subsection{Equilibrium state}
\label{sec_eq}

According to the results of the preceding section, a
stable equilibrium state of the
generalized (damped) GP equation, or of the generalized quantum (damped) Euler
equations, is the
solution of
the minimization problem
\begin{eqnarray}
\label{eq1}
F(M)=\min_{\rho,{\bf u}} \left\lbrace F[\rho,{\bf u}]\quad |\quad M\quad {\rm
fixed}\right\rbrace.
\end{eqnarray}
An extremum of free energy at fixed mass is an equilibrium
state of the
generalized (damped) GP equation, or of the generalized quantum (damped) Euler
equations. It is determined by the variational
principle
\begin{eqnarray}
\label{eq2}
\delta F-\frac{\mu}{m}\delta M=0,
\end{eqnarray}
where $\mu$ is a Lagrange multiplier (chemical potential) taking  into account
the mass constraint. This variational problem gives
${\bf u}={\bf 0}$ (the equilibrium state is static) and 
\begin{eqnarray}
\label{eq3}
m\Phi+m\Phi_{\rm ext}+m h(\rho)+Q_g=\mu.
\end{eqnarray}
This is the generalization of the Gibbs condition of constancy of 
chemical
potential (see, e.g.,  Landau and Lifshitz \cite{ll}). Indeed, if we
define the out-of-equilibrium
chemical potential by
\begin{eqnarray}
\frac{\mu({\bf r},t)}{m}=\frac{\delta F}{\delta \rho}=\Phi+\Phi_{\rm ext}+
h(\rho)+\frac{1}{m}Q_g,
\end{eqnarray}
we find that, at equilibrium, $\mu({\bf r},t)=\mu$ is a constant.
Taking the gradient of Eq.
(\ref{eq3}), and using Eq.
(\ref{mad11}), 
we recover the condition of generalized quantum hydrostatic equilibrium
(\ref{he1}). Equation
(\ref{eq3}) is also equivalent to the generalized time-independent GP equation
(\ref{tigp4}) provided that we make the identification
\begin{eqnarray}
\label{eq4}
\mu=E.
\end{eqnarray}
This shows that the Lagrange multiplier $\mu$
(chemical potential) in the variational problem (\ref{eq2}) associated with Eq.
(\ref{eq1})
can be identified with the eigenenergy $E$.  Inversely, the eigenenergy $E$ may
be
interpreted as a chemical potential. Multiplying Eq. (\ref{eq3})
by $\rho/m$, integrating over the whole domain and using Eq. (\ref{eq4})
we get 
\begin{eqnarray}
NE=q\Theta_Q^g+\int\rho h(\rho)\, d{\bf r}+2W+W_{\rm ext}.
\end{eqnarray}
Finally, according to Eq. (\ref{eq3}), an
equilibrium state of the
generalized (damped) GP equation, or of the generalized quantum (damped) Euler
equations, satisfies the relation
\begin{eqnarray}
\label{eq5}
\rho=h^{-1}\left (\frac{\mu}{m}-\frac{Q_g}{m}-\Phi-\Phi_{\rm
ext}\right ).
\end{eqnarray}
When $Q_g=\Phi=0$, this equation directly determines  the equilibrium
distribution
$\rho({\bf r})$. More generally,  Eq. (\ref{eq5}) is a differential, or an
integrodifferential, equation for $\rho({\bf r})$. An equilibrium
state which just extremizes the free energy at fixed mass may, or may
not, be stable. Considering the second
order variations of the free energy, we find that the equilibrium state is
stable
if,
and only if, 
\begin{equation}
\delta^2 F= \frac{1}{2}\int h'(\rho)(\delta\rho)^2\, d{\bf r}+\frac{1}{2}\int
\delta\rho\delta\Phi\, d{\bf r}
+\frac{\hbar^2}{8m^2}\rho_0^{1-q}q\int\frac{1}{\rho^{2-q}}\left\lbrack
(2-q)\left
(\frac{\Delta\rho}{\rho}-\frac{1}{2}(3-q)\frac{(\nabla\rho)^2}{\rho^2}\right
){(\delta\rho)^2}+{(\nabla\delta\rho)^2}\right\rbrack\, d{\bf r}>0
\end{equation}
for all perturbations that conserve mass: $\int \delta\rho\, d{\bf r}=0$. This
inequality can also be written as
\begin{equation}
\delta^2 F= \frac{1}{2}\int h'(\rho)(\delta\rho)^2\, d{\bf r}+\frac{1}{2}\int
\delta\rho\delta\Phi\, d{\bf r}
+\frac{\hbar^2}{8m^2}\rho_0^{1-q}q\int  \left \lbrack \nabla \left
(\frac{\delta\rho}{\rho^{1-q/2}}\right )\right\rbrack^2\, d{\bf
r}+\frac{\hbar^2}{8m^2}\rho_0^{1-q}(2-q)\int
\frac{\Delta( {\rho}^{q/2})}{\rho^{2-q/2}}(\delta\rho)^2\,
d{\bf r}>0.
\end{equation}

\subsection{Functional derivatives}
\label{sec_fd}

The results of Sec. 3.6 of \cite{ggp} concerning the
functional
derivatives of $F$ remain valid in the present context provided that we make the
substitution
$Q\rightarrow Q_g$.

\subsection{Virial theorem}
\label{sec_vir}

The virial of the generalized quantum force is defined by
\begin{eqnarray}
\label{vir1}
(W^Q_{ii})_g=-\int\frac{\rho}{m}{\bf r}\cdot \nabla Q_g\, d{\bf r}.
\end{eqnarray}
Using Eq. (\ref{gqf1}) we get
\begin{eqnarray}
\label{vir2}
(W^Q_{ii})_g=\int P^g_{ii}\, d{\bf r},
\end{eqnarray}
where $P^g_{ii}$ is given by Eq. (\ref{gqf2c}). It is important to note
that $(W^Q_{ii})_g\neq 2\Theta^g_Q$ when $q\neq 1$ contrary to the standard
quantum mechanics case ($q=1$) where  $W^Q_{ii}=
2\Theta_Q$ \cite{ggp}. This is because $\Theta_c+\Theta^g_Q$ does not
correspond to the standard kinetic energy $\Theta=\frac{\hbar^2}{2m^2}\int
|\nabla\psi|^2 \, d{\bf r}$ when we make the Madelung transformation. Indeed,
the generalized kinetic energy  $\Theta_g=\Theta_c+\Theta^g_Q$ is given by
\begin{eqnarray}
\Theta_g=\frac{\hbar^2}{2m^2}\int |\nabla\psi|^2 \, d{\bf
r}+\frac{\hbar^2}{8m^2}\psi_0^2 q \int
\frac{\lbrack\nabla (|\psi|^2/\psi_0^2)\rbrack^2}{(|\psi|^2/\psi_0^2)^{2-q}}\,
d{\bf
r}-\frac{\hbar^2}{8m^2} \int
\frac{\lbrack\nabla(|\psi|^2)\rbrack^2}{|\psi|^2}\,
d{\bf
r}.
\end{eqnarray}
Apart from
this difference, the results of Sec.
4 and Appendix G of \cite{ggp} remain valid. The scalar virial theorem
associated with the generalized damped GP equation, or with the generalized
quantum damped Euler equations,  writes
\begin{eqnarray}
\label{vir3}
\frac{1}{2}\ddot I+\frac{1}{2}\xi\dot I+\omega_0^2 I=2\Theta_c+\int
P_{ii}^g+d\int
P\, d{\bf r}+W_{ii},
\end{eqnarray}
where $W_{ii}$ is the virial of the mean field force and we have considered
the case of an external harmonic potential.
In the strong friction limit $\xi\rightarrow +\infty$, the  generalized quantum
damped Euler equations reduce to the generalized quantum Smoluchowski equation
and the scalar virial theorem becomes
\begin{eqnarray}
\label{vir4}
\frac{1}{2}\xi\dot I+\omega_0^2 I=\int
P_{ii}^g+d\int
P\, d{\bf r}+W_{ii}.
\end{eqnarray}
In all cases, the  equilibrium scalar virial theorem writes
\begin{eqnarray}
\label{vir5}
\int
P_{ii}^g+d\int
P\, d{\bf r}+W_{ii}-\omega_0^2 I=0.
\end{eqnarray}

For classical systems ($\hbar=0$) described
by the damped Euler equations with an isothermal equation of state $P=\rho k_B
T/m$, and in the absence of mean field force ($\Phi=0$), the scalar
virial theorem takes the form
\begin{eqnarray}
\label{vir6}
\frac{1}{2}\ddot I+\frac{1}{2}\xi\dot I+\omega_0^2 I=2\Theta_c+dNk_B T.
\end{eqnarray}
In the strong friction limit $\xi\rightarrow +\infty$, the damped Euler
equations reduce to the Smoluchowski equation and the scalar virial theorem
becomes
\begin{eqnarray}
\label{vir7}
\frac{1}{2}\xi\dot I+\omega_0^2 I=dNk_B T.
\end{eqnarray}
At equilibrium, we get
\begin{eqnarray}
\label{vir8}
dNk_B T-\omega_0^2 I=0.
\end{eqnarray}
We note that Eq. (\ref{vir7}) is a closed differential equation
for $I$. Solving this equation and using $I=M\langle r^2\rangle$ we
directly obtain Eq.
(\ref{de13j}). In the absence of external potential ($\omega_0=0$),
the foregoing equations reduce to
\begin{eqnarray}
\label{vir9}
\frac{1}{2}\ddot I+\frac{1}{2}\xi\dot I=2\Theta_c+dNk_B T
\end{eqnarray}
and
\begin{eqnarray}
\label{vir10}
\frac{1}{2}\xi\dot I=dNk_B T.
\end{eqnarray}
The solution of Eq. (\ref{vir10}) directly leads to Eq. (\ref{de10j}).

\subsection{Conservation of linear and angular momentum}
\label{sec_lmam}

Using hydrodynamic variables, the linear
momentum of the system is given by
\begin{eqnarray}
\label{lmam1}
{\bf P}=-i\frac{\hbar}{m}\int\psi^*\nabla\psi\, d{\bf r}\quad\Rightarrow \quad
{\bf P}=\int\rho {\bf u}\, d{\bf r}.
\end{eqnarray}
Taking its time derivative and using Eq. (\ref{mad14b}), we obtain
\begin{eqnarray}
\label{lmam2}
\frac{d{\bf P}}{dt}=-\xi{\bf P}\quad {\rm i.e.} \quad {\bf P}(t)={\bf P}_0
e^{-\xi t}.
\end{eqnarray}
To get this result, we have used the identity $\int \rho\nabla\Phi\, d{\bf
r}={\bf
0}$ (easily obtained from Eq. (\ref{gw1b}) by
interchanging the dummy
variables ${\bf r}$ and ${\bf r}'$) and the identity $\int (\rho/m)\nabla Q_g\,
d{\bf r}=\int
\partial_j P_{ij}^g\, d{\bf r}={\bf 0}$ obtained from Eq. (\ref{gqf1}). The
linear momentum is conserved when $\xi=0$.

Using
hydrodynamic variables, the angular
momentum of the system is given by
\begin{eqnarray}
\label{lmam3}
{\bf L}=-i\frac{\hbar}{m}\int\psi^* ({\bf r}\times \nabla) \psi\, d{\bf
r}\quad\Rightarrow \quad {\bf L}=\int\rho\, {\bf r}\times {\bf u}\, d{\bf r}.
\end{eqnarray}
Taking its time derivative and using Eq. (\ref{mad14b}), we obtain
\begin{eqnarray}
\label{lmam4}
\frac{d{\bf L}}{dt}=-\xi{\bf L}\quad\Rightarrow \quad {\bf L}(t)={\bf L}_0
e^{-\xi t}.
\end{eqnarray}
To get this result, we have used the virial identity (see Appendix G of
\cite{ggp})
\begin{eqnarray}
\label{lmam5}
\int x_i\frac{\partial}{\partial t}(\rho u_j)\, d{\bf r}=\int \rho
u_i u_j\, d{\bf r}+\delta_{ij}\int P\, d{\bf r}+W_{ij}+W_{ij}^{\rm
ext}+(W_{ij}^{Q})_{g}-\xi\int\rho x_i u_j\, d{\bf r}.
\end{eqnarray}
Since all the tensors on the right hand side except the last one are
symmetric, we get
\begin{eqnarray}
\label{lmam6}
\frac{d}{dt} \int \rho (x_iu_j-x_j u_i) \,
d{\bf r}=-\xi\int\rho
(x_i u_j-x_j u_i)\, d{\bf r},
\end{eqnarray}
leading to Eq. (\ref{lmam4}). The angular momentum is conserved when $\xi=0$.

\subsection{Generalized thermodynamics and Tsallis entropy}
\label{sec_t}

In Ref. \cite{ggp} we have shown that the damped GP
equation (\ref{qe1}) is
associated with a generalized thermodynamical formalism. The generalized
entropy
$S$ that appears in this formalism is determined by the
nonlinearity $V(|\psi|^2)$ in the damped GP equation (\ref{qe1}) or by the
equation of state
$P(\rho)$ in the quantum damped Euler equation (\ref{qe3}).
In the strong friction limit $\xi\rightarrow +\infty$, and when quantum effects
are neglected (TF approximation), the damped GP equation (\ref{qe1}) becomes
equivalent to the generalized Smoluchowski equation (\ref{de3}). In that case,
we
recover the  generalized thermodynamical formalism associated with NFP
equations considered in
Ref. \cite{nfp}. The  generalized thermodynamical
formalism of Ref. \cite{ggp} remains valid for the generalized damped GP
equation
(\ref{gw1}). The only
modification to make with respect to Ref. \cite{ggp} is to replace the von
Weizs\"acker functional (\ref{gf7}) by the generalized von Weizs\"acker
functional (\ref{gf5}) in the expression of the free energy
$F$. 

The  generalized free energy  (\ref{gf1}) can be written as
\begin{eqnarray}
\label{tq1}
F=E_*+U=E_*-T_{\rm eff}S,
\end{eqnarray}
where 
\begin{eqnarray}
\label{tq2}
E_*=\Theta_c+\Theta^g_Q+W+W_{\rm ext}
\end{eqnarray}
is the energy that includes the classical kinetic energy $\Theta_c$, the
generalized quantum
kinetic energy $\Theta_Q^g$, the mean field energy $W$, and the
external potential energy $W_{\rm ext}$.  It can be written explicitly as
\begin{eqnarray}
\label{tq3}
E_*=\int\rho \frac{{\bf u}^2}{2}\, d{\bf r}+\frac{1}{qm}\int \rho Q_g\, d{\bf r}
+\frac{1}{2}\int\rho\Phi\, d{\bf r}+\int\rho\Phi_{\rm ext}\, d{\bf r}.
\end{eqnarray}
On the other hand, we have written the internal energy as
$U=-T_{\rm eff} S$, where $T_{\rm eff}$ is an effective
temperature and $S=-\int C(\rho)\, d{\bf r}$ is a generalized entropy
determined by the potential $V(\rho)$ according to the relation
$C(\rho)=V(\rho)/T_{\rm eff}$. We note that
$h(\rho)=T_{\rm eff}C'(\rho)$ so that the equilibrium state determined
by Eq. (\ref{eq5}) can be rewritten as
\begin{eqnarray}
\label{eq5b}
\rho=(C')^{-1}\left\lbrack \frac{1}{T_{\rm eff}}\left
(\frac{\mu}{m}-\frac{Q_g}{m}-\Phi-\Phi_{\rm
ext}\right )\right\rbrack.
\end{eqnarray}
Several examples of generalized entropies associated with the
generalized damped GP
equation (\ref{qe1}) have been given in \cite{ggp}. We consider below two of
them.

The generalized logarithmic damped GP
equation
(\ref{qe6}) is equivalent to the  generalized quantum damped Euler equations
(\ref{qe2}) and
({\ref{qe3}) with the
isothermal equation of state (\ref{e5}). In that case $h(\rho)=(k_B T/m)\ln\rho$
and $V(\rho)=(k_B T/m)\rho(\ln\rho-1)$. The associated entropy
is the Boltzmann entropy 
\begin{eqnarray}
\label{t1}
S_B=-k_B \int
\frac{\rho}{m} (\ln\rho-1)\,
d{\bf r}
\end{eqnarray}
and the free energy (\ref{gf1}) can be written as
\begin{eqnarray}
\label{t2}
F_B=E_*-TS_B.
\end{eqnarray}
The equilibrium state (\ref{eq5b}) is the Boltzmann
distribution
\begin{eqnarray}
\label{t2b}
\rho=e^{-(m\Phi+m\Phi_{\rm ext}+Q_g-\mu)/k_B T}.
\end{eqnarray}

The generalized power-law damped GP equation (\ref{qe8}) is equivalent to the 
generalized quantum damped Euler
equations (\ref{qe2}) and (\ref{qe3})
with the polytropic equation of state (\ref{e4}). In that case
$h(\rho)=[K\gamma/(\gamma-1)]\rho^{\gamma-1}$ and
$V(\rho)=[K/(\gamma-1)]\rho^{\gamma}$. The
associated entropy is the Tsallis entropy\footnote{A
power-law entropic functional similar to Eq. (\ref{t3}), but written in phase
space as $S=-\int f^{q}\, d{\bf r}d{\bf v}$, was introduced by
Ipser \cite{ipser} in a dynamical
(Vlasov) context,
in relation to Eddington's stellar polytropes \cite{eddington}. Ipser also
introduced functionals of the form $S=-\int C(f)\, d{\bf r}d{\bf v}$ (where
$C(f)$ is convex) that we now call generalized entropic functionals
\cite{gen,nfp}. Actually, Ipser's
polytropic functional $S=-\int \rho^{\gamma}\, d{\bf r}$  corresponds to
the first term of Eq. (\ref{t3}). In
principle the second term  of Eq.
(\ref{t3}), which is proportional to the mass $M$, is a constant that can be
omitted in the entropy. However, an interest of the expression (\ref{t3}) given
by Tsallis \cite{tsallis} is that it reduces to the Boltzmann entropy for
$\gamma\rightarrow 1$ thanks to L'H\^ospital's rule (see Refs.
\cite{cst,cc} for more details and for a thorough discussion between
dynamical
and thermodynamical stability).}
\begin{eqnarray}
\label{t3}
S_{\gamma}=-\frac{1}{\gamma-1}\int
(\rho^{\gamma}-\rho)\, d{\bf
r}
\end{eqnarray}
and the free energy (\ref{gf1}) can be written as
\begin{eqnarray}
\label{t4}
F_{\gamma}=E_{*}-KS_{\gamma}.
\end{eqnarray}
We note that the polytropic constant $K$ plays the role of a generalized
temperature: $T_{\rm eff}=K$. 
The equilibrium state (\ref{eq5b}) 
is the Tsallis distribution 
\begin{eqnarray}
\label{t4b}
\rho=\left (\frac{\gamma-1}{\gamma}\right
)^{\frac{1}{\gamma-1}}\left\lbrack \frac{1}{K}\left (
\frac{\mu}{m}-\Phi-\Phi_{\rm
ext}-\frac{Q_g}{m}\right )\right\rbrack^{\frac{1}{\gamma-1}}.
\end{eqnarray}
For $\gamma\rightarrow 1$, the Tsallis entropy, the Tsallis free energy and
the Tsallis distribution 
(\ref{t3})-(\ref{t4b}) return the Boltzmann entropy, the Boltzmann  free energy
and the Boltzmann
distribution  (\ref{t1})-(\ref{t2b}) (see \cite{ggp} for more details).

{\it Remark:} The generalized free energy associated with the logotropic
equation of state
(\ref{logo}) can be written
as $F=E_*-AS_{L}$, where 
\begin{eqnarray}
\label{logent}
S_{L}=\int\ln\rho\, d{\bf r}
\end{eqnarray}
is the logarithmic entropy \cite{logotrope}. The
corresponding equilibrium distribution is the
Lorentzian. The logarithmic  entropy, the logarithmic free energy and the 
Lorentzian distribution can be obtained from the Tsallis entropy, the
Tsallis free energy
and the Tsallis distribution 
(\ref{t3})-(\ref{t4b}) in the limit  $\gamma\rightarrow 0$, $K\rightarrow
+\infty$  with $K\gamma=A$ (see \cite{logotrope,ggp} for more details).

\section{Self-similar solution}
\label{sec_sss}

\subsection{Generalized damped GP equation}

We consider the generalized damped power-law GP equation 
\begin{equation}
\label{sss1}
i\hbar \frac{\partial\psi}{\partial t}=-\frac{\hbar^2}{2m}\Delta\psi
+\frac{\hbar^2}{2m}\frac{\Delta|\psi|}{|\psi|}
\psi-\frac{\hbar^2}{2m}\frac{\Delta\lbrack
(|\psi|/\psi_0)^{q}\rbrack}{(|\psi|/\psi_0)^{2-q}}
\psi+\frac{ K\gamma m}{\gamma-1}
|\psi|^{2(\gamma-1)}\psi+m\Phi_{\rm ext}\psi
-i\frac{\hbar}{2}\xi\left\lbrack \ln\left (\frac{\psi}{\psi^*}\right
)-\left\langle \ln\left (\frac{\psi}{\psi^*}\right
)\right\rangle\right\rbrack\psi
\end{equation}
with the harmonic potential (\ref{e3}). This equation is associated with
generalized quantum mechanics (characterized by the parameter $q$) and
generalized Tsallis thermodynamics (characterized by the parameter $\gamma$).
We show below that Eq. (\ref{sss1}) admits
a self-similar solution with a Tsallis invariant profile when the parameters $q$
and $\gamma$ satisfy the relation
\begin{eqnarray}
\label{sss2}
q=2\gamma-1.
\end{eqnarray}
In that case, Eq. (\ref{sss1}) can be rewritten as
\begin{equation}
\label{sss3}
i\hbar \frac{\partial\psi}{\partial t}=-\frac{\hbar^2}{2m}\Delta\psi
+\frac{\hbar^2}{2m}\frac{\Delta|\psi|}{|\psi|}
\psi-\frac{\hbar^2}{2m}\frac{\Delta\lbrack
(|\psi|/\psi_0)^{2\gamma-1}\rbrack}{(|\psi|/\psi_0)^{3-2\gamma}}
\psi+\frac{ K\gamma m}{\gamma-1}
|\psi|^{2(\gamma-1)}\psi+m\Phi_{\rm ext}\psi
-i\frac{\hbar}{2}\xi\left\lbrack \ln\left (\frac{\psi}{\psi^*}\right
)-\left\langle \ln\left (\frac{\psi}{\psi^*}\right
)\right\rangle\right\rbrack\psi.
\end{equation}
For  $q=\gamma=1$, we recover the damped logarithmic GP equation
\begin{equation}
\label{sss4}
i\hbar \frac{\partial\psi}{\partial t}=-\frac{\hbar^2}{2m}\Delta\psi
+2k_B T \ln|\psi|\psi+m\Phi_{\rm ext}\psi
-i\frac{\hbar}{2}\xi\left\lbrack \ln\left (\frac{\psi}{\psi^*}\right
)-\left\langle \ln\left (\frac{\psi}{\psi^*}\right
)\right\rangle\right\rbrack\psi.
\end{equation}
This equation is associated with standard quantum mechanics ($q=1$) and
standard thermodynamics ($\gamma=1$). We show below that this equation admits
a self-similar solution with a Gaussian invariant profile.

{\it Remark:} In order to recover formally the
logarithmic GP equation from the power-law GP equation when $\gamma\rightarrow
1$, we should write the power-law nonlinearity as  $\frac{K\gamma
m}{\gamma-1}(|\psi|^{2(\gamma-1)}-1)$. However, the constant can be absorbed in
the potential energy so, for economy, we do not write it explicitly.

\subsection{Generalized quantum damped Euler equations}

Using the Madelung transformation (see Appendix \ref{sec_gw}), the generalized
damped power-law
GP equation (\ref{sss3}) is equivalent to the generalized quantum damped
polytropic Euler equations
\begin{eqnarray}
\label{sss5}
\frac{\partial\rho}{\partial t}+\nabla\cdot (\rho {\bf u})=0,
\end{eqnarray}
\begin{equation}
\label{sss6}
\frac{\partial {\bf u}}{\partial t}+({\bf u}\cdot \nabla){\bf
u}=-\frac{1}{\rho}\nabla P-\nabla \Phi_{\rm
ext}-\frac{1}{m}\nabla
Q_g-\xi{\bf u},
\end{equation}
involving the  harmonic potential (\ref{e3}), the polytropic equation of
state (\ref{e4}) and the generalized quantum potential (\ref{qe9}). For
$\gamma=1$, we recover the standard quantum damped isothermal Euler equations 
involving the  harmonic potential (\ref{e3}), the isothermal
equation of state (\ref{e5})  and the standard quantum potential (\ref{qe4}).

\subsection{Scaling ansatz}

We look for a self-similar solution of Eqs. (\ref{sss5}) and (\ref{sss6}) of
the form
\begin{eqnarray}
\label{sss7}
\rho({\bf r},t)=\frac{M}{R(t)^d} f\left\lbrack \frac{{\bf
r}-\chi(t){\bf r}_0}{R(t)}\right
\rbrack, \qquad {\bf u}({\bf r},t)=H(t){\bf r}+B(t){\bf r}_0.
\end{eqnarray}
We have assumed that the velocity field is an affine
function of  ${\bf r}$. Defining
\begin{eqnarray}
\label{sss8a}
{\bf x}=\frac{{\bf
r}-\chi(t){\bf r}_0}{R(t)}\qquad  ({\bf r}=R(t){\bf
x}+\chi(t){\bf r}_0),
\end{eqnarray}
we can rewrite Eq. (\ref{sss7}) as
\begin{eqnarray}
\label{sss8}
\rho({\bf r},t)=\frac{M}{R(t)^d} f({\bf x}), \qquad   {\bf u}({\bf
r},t)=H(t)R(t){\bf
x}+\left\lbrack H(t)\chi(t)+B(t)\right\rbrack{\bf r}_0.
\end{eqnarray}
In the foregoing equations $R(t)$ is the typical size (radius) of
the BEC at time $t$, $\langle {\bf
r}\rangle(t)=\chi(t){\bf r}_0$ represents the position of the center of the BEC,
and $f({\bf
x})$ is the
invariant density profile. We assume that the self-similar density profile
contains all the
mass ($M=\int \rho({\bf r},t)\, d{\bf r}$) so that  $\int f({\bf x})\, d{\bf
x}=1$.

The continuity equation
(\ref{sss5}) can be rewritten
as
\begin{eqnarray}
\label{sss9}
\frac{\partial\ln\rho}{\partial t}+\nabla\cdot {\bf u}+ {\bf
u}\cdot \nabla\ln\rho=0.
\end{eqnarray}
From Eq. (\ref{sss7}), we obtain
\begin{eqnarray}
\label{sss10}
\frac{\partial\ln\rho}{\partial t}=-\left (\frac{\dot R}{R}{\bf
x}+\frac{\dot\chi}{R}{\bf r}_0\right )\cdot \nabla_{\bf
x}\ln f-d\frac{\dot R}{R},\qquad \nabla\ln\rho=\frac{1}{R}\nabla_{\bf x}\ln
f,\qquad \nabla\cdot {\bf u}=dH.
\end{eqnarray}
Substituting the foregoing relations into Eq. (\ref{sss9}), we get
\begin{eqnarray}
\label{sss11}
\left (H-\frac{\dot R}{R}\right )\left (d+{\bf x}\cdot \nabla_{\bf x}\ln f\right
)+\frac{{\bf r}_0}{R}\cdot \nabla_{\bf x}\ln f\left (H\chi+B-\dot\chi\right )=0.
\end{eqnarray}
This equation must be satisfied for all ${\bf x}$. This implies
\begin{eqnarray}
\label{sss12}
H(t)=\frac{\dot R}{R}\quad {\rm and}\quad B(t)=\dot\chi-H\chi.
\end{eqnarray}
We note the formal analogy between the first term and the Hubble constant
in cosmology. Using  Eq. (\ref{sss12}) we can rewrite Eq.
(\ref{sss7}) as
\begin{eqnarray}
\label{mal1}
{\bf u}({\bf r},t)=H(t)({\bf r}-\chi(t){\bf r}_0)+\dot\chi{\bf r}_0.
\end{eqnarray}
Similarly,  Eq. (\ref{sss8}) can be rewritten as
\begin{eqnarray}
\label{sss13}
{\bf u}({\bf r},t)=\dot R{\bf x}+\dot\chi{\bf r}_0.
\end{eqnarray}
Comparing Eqs. (\ref{sss8a}) and (\ref{sss13}) we note that ${\bf u}({\bf
r},t)=(\partial{\bf r}/\partial t)_{\bf x}$ where the time derivative is taken
at fixed ${\bf
x}$.

Using Eq.
(\ref{sss7}), the left hand side of
the generalized quantum damped Euler equation (\ref{sss6}) can be
written as
\begin{eqnarray}
\label{sss14}
\frac{\partial {\bf u}}{\partial t}+({\bf u}\cdot \nabla){\bf
u}=(\dot H+H^2){\bf r}+(\dot B+HB){\bf r}_0=(\dot H+H^2)R{\bf x}+\left\lbrack
\dot B+HB+\chi(\dot H+H^2)\right\rbrack {\bf r}_0.
\end{eqnarray}
According to Eq. (\ref{sss12}), we have
\begin{eqnarray}
\label{sss15}
\dot H+H^2=\frac{\ddot R}{R},\qquad \dot B+HB=\ddot\chi-(\dot H+H^2)\chi.
\end{eqnarray}
Combining Eqs. (\ref{sss14}) and (\ref{sss15}), we obtain
\begin{eqnarray}
\label{sss16}
\frac{D{\bf u}}{Dt}\equiv \frac{\partial {\bf u}}{\partial t}+({\bf u}\cdot
\nabla){\bf
u}=\ddot R {\bf x}+\ddot \chi {\bf r}_0.
\end{eqnarray}
Comparing Eqs. (\ref{sss8a}) and (\ref{sss16}) we note that $D{\bf
u}/Dt=(\partial^2{\bf r}/\partial t^2)_{\bf x}$ where the time derivative is
taken
at fixed ${\bf
x}$.

For the polytropic equation of state (\ref{e4}), the pressure term in the
right hand side of the generalized quantum damped Euler equation
(\ref{sss6}) is given by
\begin{eqnarray}
\label{sss17}
-\frac{1}{\rho}\nabla P=-K\gamma\rho^{\gamma-2}\nabla\rho.
\end{eqnarray}
With the scaling ansatz from Eq. (\ref{sss7}),   we obtain
\begin{eqnarray}
\label{sss18}
-\frac{1}{\rho}\nabla
P=-K\gamma\frac{M^{\gamma-1}}{R^{d\gamma-d+1}}f^{\gamma-2}\nabla_{\bf x} f.
\end{eqnarray}
Assuming that $f$ depends only on $x=|{\bf x}|$, we get
\begin{eqnarray}
\label{sss18a}
-\frac{1}{\rho}\nabla
P=-K\gamma\frac{M^{\gamma-1}}{R^{d\gamma-d+1}}f^{\gamma-2}\frac{f'(x)}{x}{\bf
x}.
\end{eqnarray}
On the other hand, we have
\begin{eqnarray}
\label{sss19}
-\nabla\Phi_{\rm ext}=-\omega_0^2 {\bf r}=-\omega_0^2 (R {\bf x}+\chi {\bf r}_0)
\end{eqnarray}
and
\begin{eqnarray}
\label{sss20}
-\xi {\bf u}=-\xi \dot R {\bf x}-\xi\dot\chi {\bf r}_0.
\end{eqnarray}
The generalized quantum force $-(1/m)\nabla Q_g$ will be considered later (see
Appendix \ref{sec_acc}). For
the moment we do not take it into account. Substituting Eqs. (\ref{sss16}),
(\ref{sss18a}), (\ref{sss19})  and (\ref{sss20}) into the generalized quantum
damped polytropic Euler equation
(\ref{sss6}), we get
\begin{eqnarray}
\label{sss21}
\ddot \chi+\xi\dot\chi+\omega_0^2\chi=0
\end{eqnarray}
and
\begin{eqnarray}
\label{sss22}
\ddot R+\xi\dot
R+\omega_0^2 R = -K\gamma\frac{M^{\gamma-1}}{R^{d\gamma-d+1}}f^{\gamma-2}\frac{
f'(x)}{x}.
\end{eqnarray}
The first equation determines the evolution of the center
$\langle {\bf r}\rangle=\chi(t){\bf r}_0$ of the BEC. We see that it follows the
classical equation of motion of a damped particle in a harmonic potential. The
general solution the differential equation (\ref{sss21}) is
given in Appendix \ref{sec_chi}.
 Below,
we focus on
Eq. (\ref{sss22}). The variables of position and time separate provided that
\begin{eqnarray}
\label{sss23}
f^{\gamma-2}\frac{df}{dx}+2Ax=0
\end{eqnarray}
and
\begin{eqnarray}
\label{sss24}
\ddot R+\xi\dot R+\omega_0^2
R=2AK\gamma\frac{M^{\gamma-1}}{R^{d\gamma-d+1}},
\end{eqnarray}
where $A$ is a constant (the factor $2$ has been introduced for
convenience). These differential equations determine the invariant density
profile $f(x)$ of the BEC and the evolution of its radius  $R(t)$.

{\it Remark:} We can add in the external potential a term of the form $\Phi_{\rm
ext}=-{\bf A}\cdot {\bf r}$ corresponding to a constant force. The self-similar
solution remains valid provided that ${\bf r}_0$ lies in the direction of ${\bf
A}$. In that case, Eq. (\ref{sss21}) is replaced by $\ddot
\chi+\xi\dot\chi+\omega_0^2\chi=A/r_0$. The other equations are not altered. We
also note that the self-similar solution remains valid when $\omega_0(t)$ and
$A(t)$ depend on time.

\subsection{Invariant density profile and system's size}

The differential equation (\ref{sss23}) determining the invariant
density profile of the system\footnote{This differential
equation was enlightened in Sec. VI.A. of \cite{langcrit}.} can
be integrated into
\begin{eqnarray}
\label{sss25}
f(x)=\left\lbrack C-(\gamma-1)A x^2\right\rbrack_+^{1/(\gamma-1)},
\end{eqnarray}
where $[x]_+=x$ if $x\ge 0$ and $[x]_+=0$ if $x\le 0$. Therefore, the invariant
profile is given by a Tsallis distribution of
index $\gamma$. We can take $C=A$ 
without loss of generality. Denoting this constant by $Z^{1-\gamma}$, we get
\begin{eqnarray}
\label{sss26}
f(x)=\frac{1}{Z}\left\lbrack 1-(\gamma-1)x^2\right\rbrack_+^{1/(\gamma-1)},
\end{eqnarray}
where $Z$ is determined by the normalization condition $\int f({\bf x})\, d{\bf
x}=1$. This yields 
\begin{eqnarray}
\label{sss27}
Z=\int_0^{x_{\rm max}}\left\lbrack
1-(\gamma-1)x^2\right\rbrack_+^{1/(\gamma-1)}\, S_d x^{d-1}\, dx,
\end{eqnarray}
where $x_{\rm max}=1/\sqrt{\gamma-1}$ if $\gamma\ge 1$ and $x_{\rm max}=+\infty$
if $(d-2)/d<\gamma\le 1$ (the distribution is not normalizable when $\gamma\le
(d-2)/d$). We recall that $S_d=2\pi^{d/2}/\Gamma(d/2)$ is the surface of the
unit sphere in $d$ dimensions.
The integrals can be expressed in
terms of Gamma functions leading to Eqs. (\ref{e8a}) and (\ref{e8b}). On the
other hand, the differential equation (\ref{sss24})
determining the evolution of the system's radius becomes
\begin{eqnarray}
\label{sss28}
\ddot R+\xi\dot R+\omega_0^2
R=2Z^{1-\gamma}K\gamma\frac{M^{\gamma-1}}{R^{d\gamma-d+1}}.
\end{eqnarray}

For $\gamma=1$, corresponding to the isothermal equation of state (\ref{e5}),
the differential equation (\ref{sss23}) reduces to
\begin{eqnarray}
\label{sss29}
\frac{df}{dx}+2Axf(x)=0.
\end{eqnarray}
We can take $A=1$ without loss of generality. In that case, Eq.
(\ref{sss29}) leads to the Gaussian invariant profile 
\begin{eqnarray}
\label{sss29b}
f(x)=\frac{1}{\pi^{d/2}}e^{-x^2},
\end{eqnarray}
where we have used the normalization condition $\int f({\bf x})\, d{\bf
x}=1$. We can check that the Gaussian distribution (\ref{sss29b}) is the
limiting form of the Tsallis distribution (\ref{sss26}) for
$\gamma\rightarrow 1$. On the other hand, the differential equation
(\ref{sss24}) determining the evolution of the system's radius reduces to
\begin{eqnarray}
\label{sss30}
\ddot R+\xi\dot R+\omega_0^2
R=\frac{2k_B T}{mR}.
\end{eqnarray}
This is the
limiting form of the differential equation  (\ref{sss28}) for
$\gamma\rightarrow 1$.

\subsection{Accounting for the quantum potential}
\label{sec_acc}

When $q=2\gamma-1$, as assumed at the begining of this Appendix, the generalized
quantum potential is given by Eq. (\ref{qe9}). When
$\rho({\bf r},t)$ is given by Eq. (\ref{sss7}) where $f({\bf
x})$ is the Tsallis
distribution (\ref{sss26}), we have the relation (see Appendix \ref{sec_gqA})
\begin{eqnarray}
\label{sss31}
-\frac{1}{m}\nabla
Q_g=\frac{\hbar^2}{m^2}\rho_0^{2(1-\gamma)}\frac{M^{
2(\gamma-1)}}{R^{ 2d(\gamma-1)+3}}\frac { 2\gamma-1} { Z^ { 2(\gamma-1) } }
\lbrack 1+d(\gamma-1)\rbrack {\bf x}.
\end{eqnarray}
Since the generalized quantum force (\ref{sss31}) is proportional to ${\bf x}$
with a proportionality constant depending only on $t$, and not on $x$, we
conclude that the
generalized quantum damped polytropic Euler equations (\ref{sss5}) and
(\ref{sss6}) admit a Tsallis
self-similar solution. The differential equation determining the evolution of
the radius $R(t)$ is given by
\begin{eqnarray}
\label{sss32}
\ddot R+\xi\dot R+\omega_0^2
R=2Z^{1-\gamma}K\gamma\frac{M^{\gamma-1}}{R^{d\gamma-d+1}}+\frac{\hbar^2}{m^2}
\rho_0^{2(1-\gamma)}\frac{M^{
2(\gamma-1)}}{R^{ 2d(\gamma-1)+3}}\frac { 2\gamma-1} { Z^ { 2(\gamma-1) } }
\lbrack 1+d(\gamma-1)\rbrack.
\end{eqnarray}

For $q=\gamma=1$,  the standard quantum potential is given by Eq. (\ref{qe4}).
When $\rho({\bf r},t)$ is given by Eq. (\ref{sss7}) where $f({\bf x})$ is 
the Boltzmann
distribution (\ref{sss29b}), we have the relation (see Appendix \ref{sec_gqA})
\begin{eqnarray}
\label{sss33}
-\frac{1}{m}\nabla
Q=\frac{\hbar^2}{m^2}\frac{1}{R^{3}}{\bf x}.
\end{eqnarray}
Using the same argument as before, we conclude that the standard quantum
damped
isothermal Euler equations (\ref{sss5}) and
(\ref{sss6}) admit a Boltzmann self-similar solution. The
differential equation  determining the evolution of
the radius $R(t)$ is given by
\begin{eqnarray}
\label{sss34}
\ddot R+\xi\dot R+\omega_0^2
R=\frac{2k_B T}{mR}+\frac{\hbar^2}{m^2R^3}.
\end{eqnarray}
This is the limiting form of  the differential equation 
(\ref{sss32}) for $\gamma\rightarrow 1$.

\subsection{The function $\chi(t)$}
\label{sec_chi}

The differential equation determining the evolution of $\chi(t)$ 
is given by
\begin{eqnarray}
\label{chi1}
\ddot \chi+\xi\dot\chi+\omega_0^2\chi=0.
\end{eqnarray}
This is the classical equation of motion of a damped particle submitted to a
one dimensional harmonic potential. The solution of this equation is well-known.
We have to
distinguish three cases:

(i) When $\xi>2\omega_0$,
\begin{eqnarray}
\label{chi2}
\chi(t)=e^{-\xi t/2}\left
(Ae^{\frac{1}{2}\sqrt{\xi^2-4\omega_0^2}t}+B e^{-\frac{1}{2}\sqrt{
\xi^2-4\omega_0^2}t}\right ).
\end{eqnarray}

(ii) When $\xi<2\omega_0$,
\begin{eqnarray}
\label{chi3}
\chi(t)=e^{-\xi t/2}\left
\lbrack A \cos\left (\frac{1}{2}\sqrt{4\omega_0^2-\xi^2}t\right )+B
\sin\left (\frac{1}{2}\sqrt{4\omega_0^2-\xi^2}t\right ) \right \rbrack.
\end{eqnarray}

(iii) When $\xi=2\omega_0$,
\begin{eqnarray}
\label{chi4}
\chi(t)=e^{-\xi t/2}\left (At+B\right ).
\end{eqnarray}

In these expressions, $A$ and $B$ are two constants of integration determined by
the initial
conditions. In particular, for $\xi=\omega_0=0$ we have
\begin{eqnarray}
\label{mal2}
\chi(t)=\frac{v}{r_0}t+1.
\end{eqnarray}

In the strong friction limit
$\xi\rightarrow +\infty$, the differential equation determining the evolution of
$\chi(t)$ reduces to
\begin{eqnarray}
\label{chi5}
\dot\chi+\frac{\omega_0^2}{\xi}\chi=0.
\end{eqnarray}
Its solution is
\begin{eqnarray}
\label{chi6}
\chi(t)=e^{-\frac{\omega_0^2}{\xi}t},
\end{eqnarray}
where we have assumed that the density
profile is initially centered on ${\bf r}_0$ so that $\chi(0)=1$. In the absence
of
external force ($\omega_0=0$), Eq. (\ref{chi5}) reduces to
\begin{eqnarray}
\label{chi7}
\dot\chi=0
\end{eqnarray}
and its solution is
\begin{eqnarray}
\label{chi8}
\chi(t)=1.
\end{eqnarray}

\subsection{Moments of the distribution}
\label{sec_rr}

We define the first and  second moments of the
distribution $\rho({\bf r},t)$ by
\begin{eqnarray}
\label{rr1n}
\langle {\bf r}\rangle=\frac{1}{M}\int\rho {\bf r}\, d{\bf r},
\end{eqnarray}
\begin{eqnarray}
\label{rr1}
\langle r^2\rangle=\frac{1}{M}\int\rho r^2\, d{\bf r}=\frac{I}{M},
\end{eqnarray}
where $I$ is the moment of inertia (\ref{gf10b}). Using Eqs. (\ref{sss7})
and (\ref{sss8a}), and recalling that $f({\bf x})$ is spherically symmetric,  we
get
\begin{eqnarray}
\label{rr2n}
\langle {\bf r}\rangle=\chi(t){\bf r}_0,\qquad \langle {\bf
v}\rangle=\frac{d\langle {\bf
r}\rangle}{dt}=\dot\chi{\bf r}_0,
\end{eqnarray}
and
\begin{eqnarray}
\label{rr2}
\langle r^2\rangle=R^2(t) \langle x^2\rangle +\chi^2(t)r_0^2,\qquad \langle
({\bf r}-\chi(t){\bf r}_0)^2\rangle=R^2(t) \langle x^2\rangle.
\end{eqnarray}
For the Gaussian distribution (\ref{e10}), we have
\begin{eqnarray}
\label{rr3}
\langle x^2\rangle =\frac{d}{2}.
\end{eqnarray}
For the Tsallis distribution (\ref{e8}), we have
\begin{eqnarray}
\label{rr4}
\langle x^2\rangle =\frac{d}{d(\gamma-1)+2\gamma}.
\end{eqnarray}
The variance exists provided that $\gamma>d/(d+2)$. In particular, it is not
defined when $\gamma<0$.

\subsection{The wavefunction}
\label{sec_wf}

From Eqs. (\ref{mad5}) and  (\ref{sss13}) we find that the action (phase) is
given by
\begin{eqnarray}
\label{wf1}
S({\bf r},t)=\frac{1}{2}m\dot R R x^2+m R \dot\chi{\bf r}_0\cdot {\bf x}+S_0(t),
\end{eqnarray}
where $S_0(t)$ is a ``constant'' of integration that can depend on time.  
Using Eqs. (\ref{mad1}),  (\ref{sss7}) and (\ref{wf1}), the wavefunction can be
written as 
\begin{equation}
\label{wf2}
\psi({\bf r},t)=\frac{\sqrt{M}}{R(t)^{d/2}}f\left\lbrack \frac{{\bf
r}-\chi(t){\bf r}_0}{R(t)}\right
\rbrack^{1/2}  e^{i\frac{m}{\hbar}\left\lbrack \frac{1}{2}\frac{\dot R}{R}
({\bf r}-\chi(t){\bf r}_0)^2+\dot\chi {\bf
r}_0\cdot ({\bf r}-\chi(t){\bf r}_0)+\frac{S_0(t)}{m}\right\rbrack}.
\end{equation}
It can either represent the condensate wavefunction of a BEC or the wave
packet of a single quantum particle (see footnote 16).
Depending on the interpretation $\langle {\bf
r}\rangle(t)=\chi(t){\bf r}_0$ represents either the position of the center of
the BEC
or the position of the center of the wavepacket associated with the quantum
particle, and $\langle {\bf
v}\rangle(t)=\dot\chi(t){\bf r}_0$ represents its velocity (see
Appendix \ref{sec_rr}). According to Eq. (\ref{chi1}), these quantities follow
the
classical equations of motion of a damped particle in a harmonic potential. This
is a particular case of the Ehrenfest theorem (see Appendix
C of \cite{chavnot}). On the other hand, $R(t)$ is a measure of the size of
the BEC or a
measure of the width of the wavepacket (see Appendix \ref{sec_rr}). Stationary
solutions
can be interpreted as the equilibrium state of the BEC or the localization of
the wave packet of a quantum particle.

For the Gaussian
self-similar solution, using Eq. (\ref{e10}), we get
\begin{equation}
\label{wf3}
\psi({\bf r},t)=\frac{\sqrt{M}}{R(t)^{d/2}\pi^{d/4}}e^{-\frac{({\bf
r}-\chi(t){\bf r}_0)^2}{2R(t)^2}}  e^{i\frac{m}{\hbar}\left\lbrack
\frac{1}{2}\frac{\dot R}{R}
({\bf r}-\chi(t){\bf r}_0)^2+\dot\chi {\bf
r}_0\cdot ({\bf r}-\chi(t){\bf r}_0)+\frac{S_0(t)}{m}\right\rbrack}.
\end{equation}
For a free quantum particle with $\omega_0=\xi=T=0$, $R(t)$
is given by Eq. (\ref{qe7b}), $\chi(t)$ is given by Eq. (\ref{mal2}), and the
wavefunction (\ref{wf3}) takes the form
\begin{eqnarray}
\label{wf3j}
\psi({\bf r},t)&=&\frac{\sqrt{M}}{R(t)^{d/2}\pi^{d/4}}e^{-\frac{({\bf
r}-{\bf r}_0-{\bf v} t)^2}{2R(t)^2}}  e^{i\frac{m}{2\hbar}\frac{\dot R}{R}
({\bf r}-{\bf r}_0-{\bf v} t)^2}e^{i\frac{1}{\hbar}\left ({\bf p}\cdot {\bf
r}-Et\right )}\nonumber\\
&=&\frac{\sqrt{M}}{\left\lbrack\pi R_0^2\left
(1+\frac{\hbar^2t^2}{m^2R_0^4}\right
)\right\rbrack^{d/4}}e^{-\frac{({\bf
r}-{\bf r}_0-{\bf v} t)^2}{2R_0^2\left (1+\frac{i\hbar t}{mR_0^2}\right )}} 
e^{i\frac{1}{\hbar}\left ({\bf p}\cdot {\bf
r}-Et\right )},
\end{eqnarray}
with ${\bf p}=m{\bf v}$ and $E=p^2/2m$. For
$t\rightarrow +\infty$, the radius $R(t)\rightarrow +\infty$  [see Eq.
(\ref{qe7b})] because of the spreading of the
wave packet. In that case, Eq.
(\ref{wf3j}) reduces to a pure plane wave $\psi({\bf r},t)\propto
e^{i\frac{1}{\hbar}\left ({\bf
p}\cdot {\bf
r}-Et\right )}$ and the quantum particle is completely delocalized. On the
other hand, when the
differential equation (\ref{sss34}) determining the evolution of the
radius $R(t)$ has a stable stationary solution, $R(t)=R_e$,\footnote{In that
case we have to assume that $T<0$ (see Sec. 6.2 of
\cite{chavnot}).} the logarithmic GP equation (\ref{qe6}) admits a solitonic
solution of the
form 
\begin{equation}
\label{wf3s}
\psi({\bf r},t)=\frac{\sqrt{M}}{R_e^{d/2}\pi^{d/4}}e^{-\frac{({\bf
r}-\chi(t){\bf r}_0)^2}{2R_e^2}}  e^{i\frac{m}{\hbar}\left\lbrack
\dot\chi {\bf
r}_0\cdot ({\bf r}-\chi(t){\bf r}_0)+\frac{S_0(t)}{m}\right\rbrack}.
\end{equation}
In particular, when $\xi=\omega_0=0$, the foregoing
equation reduces to
\begin{equation}
\label{wf3sb}
\psi({\bf r},t)=\frac{\sqrt{M}}{R_e^{d/2}\pi^{d/4}}e^{-\frac{({\bf
r}-{\bf r}_0-{\bf v}t)^2}{2R_e^2}}  e^{i\frac{1}{\hbar}\left ({\bf p}\cdot {\bf
r}-Et\right )}.
\end{equation}
This solution is called a
gausson \cite{bbm}. It corresponds to a uniformly moving Gaussian wave packet
modulated by the de Broglie plane wave (${\bf p}=\hbar {\bf k}$, $E=\hbar
\omega$). The gausson arises from the invariance of the logarithmic GP equation
under Galilean transformation. In this manner, we can use the static solution of
the logarithmic GP equation to generate uniformly moving solutions characterized
by the velocity ${\bf v}$ and the initial position ${\bf r}_0$. The particle is
localized in a region of size [see Eq. (\ref{sss34})]
\begin{equation}
R_e=\frac{\hbar}{\sqrt{2mk_B
|T|}},
\end{equation}
which has the form of a de Broglie length (with a negative
temperature). If, for curiosity, we compare $R_e$ with the classical radius of
the electron
$r_e=e^2/m_e c^2=2.82\times 10^{-15}\, {\rm m}$, we obtain a temperature scale
$k_B |T|=m_e
c^4\hbar^2/2e^4= m_e c^2/2\alpha^2=4.80\, {\rm GeV}$, where
$\alpha=e^2/\hbar c\simeq
1/137$ is the fine-structure constant.

For the Tsallis
self-similar solution, using Eq. (\ref{e8}), we get
\begin{equation}
\label{wf4}
\psi({\bf r},t)=\frac{\sqrt{M}}{Z^{1/2}R(t)^{d/2}}
\left \lbrack 1-(\gamma-1)\left ( \frac{{\bf
r}-\chi(t){\bf r}_0}{R(t)}\right )^2\right\rbrack_+^{1/[2(\gamma-1)]}
e^{i\frac{m}{\hbar}\left\lbrack \frac{1}{2}\frac{\dot R}{R}
({\bf r}-\chi(t){\bf r}_0)^2+\dot\chi {\bf
r}_0\cdot ({\bf r}-\chi(t){\bf r}_0)+\frac{S_0(t)}{m}\right\rbrack}.
\end{equation}
When the differential equation (\ref{sss32}) determining the evolution of
the
radius $R(t)$ has a stable stationary solution, $R(t)=R_e$, the generalized
power-law GP equation (\ref{qe11}) admits a
solitonic
solution of the
form 
\begin{equation}
\label{wf4q}
\psi({\bf r},t)=\frac{\sqrt{M}}{Z^{1/2}R_e^{d/2}}
\left \lbrack 1-(\gamma-1)\left ( \frac{{\bf
r}-\chi(t){\bf r}_0}{R_e}\right )^2\right\rbrack_+^{1/[2(\gamma-1)]}
e^{i\frac{m}{\hbar}\left\lbrack \dot\chi {\bf
r}_0\cdot ({\bf r}-\chi(t){\bf r}_0)+\frac{S_0(t)}{m}\right\rbrack}.
\end{equation}
In particular, when $\xi=\omega_0=0$, the foregoing equation reduces to
\begin{equation}
\label{wf4b}
\psi({\bf r},t)=\frac{\sqrt{M}}{Z^{1/2}R_e^{d/2}}
\left \lbrack 1-(\gamma-1)\left ( \frac{{\bf
r}-{\bf r}_0-{\bf v}t}{R_e}\right )^2\right\rbrack_+^{1/[2(\gamma-1)]}
 e^{i\frac{1}{\hbar}\left ({\bf p}\cdot {\bf
r}-\frac{p^2}{2m}t\right )}.
\end{equation}
This solution could be called a $\gamma$-gausson. It is
localized in a region of size [see Eq. (\ref{sss32})]
\begin{equation} 
R_e=\left\lbrace \frac{
\hbar^2}{2m^2|K|}
\rho_0^{2(1-\gamma)}\frac{M^{\gamma-1}}{Z^ {\gamma-1}}\frac { 2\gamma-1}
{\gamma}
\lbrack 1+d(\gamma-1)\rbrack\right\rbrace^{1/\lbrack 2+d(\gamma-1)\rbrack}.
\end{equation}

{\it Remark:} The self-similar solution obtained in the previous subsections can
also be obtained by substituting the Ansatz (\ref{wf2}) for the wavefunction
into the generalized damped power-law GP equation (\ref{sss1}) and separating
real and
imaginary parts. However, the algebra is more complicated (and more obscure)
than using the hydrodynamic representation of the GP equation.

\subsection{Bohm's Lagrangian point of view}
\label{sec_lag}

The fluid equations (\ref{sss5}) and (\ref{sss6}) arising from the Madelung
transformation have been written in the Eulerian point of view, i.e., the
velocity field ${\bf u}({\bf r},t)$ is calculated at a fixed position ${\bf r}$.
Alternatively, following Bohm \cite{bohm1,bohm2}, we can adopt a Lagrangian
point of
view and
consider the motion of ``fluid particles'', which form an ensemble of
``Bohmian particles'', with position ${\bf r}_{\rm B}(t)$. Their equation of
motion is
obtained by
writing
\begin{eqnarray}
\label{lag1}
\frac{d{\bf r}_{\rm B}}{dt}={\bf u}({\bf r}_{\rm B}(t),t).
\end{eqnarray}
A given Bohmian particle can be labeled by its initial position ${\bf r}_{\rm
B,0}$. Now, for the self-similar solution considered in this Appendix, the
velocity
field ${\bf u}({\bf r},t)$ is given by Eq. (\ref{sss13}). Using Eq.
(\ref{sss8a}) it can be
rewritten as
\begin{eqnarray}
\label{lag2}
{\bf u}({\bf r},t)=\frac{\dot R}{R}({\bf r}-\langle {\bf r}\rangle)+\dot{\langle
{\bf r}\rangle},
\end{eqnarray}
where we recall that $\langle {\bf r}\rangle=\chi(t){\bf r}_0$ denotes the
position of the center of the wavepacket, which may also be interpreted as the
position of a ``classical particle'' following the Newtonian equation of
motion (\ref{chi1}) (see Appendix \ref{sec_wf}). Combining
Eqs. (\ref{lag1}) and (\ref{lag2}), we obtain
\begin{eqnarray}
\label{lag3}
\frac{d}{dt}({\bf r}_{\rm B}-\langle {\bf r}\rangle)=\frac{\dot R}{R}({\bf
r}_{\rm B}-\langle
{\bf r}\rangle),
\end{eqnarray}
which is immediately integrated into
\begin{eqnarray}
\label{lag4}
({\bf r}_{\rm B}-\langle {\bf r}\rangle)(t)= \frac{R(t)}{R_0} ({\bf r}_{\rm
B}-\langle {\bf
r}\rangle)_0.
\end{eqnarray}
Interestingly, this equation gives the evolution of the separation between a
``Bohmian  particle'' with position ${\bf r}_{\rm B}(t)$ and
the ``classical
particle'' (or the center of the wave packet) with position  $\langle {\bf
r}\rangle (t)$. We note
that the Bohmian particle which coincides with the classical particle
initially
(${\bf r}_{\rm B,0}=\langle {\bf r}\rangle_0$), coincides with it for all
times (${\bf
r}_{\rm B}(t)=\langle {\bf r}\rangle (t)$).

As a simple example, let us consider the
case of a free quantum particle with $\omega_0=T=0$ described by the standard
Schr\"odinger equation. In the nondissipative
limit $\xi=0$, using Eqs.
(\ref{qe7b}), (\ref{mal1}) and Eq. (\ref{mal2}), we obtain 
\begin{eqnarray}
\label{lag5}
{\bf u}({\bf r},t)=\frac{\frac{\hbar^2 t}{m^2R_0^2}}{R_0^2+\frac{\hbar^2
t^2}{m^2R_0^2}}({\bf r}-{\bf v} t-{\bf r}_0)+{\bf v}.
\end{eqnarray}
Solving Eq. (\ref{lag1}) with this velocity field, we find that
\begin{eqnarray}
\label{lag6}
{\bf r}_{\rm B}(t)-{\bf r}_0=({\bf r}_{\rm B,0}-{\bf r}_0)\sqrt{1+\frac{\hbar^2
t^2}{m^2R_0^4}}+{\bf v} t,
\end{eqnarray}
which gives the evolution of the Bohmian particle with initial
position ${\bf r}_{\rm
B,0}$.
We can check that Eq.
(\ref{lag6}) is equivalent to Eq. (\ref{lag4}). The velocity of the
Bohmian particle is\footnote{For a pure plane wave, the Bohmian particles all
have the same
velocity ${\bf v}_{\rm B}={\bf u}=\nabla S/m=\hbar {\bf k}/m={\bf p}/m={\bf v}$.
On the other hand, the
wave moves with a phase velocity $v_{\phi}=\omega/k=E/p=p/2m$. They
differ by a factor $2$.}
\begin{eqnarray}
\label{lag7}
{\bf v}_{\rm B}(t)=({\bf r}_{\rm B,0}-{\bf
r}_0)\frac{\frac{\hbar^2t}{m^2R_0^4}}{\sqrt{1+\frac{\hbar^2
t^2}{m^2R_0^4}}}+{\bf v}.
\end{eqnarray}
In Bohm's interpretation of quantum mechanics, a wavepacket is represented by
an ensemble of particles. The Bohmian particles with $r_{\rm B,0}<r_0$
have a smaller velocity than the classical particle ($v_B<v$) while the Bohmian
particles with $r_{\rm B,0}>r_0$ have a larger
velocity than the classical particle ($v_B<v$). Therefore, relative to the
classical particle (center of the wavepacket) the Bohmian particles with $r_{\rm
B}<\langle r\rangle$ move to the left and the particles with $r_{\rm
B}>\langle r\rangle$ move to the
right. This accounts for the spreading of the wavepacket ($r_{\rm B}-\langle
r\rangle\rightarrow \pm\infty$ for $t\rightarrow +\infty$). In the strong
friction limit $\xi\rightarrow
+\infty$, using
Eqs.
(\ref{qe7bb}), (\ref{mal1}) and Eq. (\ref{chi8}), we obtain 
\begin{eqnarray}
\label{lag8}
{\bf u}({\bf r},t)=\frac{\frac{\hbar^2}{\xi m^2}}{R_0^4+\frac{4\hbar^2
t}{\xi m^2}}({\bf r}-{\bf r}_0),
\end{eqnarray}
\begin{eqnarray}
\label{lag9}
{\bf r}_{\rm B}(t)-{\bf r}_0=({\bf r}_{\rm B,0}-{\bf r}_0)\left
(1+\frac{4\hbar^2
t}{\xi m^2R_0^4}\right )^{1/4},
\end{eqnarray}
\begin{eqnarray}
\label{lag10}
{\bf v}_{\rm B}(t)=({\bf r}_{\rm B,0}-{\bf
r}_0)  \left
(1+\frac{4\hbar^2
t}{\xi m^2R_0^4}\right )^{-3/4}\frac{\hbar^2}{\xi m^2R_0^4}.
\end{eqnarray}
The interpretation is essentially the same as given previously, except that the
wavepacket does not move ($v=0$) in the present case. The extension of the
preceding results to the generalized Schr\"odinger equation associated with
Tsallis distributions will be given in a future contribution \cite{prep}.

\subsection{Another self-similar solution}
\label{sec_oss}

Let us look for a self-similar solution of Eqs. (\ref{sss5}) and (\ref{sss6})
of the form of Eq. (\ref{sss7}) without assuming the relation (\ref{sss2}),
i.e., using the general expression (\ref{gw4}) of the quantum potential.
Substituting Eq.  (\ref{sss7}) into Eq. (\ref{gw4}) and taking its gradient,
we get
\begin{eqnarray}
\label{oss1}
-\frac{1}{m}\nabla
Q_g=\frac{\hbar^2}{2m^2}\frac{1}{\rho_0^{q-1}}\frac{1}{R^3}\left
(\frac{M}{R^d}\right )^{q-1}\frac{1}{x}\frac{d}{dx}\left\lbrack
\frac{\Delta_{\bf x}f(x)^{q/2}}{f(x)^{1-q/2}}\right\rbrack {\bf x}.
\end{eqnarray}
Substituting Eqs. (\ref{sss16}),
(\ref{sss18a}), (\ref{sss19}),  (\ref{sss20}) and (\ref{oss1}) into the
generalized quantum
damped polytropic Euler equation
(\ref{sss6}), we get Eq. (\ref{sss21}) and
\begin{eqnarray}
\label{oss2}
\ddot R+\xi \dot R+\omega_0^2
R=-K\gamma\frac{M^{\gamma-1}}{R^{d\gamma-d+1}}f^{\gamma-2}(x)\frac{f'(x)}{x}
+\frac {\hbar^2}{2m^2}\frac{1}{\rho_0^{q-1}}\frac{1}{R^3}
\left
(\frac{M}{R^d}\right )^{q-1}\frac{1}{x}\frac{d}{dx}\left\lbrack
\frac{\Delta_{\bf x}f(x)^{q/2}}{f(x)^{1-q/2}}\right\rbrack.
\end{eqnarray}
In general, the variables of position and time do not separate so that Eqs.
(\ref{sss5}) and (\ref{sss6}) do not systematically admit a self-similar
solution. An exception, as we have seen, is when $q=2\gamma-1$ because, in that
case, the terms depending on $x$ in Eq. (\ref{oss2}) become constant for the
Tsallis distribution (\ref{sss26}). Indeed, one has (see Appendix
\ref{sec_gqA}):
\begin{eqnarray}
\label{oh}
f^{\gamma-2}(x)\frac{f'(x)}{x}=-2Z^{1-\gamma},\qquad
\frac{1}{x}\frac{d}{dx}\left\lbrack
\frac{\Delta_{\bf
x}f(x)^{\gamma-1/2}}{f(x)^{3/2-\gamma}}\right\rbrack=\frac{4}{Z^{2(\gamma-1)}}
\left (\gamma-\frac{1}{2}\right )\left\lbrack d(\gamma-1)+1\right\rbrack.
\end{eqnarray}
In that case, we recover Eq. (\ref{sss32}). Another exception is when the
two terms on the right hand side of Eq. (\ref{oss2}) have the same
time dependence. This imposes $d\gamma-d+1=3+d(q-1)$, i.e., 
\begin{eqnarray}
\label{oss3}
\gamma=q+\frac{2}{d}.
\end{eqnarray}
For $q=1$ this condition corresponds to $\gamma=1+2/d$.\footnote{Interestingly,
this is the polytropic index of a nonrelativistic Fermi gas at
$T=0$ in dimension $d$ \cite{prdwd}.} In that case, the variables of position
and
time separate provided that 
\begin{eqnarray}
\label{oss4}
f(x)^{2/d+q-2}\frac{df}{dx}-\frac{1}{K}\frac{1}{q+\frac{2}{d}}\frac{1}{M^{
2/d}}\frac
{\hbar^2}{2m^2}\frac{1}{\rho_0^{q-1}}\frac{d}{dx}\left\lbrack
\frac{\Delta_{\bf x}f(x)^{q/2}}{f(x)^{1-q/2}}\right\rbrack+2Ax=0
\end{eqnarray}
and
\begin{eqnarray}
\label{oss5}
\ddot R+\xi \dot R+\omega_0^2
R=2AK\left
(q+\frac{2}{d}\right )\frac{M^{2/d}}{R^{3}} \left (\frac{M}{R^d}\right )^{q-1},
\end{eqnarray}
where $A$ is a constant. These
differential equations determine the invariant density
profile $f(x)$ and the evolution of the radius $R(t)$ of the system.
For $q=1$ they
reduce to
\begin{eqnarray}
\label{oss6}
f(x)^{2/d-1}\frac{df}{dx}-\frac{1}{K}\frac{1}{1+\frac{2}{d}}\frac{1}{M^{
2/d}}\frac
{\hbar^2}{2m^2}\frac{d}{dx}\left\lbrack
\frac{\Delta_{\bf x}\sqrt{f(x)}}{\sqrt{f(x)}}\right\rbrack+2Ax=0
\end{eqnarray}
and
\begin{eqnarray}
\label{oss7}
\ddot R+\xi \dot R+\omega_0^2
R=2AK\left
(1+\frac{2}{d}\right )\frac{M^{2/d}}{R^{3}}.
\end{eqnarray}
The  invariant density profile $f(x)$ is different from the Gaussian or
the Tsallis profiles considered previously.\footnote{Such profiles would be a
particular solution of Eqs. (\ref{oss4}) and (\ref{oss6}) provided that
$q=2\gamma-1$ and $\gamma=q+2/d$, leading to $\gamma=(d-2)/d$. But, in that
case, the profile is not normalizable.}

{\it Remark:} In the case where $K=0$, Eq. (\ref{oss2}) reduces to 
\begin{eqnarray}
\label{oss2b}
\ddot R+\xi \dot R+\omega_0^2
R=\frac {\hbar^2}{2m^2}\frac{1}{\rho_0^{q-1}}\frac{1}{R^3}
\left
(\frac{M}{R^d}\right )^{q-1}\frac{1}{x}\frac{d}{dx}\left\lbrack
\frac{\Delta_{\bf x}f(x)^{q/2}}{f(x)^{1-q/2}}\right\rbrack.
\end{eqnarray}
The variables of position and time separate provided that
\begin{eqnarray}
\label{mae}
\frac{1}{x}\frac{d}{dx}\left\lbrack
\frac{\Delta_{\bf
x}f(x)^{q/2}}{f(x)^{1-q/2}}\right\rbrack=\frac{2q}{Z^{q-1}}\left\lbrack
\frac{d}{2}(q-1)+1\right\rbrack
\end{eqnarray}
and
\begin{eqnarray}
\label{oss5b}
\ddot R+\xi \dot R+\omega_0^2
R=\frac{\hbar^2}{m^2} \frac{1}{\rho_0^{q-1}}\frac{1}{R^3}\left
(\frac{M}{R^d}\right )^{q-1}\frac{1}{Z^{q-1}}q\left\lbrack
\frac{d}{2}(q-1)+1\right\rbrack,
\end{eqnarray}
where $Z$ is a constant (the right hand side of Eq. (\ref{mae}) has been written
under that form for commodity). A particular solution of Eq. (\ref{mae}) is the
Tsallis distribution (\ref{sss26}) with index $\gamma=(q+1)/2$ [see Eq.
(\ref{oh})], reducing to the Gaussian for $q=1$. However, the Tsallis
distribution may not be the only solution of  Eq. (\ref{mae}).

\subsection{Eigenenergy}
\label{sec_eigen}

The eigenenergy $E$ of the generalized time-independent GP
equation can be obtained by applying Eq. (\ref{tigp4}) at $r=0$. This
yields
\begin{equation}
\label{ei1}
E=mh\lbrack \rho(0)\rbrack+Q_g(0).
\end{equation}
In the isothermal case [see Eq. (\ref{lp})], we
have
\begin{equation}
\label{ei2}
V(\rho)=\frac{k_B T}{m}\rho (\ln\rho -1),\qquad h(\rho)=V'(\rho)=\frac{k_B
T}{m}\ln\rho.
\end{equation}
In the polytropic case  [see Eq. (\ref{pp})], we have
\begin{equation}
\label{ei3}
V(\rho)=\frac{K}{\gamma-1}\rho^{\gamma},\qquad
h(\rho)=V'(\rho)=\frac{K\gamma}{\gamma-1}\rho^{\gamma-1}.
\end{equation}
Let us first consider, for simplicity, the
case $T=K=0$ for which Eq. (\ref{ei1}) reduces to
\begin{equation}
\label{ei4}
E=Q_g(0).
\end{equation}
For the standard Schr\"odinger equation ($q=1$), the equilibrium density is
Gaussian. Using Eqs. (\ref{sss34}), (\ref{ei4}) and (\ref{gq5}),
we recover the standard results
\begin{equation}
\label{ei5}
R_e=\left (\frac{\hbar}{m\omega_0}\right
)^{1/2},\qquad E=\frac{d}{2}\hbar\omega_0
\end{equation}
of the quantum harmonic oscillator's ground state. For the generalized
Schr\"odinger equation
($q=2\gamma-1$), the equilibrium density is a  Tsallis distribution.
Using Eqs. (\ref{sss32}), (\ref{ei4}) and (\ref{gq10bb}), we find that
\begin{equation}
\label{ei6}
R_e=\left \lbrace
\frac{\hbar^2}{m^2\omega_0^2}\rho_0^{2(1-\gamma)}M^{2(\gamma-1)}
\frac{2\gamma-1}{Z^{2(\gamma-1)}}\lbrack
1+d(\gamma-1)\rbrack\right\rbrace^{1/\lbrack
2d(\gamma-1)+4\rbrack},\qquad
E=\frac{d}{2}m\omega_0^2R_e^2\frac{1}{1+d(\gamma-1)}.
\end{equation}
For $\gamma=1$, we recover Eq. (\ref{ei5}). The general case ($T\neq 0$ or
$K\neq 0$) can be treated similarly by
substituting the expression (\ref{ei2}) or (\ref{ei3}) of the enthalpy  in the
eigenenergy (\ref{ei1}) and using $\rho(0)=(M/R_e^d)f(0)$ with
$f(0)=1/\pi^{d/2}$ for the Gaussian distribution and $f(0)=1/Z$ for the Tsallis
distribution.

\section{Generalized quantum mechanics}
\label{sec_gq}

In this Appendix, we expose the ideas leading to the generalized 
quantum damped  Euler equations (\ref{mad12}) and (\ref{mad13}) which are
equivalent to the generalized damped GP equation (\ref{gw1}). Our aim
is to
motivate the introduction of the  generalized quantum potential (\ref{gw4})
associated with the nonlinear Laplacian operator (\ref{gw2b}).

\subsection{Generalized quantum potential}
\label{sec_gqA}

Let us first consider the standard quantum potential
\begin{equation}
\label{gq1}
Q=-\frac{\hbar^2}{2m}\frac{\Delta
\sqrt{\rho}}{\sqrt{\rho}}=-\frac{\hbar^2}{4m}\left\lbrack
\frac{\Delta\rho}{\rho}-\frac{1}{2}\frac{(\nabla\rho)^2}{\rho^2}\right\rbrack.
\end{equation}
Applying the Laplacian operator to the Gaussian distribution (\ref{e10}), we 
find that
\begin{eqnarray}
\label{gq2}
\Delta_{\bf x}\sqrt{f}=\sqrt{A}(x^2-d)
e^{-x^2/2}.
\end{eqnarray}
Therefore, the ratio 
\begin{eqnarray}
\label{gq3}
\frac{\Delta_{\bf
x}\sqrt{f}}{\sqrt{f}}=x^2-d
\end{eqnarray}
is a quadratic fonction of $x$. As a result, the self-similar Gaussian profile
defined by Eqs. (\ref{e6}) and (\ref{e10}) satisfies
\begin{eqnarray}
\label{gq4}
\frac{\Delta\sqrt{\rho}}{\sqrt{\rho}}=\frac{1}{R^2}\left
(\frac{r^2}{R^2}-d\right ).
\end{eqnarray}
Substituting this result into Eq. (\ref{gq1}), we obtain
\begin{eqnarray}
\label{gq5}
Q=-\frac{\hbar^2}{2mR^2}\left (\frac{r^2}{R^2}-d\right ),
\end{eqnarray}
implying
\begin{eqnarray}
\label{gq6}
-\frac{1}{m}\nabla
Q=\frac{\hbar^2}{m^2}\frac{1}{R^3} {\bf x}.
\end{eqnarray}
Therefore, for a Gaussian self-similar distribution, the standard quantum force
$-(1/m)\nabla Q$ is proportional to ${\bf x}$ with a prefactor depending on
time but not on $x$. According to the calculations of
Appendix \ref{sec_sss}, this implies that the standard damped logarithmic GP
equation
(\ref{qe6}),
or the standard quantum damped isothermal
Euler equations (\ref{e5}), (\ref{qe2}) and ({\ref{qe3}), admit an
exact self-similar solution with a Gaussian invariant
profile.

We now generalize this procedure to the case of polytropes. We
want to find a generalized form of quantum potential such that the generalized
damped power-law GP equation (\ref{qe11}), or the generalized quantum damped 
polytropic Euler equations (\ref{e4}), (\ref{qe2gen}) and (\ref{qe3gen}), admit
an exact self-similar solution
with a Tsallis invariant profile. Applying the Laplacian operator to a power
$\alpha$ of the Tsallis distribution (\ref{e8}) we find that
\begin{eqnarray}
\label{gq7}
\Delta_{\bf x}(f^{\alpha})=\frac{2\alpha}{Z^{\alpha}}\left\lbrack
1-(\gamma-1)x^2\right\rbrack_+^{\frac{\alpha-2(\gamma-1)}{\gamma-1}}\left\lbrack
(2\alpha+(d-2)(\gamma-1))x^2-d\right\rbrack.
\end{eqnarray}
Therefore, the ratio  
\begin{eqnarray}
\label{gq8}
\frac{\Delta_{\bf
x}(f^{\alpha})}{f^{\alpha-2(\gamma-1)}}=\frac{2\alpha}{Z^{2(\gamma-1)}}
\left\lbrack
(2\alpha+(d-2)(\gamma-1))x^2-d\right\rbrack
\end{eqnarray}
is a quadratic fonction of $x$. As a result,  the self-similar Tsallis profile
defined by Eqs.  (\ref{e6}) and (\ref{e8}) satisfies
\begin{eqnarray}
\label{gq9}
\frac{\Delta(\rho^{\alpha})}{\rho^{\alpha-2(\gamma-1)}}=\frac{M^{
2(\gamma-1)}}{R^{ 2d(\gamma-1)+2}}\frac { 2\alpha } { Z^ { 2(\gamma-1) } }
\left\lbrack
(2\alpha+(d-2)(\gamma-1))\frac{r^2}{R^2}-d\right\rbrack.
\end{eqnarray}
This suggests introducing the generalized quantum potential
\begin{eqnarray}
\label{gq10}
Q_g=-\frac{\hbar^2}{2m}\frac{\Delta\lbrack(\rho/\rho_0)^{\alpha}\rbrack}{
(\rho/\rho_0)^ { \alpha-2(\gamma-1) } },
\end{eqnarray}
which reduces to the standard quantum potential (\ref{gq1}) for $\gamma=1$
and $\alpha=1/2$. Substituting Eq. (\ref{gq9}) into Eq. (\ref{gq10}), we obtain
\begin{eqnarray}
\label{gq10b}
Q_g=-\frac{\hbar^2}{2m}\rho_0^{2(1-\gamma)}\frac{M^{
2(\gamma-1)}}{R^{ 2d(\gamma-1)+2}}\frac { 2\alpha } { Z^ { 2(\gamma-1) } }
\left\lbrack
(2\alpha+(d-2)(\gamma-1))\frac{r^2}{R^2}-d\right\rbrack
\end{eqnarray}
implying
\begin{eqnarray}
\label{gq11}
-\frac{1}{m}\nabla
Q_g=\frac{\hbar^2}{m^2}\rho_0^{2(1-\gamma)}\frac{M^{
2(\gamma-1)}}{R^{ 2d(\gamma-1)+3}}\frac { 2\alpha } { Z^ { 2(\gamma-1) } }
\left\lbrack 2\alpha+(d-2)(\gamma-1)\right\rbrack {\bf x}.
\end{eqnarray}
Therefore, for a Tsallis self-similar distribution, the generalized quantum
force
$-(1/m)\nabla Q_g$ is proportional to ${\bf x}$ with a prefactor depending
on time but not on $x$. According to the calculations of
Appendix \ref{sec_sss}, this implies that the generalized damped power-law GP
equation (\ref{qe11}), or
the generalized quantum damped polytropic Euler equations (\ref{e4}),
(\ref{qe2gen}) and (\ref{qe3gen}), admit an
exact self-similar solution with a Tsallis invariant profile.

We note that the generalized
quantum potential (\ref{gq10}) depends on two parameters
$\alpha$ and $\gamma$. We shall impose
the
relation
\begin{eqnarray}
\label{qg12}
\alpha=\gamma-\frac{1}{2}.
\end{eqnarray}
This relation allows us to write the functional associated with the 
generalized quantum potential (\ref{gq10}) as a generalized von
Weizs\"acker functional (see Appendix \ref{sec_gvw}). With the relation
(\ref{qg12}), the generalized
quantum potential (\ref{gq10}) takes the form
\begin{eqnarray}
\label{gq13}
Q_g=-\frac{\hbar^2}{2m}\frac{\Delta\lbrack(\rho/\rho_0)^{\gamma-1/2}
\rbrack } {
(\rho/\rho_0)^{3/2-\gamma} }.
\end{eqnarray}
It can also be written as
\begin{eqnarray}
\label{gq14}
Q_g=-\frac{\hbar^2}{4m}(2\gamma-1)\left\lbrack
\frac{\Delta(\rho/\rho_0)}{(\rho/\rho_0)^{3-2\gamma}}-\frac{1}{2}
(3-2\gamma)\frac{\lbrack \nabla
(\rho/\rho_0)\rbrack^2}{(\rho/\rho_0)^{4-2\gamma}}\right\rbrack.
\end{eqnarray}
Setting $q=2\gamma-1$, we obtain Eq. (\ref{gw4}).
For $\gamma=1$ we recover the standard quantum potential 
(\ref{gq1}). On the other hand, with the relation (\ref{qg12}),
Eqs. (\ref{gq10b}) and (\ref{gq11})
reduce to
\begin{eqnarray}
\label{gq10bb}
Q_g=-\frac{\hbar^2}{2m}\rho_0^{2(1-\gamma)}\frac{M^{
2(\gamma-1)}}{R^{ 2d(\gamma-1)+2}}\frac { 2\gamma-1 } { Z^ { 2(\gamma-1) } }
\left\lbrack
(2\gamma-1+(d-2)(\gamma-1))\frac{r^2}{R^2}-d\right\rbrack,
\end{eqnarray}
\begin{eqnarray}
\label{gq15}
-\frac{1}{m}\nabla
Q_g=\frac{\hbar^2}{m^2}\rho_0^{2(1-\gamma)}\frac{M^{
2(\gamma-1)}}{R^{ 2d(\gamma-1)+3}}\frac { 2\gamma-1} { Z^ { 2(\gamma-1) } }
(1+d(\gamma-1)) {\bf x}.
\end{eqnarray}
For $\gamma=1$, we recover the results from Eqs. (\ref{gq5}) and (\ref{gq6}).

\subsection{Generalized von Weizs\"acker functional}
\label{sec_gvw}

We want to find a functional $\Theta^g_Q$ such
that
\begin{eqnarray}
\label{gvw1}
\delta \Theta^g_Q= \int \frac{Q_g}{m}\delta\rho\, d{\bf r},
\end{eqnarray}
where $Q_g$ is the generalized quantum potential defined by Eq. (\ref{gq10}).
The identity (\ref{gvw1}) is necessary to obtain the
equilibrium condition (\ref{eq3}) from the extremization of the generalized free
energy functional (\ref{gf1}). In this manner, a stationary solution of the
generalized damped GP equation (\ref{gw1}) is guaranteed to be an extremum of
free energy at
fixed mass
and the $H$-theorem from Appendix \ref{sec_gh} is satisfied. In standard quantum
mechanics, the functional $\Theta_Q$ associated with the standard quantum
potential (\ref{gq1}) is the von Weizs\"acker functional
(\ref{gf7}).\footnote{It corresponds to the quantum kinetic energy term
$\Theta_Q=(\hbar^2/8m^2)\int [(\nabla\rho)^2/\rho]\, d{\bf r}$ when we perform
the Madelung
transformation in the total kinetic energy $\Theta=\langle
\psi|H|\psi\rangle=(\hbar^2/2m^2)\int |\nabla\psi|^2\, d{\bf r}$ of a quantum
particle, the other term
being the classical kinetic energy $\Theta_c=(1/2)\int \rho {\bf u}^2\, d{\bf
r}$. We thus have $\Theta=\Theta_c+\Theta_Q$ (see \cite{ggp} for more details).}
The object of this Appendix
is to find the proper generalization of this functional in relation to the
generalized quantum potential (\ref{gq10}).

Let us consider a functional of the
form
\begin{eqnarray}
\label{gvw2}
\Theta_Q^g=C\int \lbrack\nabla (\rho^{\alpha})\rbrack^2\, d{\bf r}=\alpha^2 C
\int \frac{(\nabla
\rho)^2}{\rho^{2(1-\alpha)}}\, d{\bf r},
\end{eqnarray}
where $C$ is a constant. Its first variations are
\begin{eqnarray}
\label{gvw3}
\delta \Theta_Q^g=-2\alpha C \int
\frac{\Delta(\rho^{\alpha})}{\rho^{1-\alpha}}\delta\rho\, d{\bf r}.
\end{eqnarray}
Considering expression (\ref{gq10}) of the generalized quantum potential,
we
obtain Eq. (\ref{gvw1}) provided that
$1-\alpha=\alpha-2(\gamma-1)$, i.e., $\alpha=\gamma-{1}/{2}$.
This is the relation announced in the preceding section [see Eq. (\ref{qg12})].
We have therefore found a functional $\Theta^g_Q$ such
that Eq. (\ref{gvw1}) is satisfied when $Q_g$ is the generalized quantum
potential defined by Eq. (\ref{gq13}). Using relation (\ref{qg12}), we
can rewrite
Eq. (\ref{gvw2}) as
\begin{eqnarray}
\label{gvw5}
\Theta_Q^g=C\int \lbrack\nabla (\rho^{\gamma-1/2})\rbrack^2\, d{\bf
r}=\left (\gamma-\frac{1}{2}\right )^2 C
\int \frac{(\nabla
\rho)^2}{\rho^{3-2\gamma}}\, d{\bf r}
\end{eqnarray}
and Eq. (\ref{gvw3}) as
\begin{eqnarray}
\label{gvw6}
\delta \Theta_Q^g=-(2\gamma-1) C \int
\frac{\Delta(\rho^{\gamma-1/2})}{\rho^{3/2-\gamma}}\delta\rho\, d{\bf r}.
\end{eqnarray}
Considering Eqs. (\ref{gq13}) and (\ref{gvw6}), we find that relation
(\ref{gvw1})
is exactly satisfied by taking 
\begin{eqnarray}
\label{gvw7}
C=\frac{\hbar^2}{2m^2}\frac{1}{2\gamma-1}\frac{1}{\rho_0^{2(\gamma-1)}}.
\end{eqnarray}
Therefore, the final expression of the generalized von Weizs\"acker functional
is
\begin{eqnarray}
\label{gvw8}
\Theta_Q^g=\frac{\hbar^2}{2m^2}\frac{\rho_0}{2\gamma-1} \int
\lbrace\nabla \lbrack(\rho/\rho_0)^{\gamma-1/2}\rbrack\rbrace^2\, d{\bf
r}.
\end{eqnarray}
It can be written under the equivalent forms
\begin{eqnarray}
\label{gvw9}
\Theta_Q^g=-\frac{\hbar^2}{2m^2}\frac{\rho_0}{2\gamma-1} \int \left
(\frac{\rho}{\rho_0}\right )^{\gamma-1/2}\Delta \left
\lbrack  \left
(\frac{\rho}{\rho_0}\right )^{\gamma-1/2}  \right\rbrack\, d{\bf
r},
\end{eqnarray}
\begin{eqnarray}
\label{gvw10}
\Theta_Q^g=\frac{1}{2\gamma-1}\int\rho\frac{Q}{m}\, d{\bf r},
\end{eqnarray}
\begin{eqnarray}
\label{gvw11}
\Theta_Q^g=\frac{\hbar^2}{8m^2}\rho_0(2\gamma-1) \int
\frac{\lbrack\nabla (\rho/\rho_0)\rbrack^2}{(\rho/\rho_0)^{3-2\gamma}}\, d{\bf
r}.
\end{eqnarray}
Setting $q=2\gamma-1$, we obtain Eqs. (\ref{gf3})-(\ref{gf6}).

\subsection{Generalized Fisher functional}
\label{sec_gfi}

Let us first consider the standard diffusion equation
\begin{eqnarray}
\label{gfi1}
\frac{\partial\rho}{\partial t}=D\Delta\rho.
\end{eqnarray}
If we compute the rate of change of the Boltzmann entropy (\ref{t1}) and use
Eq. (\ref{gfi1}), we
obtain
\begin{eqnarray}
\label{gfi2}
\dot S_{B}=k_B D S_F,
\end{eqnarray}
where
\begin{eqnarray}
\label{gfi3}
S_F=\frac{1}{m}\int \frac{(\nabla\rho)^2}{\rho}\, d{\bf r}
\end{eqnarray}
is the Fisher entropy \cite{fisher}. We note that the Fisher entropy is
related to the von Weizs\"acker functional (\ref{gf7}) by 
\begin{eqnarray}
\label{gfi4}
\Theta_Q=\frac{\hbar^2}{8m} S_F.
\end{eqnarray}
Therefore, we can write 
\begin{eqnarray}
\label{gfi5}
\dot S_{B}=k_B D \frac{8m}{\hbar^2}\Theta_Q.
\end{eqnarray}
This equation relates the Boltzmann entropy to the von Weizs\"acker  functional.
It provides an intriguing relation
between standard thermodynamics (Boltzmann)
and standard quantum mechanics (Schr\"odinger).

In connection to the  generalized  von Weizs\"acker functional
(\ref{gvw11}), we introduce a generalized Fisher entropy so as to preserve the
relation (\ref{gfi4}). Writing
\begin{eqnarray}
\label{gfi4b}
\Theta_Q^g=\frac{\hbar^2}{8m} S_F^g,
\end{eqnarray}
we obtain a generalized Fisher entropy of the form
\begin{eqnarray}
\label{gfi6}
S_F^g= \frac{\rho_0}{m}(2\gamma-1) \int
\frac{\lbrack\nabla (\rho/\rho_0)\rbrack^2}{(\rho/\rho_0)^{3-2\gamma}}\, d{\bf
r}.
\end{eqnarray}
It reduces to Eq. (\ref{gfi3}) when $\gamma=1$.
Let us now consider the anomalous diffusion equation 
\begin{eqnarray}
\label{gfi7}
\xi\frac{\partial\rho}{\partial t}=K\Delta\rho^{\gamma}.
\end{eqnarray}
If we compute the rate of change of the Tsallis entropy (\ref{t3}) and use
Eqs. (\ref{gfi6}) and (\ref{gfi7}) we obtain
\begin{eqnarray}
\label{gfi8}
\dot S_{\gamma}=
\frac{Km}{\xi}\frac{\gamma^2}{2\gamma-1}\rho_0^{2(\gamma-1)}S^g_{F}.
\end{eqnarray}
Using Eq. (\ref{gfi4b}), we can write
\begin{eqnarray}
\label{gfi9}
\dot
S_{\gamma}=\frac{Km}{\xi}\frac{\gamma^2}{2\gamma-1}\rho_0^{2(\gamma-1)}\frac{8m}
{\hbar^2}\Theta_Q^g.
\end{eqnarray}
This equation relates the Tsallis entropy to the generalized von Weizs\"acker 
functional. It provides an intriguing relation
between generalized thermodynamics (Tsallis)
and the form of generalized quantum mechanics introduced in this paper.

\subsection{Generalized quantum pressure tensor}
\label{sec_gp}

In standard quantum mechanics, the quantum force can be written as the gradient
of a quantum pressure tensor $P_{ij}$ \cite{ggp}. Indeed, we have
\begin{eqnarray}
\label{gp1a}
(F_Q)_i=-\frac{1}{m}\partial_iQ=-\frac{1}{\rho}\partial_jP_{ij}
\end{eqnarray} 
with
\begin{equation}
\label{gp1b}
P_{ij}^{(1)}=-\frac{\hbar^2}{4m^2}\rho\,
\partial_i\partial_j\ln\rho=\frac{\hbar^2}{4m^2}\left (\frac{1}{\rho}
\partial_i\rho\partial_j\rho-\partial_i\partial_j\rho\right )\qquad {\rm
or}\qquad
P_{ij}^{(2)}=\frac{\hbar^2}{4m^2}\left
(\frac{1}{\rho}\partial_i\rho\partial_j\rho-\delta_{ij}\Delta\rho\right ).
\end{equation}
This tensor is manifestly symmetric: $P_{ij}=P_{ji}$. The tensors defined
by Eq. (\ref{gp1b}) are related to each other by
\begin{equation}
\label{gp1c}
P_{ij}^{(1)}=P_{ij}^{(2)}+\frac{\hbar^2}{4m^2}(\delta_{ij}
\Delta\rho-\partial_i\partial_j\rho).
\end{equation}
They differ by a tensor
$\chi_{ij}=\delta_{ij}\Delta\rho-\partial_i\partial_j\rho$ satisfying 
$\partial_j\chi_{ij}=0$. 

We
show below
that Eq. (\ref{gp1a}) remains valid for the generalized quantum force, i.e., we
show that the generalized quantum force can be written as
\begin{eqnarray}
\label{gp1}
(F_Q^g)_i=-\frac{1}{m}\partial_iQ_g=-\frac{1}{\rho}\partial_jP^g_{ij},
\end{eqnarray} 
where $P^g_{ij}$ is a generalized quantum potential tensor. To determine
$P^g_{ij}$, we
first note that the generalized quantum potential (\ref{gq14}) is
proportional
to
\begin{eqnarray}
\label{gp2}
\overline{Q}_g=\frac{\Delta\rho}{\rho^{3-2\gamma}}-\frac{1}{2}
(3-2\gamma)\frac{(\nabla\rho)^2}{\rho^{4-2\gamma}}
\end{eqnarray}
so that its gradient is proportional to
\begin{eqnarray}
\label{gp3}
-\partial_i\overline{Q}_g=(3-2\gamma)\frac{1}{\rho^{4-2\gamma}}\partial_i\rho
\Delta\rho+(3-2\gamma)\frac{1}{\rho^{4-2\gamma}}\partial_{ij}\rho
\partial_j\rho-(3-2\gamma)(2-\gamma)\frac{1}{\rho^{5-2\gamma}}\partial_{i}\rho
(\nabla\rho)^2-\frac{1}{\rho^{3-2\gamma}}\partial_{i}\Delta\rho.
\end{eqnarray}
We want to find a tensor $\overline{P}_{ij}$ such that
\begin{eqnarray}
\label{gp4}
-k\partial_i\overline{Q}_g=\frac{1}{\rho}\partial_j\overline{P}^g_{ij},
\end{eqnarray}
where $k$ is a constant. To find the most general form of 
$\overline{P}^g_{ij}$, we consider an expression of the form
\begin{eqnarray}
\label{gp5}
\overline{P}^g_{ij}=\frac{1}{\rho^{3-2\gamma}}
\partial_i\rho\partial_j\rho-A\frac{1}{\rho^{2-2\gamma}}\delta_{ij}\Delta\rho-B
\frac{1}{\rho^{2-2\gamma}}\partial_{ij}\rho+C\frac{1}{\rho^{3-2\gamma}}
(\nabla\rho)^2\delta_{ij},
\end{eqnarray}
where $A$, $B$ and $C$ are some constants. We have
\begin{eqnarray}
\label{gp6}
\frac{1}{\rho}\partial_j\overline{P}^g_{ij}=\frac{1}{\rho^{4-2\gamma}}
\partial_i\rho\Delta\rho\left\lbrack
1+2A(1-\gamma)\right\rbrack +\frac{1}{\rho^{4-2\gamma}}
\partial_{ij}\rho\partial_j\rho \left\lbrack
1+2B(1-\gamma)+2C\right\rbrack\nonumber\\
-\frac{1}{\rho^{5-2\gamma}}
\partial_{i}\rho (\nabla\rho)^2 (3-2\gamma)(1+C)-\frac{1}{\rho^{3-2\gamma}}
\partial_{i}\Delta \rho (A+B).
\end{eqnarray}
Substituting Eqs. (\ref{gp3}) and (\ref{gp6}) into Eq. (\ref{gp4}), we
obtain
the system of equations
\begin{eqnarray}
\label{gp7}
k(3-2\gamma)=1+2A(1-\gamma),
\end{eqnarray}
\begin{eqnarray}
\label{gp8}
k(3-2\gamma)=1+2B(1-\gamma)+2C,
\end{eqnarray}
\begin{eqnarray}
\label{gp9}
k(3-2\gamma)(2-\gamma)=(3-2\gamma)(1+C),
\end{eqnarray}
\begin{eqnarray}
\label{gp10}
k=A+B.
\end{eqnarray}
Coming back to the original variables, we find that the generalized quantum
potential
(\ref{gq13}) is
related to $\overline{Q}_g$ by
\begin{eqnarray}
\label{gp14}
Q_g=-\frac{\hbar^2}{4m}(2\gamma-1)\rho_0^{2-2\gamma}\overline{Q}_g.
\end{eqnarray}
According to Eqs. (\ref{gp1}), (\ref{gp4}) and (\ref{gp14}), the
generalized pressure tensor is given by
\begin{eqnarray}
\label{gp15}
P_{ij}^g=\frac{\hbar^2}{4m^2}(2\gamma-1)\frac{1}{k}\rho_0^{2-2\gamma}
\overline{P}^g_{ ij},
\end{eqnarray}
where $\overline{P}^g_{ ij}$ is given by Eq. (\ref{gp5}) with the
coefficients  $A$, $B$, $C$ and $k$ determined by Eqs.
(\ref{gp7})-(\ref{gp10}).

For $\gamma=1$, Eqs. (\ref{gp7})-(\ref{gp10}) reduce to
\begin{eqnarray}
\label{gp11}
k=1, \qquad C=0,\qquad  A+B=1.
\end{eqnarray}
We note that there is a freedom since $A$ and $B$ are not individually
determined. Only their sum is fixed to unity. As a result, the general
expression of the standard quantum pressure tensor is
\begin{eqnarray}
\label{gp12}
P_{ij}=\frac{\hbar^2}{4m^2}
\left\lbrack \frac { 1 } { \rho}
\partial_i\rho\partial_j\rho-A\delta_{ij}\Delta\rho-(1-A)\partial_{ij}\rho
\right\rbrack.
\end{eqnarray}
It returns the usual expressions $P_{ij}^{(1)}$ and
$P_{ij}^{(2)}$ given by Eq. (\ref{gp1b}) when we take $A=0$
or $A=1$ respectively. We
can also take $A=B=1/2$ leading to
\begin{eqnarray}
\label{gp12b}
P_{ij}=\frac{\hbar^2}{4m^2}
\left\lbrack \frac { 1 } { \rho}
\partial_i\rho\partial_j\rho-\frac{1}{2}\delta_{ij}\Delta\rho-\frac{1}{
2}\partial_ { ij } \rho
\right\rbrack.
\end{eqnarray}

For
$\gamma\neq 1$, the solution of Eqs.
(\ref{gp7})-(\ref{gp10}) is
\begin{eqnarray}
\label{gp13}
A=\frac{1-k(3-2\gamma)}{2(\gamma-1)}, \qquad B=\frac{k-1}{2(\gamma-1)}, \qquad
C=k(2-\gamma)-1.
\end{eqnarray}
Therefore, the generalized pressure tensor is given by
\begin{eqnarray}
\label{gp16}
P_{ij}=\frac{\hbar^2}{4m^2}(2\gamma-1)\frac{1}{k}\rho_0^{2-2\gamma}
\left\lbrack \frac { 1 } { \rho^{3-2\gamma}}
\partial_i\rho\partial_j\rho-A\frac{1}{\rho^{2-2\gamma}}\delta_{ij}\Delta\rho-B
\frac{1}{\rho^{2-2\gamma}}\partial_{ij}\rho+C\frac{1}{\rho^{3-2\gamma}}
(\nabla\rho)^2\delta_{ij}\right\rbrack.
\end{eqnarray}
We note that $C\neq 0$ in general. This brings a new term in the
generalized quantum pressure tensor that is absent in the expression of the
standard quantum
pressure tensor (\ref{gp12}) when $\gamma=1$. On the other hand, we are free to
take $k$ as we please
provided that $k=1$ when $\gamma=1$. Let us consider particular cases.

(i) For $k=1$ we get $A=1$, $B=0$ and $C=1-\gamma$ yielding
\begin{eqnarray}
\label{gp17}
P_{ij}=\frac{\hbar^2}{4m^2}(2\gamma-1)\rho_0^{2-2\gamma}
\left\lbrack \frac { 1 } { \rho^{3-2\gamma}}
\partial_i\rho\partial_j\rho-\frac{1}{\rho^{2-2\gamma}}\delta_{ij}
\Delta\rho+(1-\gamma)\frac{1}{\rho^{3-2\gamma}}
(\nabla\rho)^2\delta_{ij}\right\rbrack.
\end{eqnarray}
When $\gamma=1$ this returns the expression $P_{ij}^{(2)}$ from
Eq. (\ref{gp1b}).

(ii) For $k=1/(3-2\gamma)$ we get $A=0$, $B=1/(3-2\gamma)$ and
$C=(\gamma-1)/(3-2\gamma)$ yielding
\begin{eqnarray}
\label{gp18}
P_{ij}=\frac{\hbar^2}{4m^2}\frac{2\gamma-1}{3-2\gamma}\rho_0^{
2-2\gamma }
\left\lbrack \frac { 1 } { \rho^{3-2\gamma}}
\partial_i\rho\partial_j\rho-\frac{1}{3-2\gamma}
\frac{1}{\rho^{2-2\gamma}}\partial_{ij}\rho+\frac{\gamma-1}{3-2\gamma}\frac{1}{
\rho^ { 3-2\gamma } }
(\nabla\rho)^2\delta_{ij}\right\rbrack.
\end{eqnarray}
When $\gamma=1$ this returns the expression $P_{ij}^{(1)}$ from Eq.
(\ref{gp1b}).

(iii) For $k=1/(2-\gamma)$ we get $A=B=1/[2(2-\gamma)]$ and
$C=0$ yielding
\begin{eqnarray}
\label{gp19}
P_{ij}=\frac{\hbar^2}{4m^2}\frac{2\gamma-1}{2-\gamma}\rho_0^{
2-2\gamma }
\left\lbrack \frac { 1 } { \rho^{3-2\gamma}}
\partial_i\rho\partial_j\rho-\frac{1}{2(2-\gamma)}\frac{1}{\rho^{2-2\gamma}}
\delta_ { ij } \Delta\rho-\frac{1}{2(2-\gamma)}
\frac{1}{\rho^{2-2\gamma}}\partial_{ij}\rho
\right\rbrack.
\end{eqnarray}
When $\gamma=1$ this returns the expression from Eq.
(\ref{gp12b}).

\end{document}